\documentclass[fleqn,usenatbib]{mnras}

\usepackage{newtxtext,newtxmath}
\usepackage[T1]{fontenc}
\DeclareRobustCommand{\VAN}[3]{#2}
\let\VANthebibliography\thebibliography
\def\thebibliography{\DeclareRobustCommand{\VAN}[3]{##3}\VANthebibliography}

\usepackage{booktabs, multirow, makecell}
\usepackage{enumitem}
\usepackage{xcolor}
\usepackage{graphicx}	% Including figure files
\usepackage{amsmath}	% Advanced maths commands
\usepackage{threeparttable}
\title[Improved 21-cm upper limits with NenuFAR]{Improved upper limits on the 21-cm signal power spectrum at $z=17.0$ and $z=20.3$ from an optimal field observed with NenuFAR}

\author[S. Munshi et al.]{
S. Munshi,$^{1}$\thanks{E-mail: munshi@astro.rug.nl (SM)}
 F. G. Mertens,$^{2,1}$ 
 J. K. Chege,$^{1,3}$ 
 L. V. E. Koopmans,$^{1}$ 
 A. R. Offringa,$^{3}$ 
 B. Semelin,$^{2}$
\newauthor
 R. Barkana,$^{4}$ 
 J. Dhandha,$^{5}$ 
 A. Fialkov,$^{5,6}$ 
 R. M\'{e}riot,$^{2,7}$ 
 S. Sikder,$^{4}$ 
 A. Bracco,$^{8,9}$ 
 S. A. Brackenhoff,$^{1}$ 
\newauthor
 E. Ceccotti,$^{1,10}$ 
 R. Ghara,$^{11}$ 
 S. Ghosh,$^{1}$ 
 I. Hothi,$^{9}$ 
 M. Mevius,$^{3}$ 
 P. Ocvirk,$^{12}$ 
 A. K. Shaw,$^{13}$ 
 S. Yatawatta,$^{3}$
\newauthor
and
 P. Zarka$^{14,15}$
\\
$^{1}$Kapteyn Astronomical Institute, University of Groningen, P.O. Box 800, 9700 AV Groningen, The Netherlands\\ 
 $^{2}$LUX, Observatoire de Paris, PSL Research University, CNRS, Sorbonne Université, F-75014 Paris, France\\ 
 $^{3}$ASTRON, PO Box 2, 7990 AA Dwingeloo, The Netherlands\\ 
 $^{4}$School of Physics and Astronomy, Tel Aviv University, Tel Aviv, 69978, Israel\\ 
 $^{5}$Institute of Astronomy, University of Cambridge, Madingley Road, Cambridge CB3 0HA, UK\\ 
 $^{6}$Kavli Institute for Cosmology, Madingley Road, Cambridge CB3 0HA, UK\\ 
 $^{7}$Blackett Laboratory, Imperial College London, Prince Consort Road, London, SW7 2AZ, UK\\ 
 $^{8}$INAF -- Osservatorio Astrofisico di Arcetri, Largo E. Fermi 5, 50125 Firenze, Italy\\ 
 $^{9}$Laboratoire de Physique de l'Ecole Normale Sup\'erieure, ENS, Universit\'e PSL, CNRS, Sorbonne Universit\'e, Universit\'e de Paris, F-75005 Paris, France\\ 
 $^{10}$INAF -- Istituto di Radioastronomia, Via P.~Gobetti 101, 40129 Bologna, Italy\\ 
 $^{11}$Department of Physical Sciences, Indian Institute of Science Education and Research Kolkata, Mohanpur, WB 741 246, India\\ 
 $^{12}$Universit\'e de Strasbourg, CNRS, Observatoire astronomique de Strasbourg (ObAS), 11 rue de l'Universit\'e, Strasbourg, France\\ 
 $^{13}$Department of Computer Science, University of Nevada Las Vegas, 4505 S. Maryland Pkwy., Las Vegas, NV 89154, USA\\ 
 $^{14}$ORN, Observatoire de Paris, Université PSL, Univ Orléans, CNRS, 18330 Nançay, France\\ 
 $^{15}$LIRA, Observatoire de Paris, Université PSL, CNRS, Sorbonne Université, Université Paris Cité, 5 place Jules Janssen, 92195 Meudon, France
}

\date{Accepted XXX. Received YYY; in original form ZZZ}

\pubyear{\the\year{}}

\begin{document}
\label{firstpage}
\pagerange{\pageref{firstpage}--\pageref{lastpage}}
\maketitle

\begin{abstract}
We report the deepest upper limits to date on the 21-cm signal power spectrum during the Cosmic Dawn (redshifts: $z>15$), using four nights of observations with NenuFAR. The limits are derived from two redshift bins, centred at $z=20.3$ and $z=17.0$, with integration times of 26.1 h and 23.6 h, from observations of an optimal target field chosen to minimise sidelobe leakage from bright sources. Our analysis incorporates improvements to the data processing pipeline, particularly in subtracting strong radio sources in the primary beam sidelobes and mitigating low-level radio frequency interference, yielding a 50-fold reduction in the excess variance compared to a previous analysis of the north celestial pole field. At $z=20.3$, we achieve a best $2\sigma$ upper limit of $\Delta^{2}_{21}<4.6 \times 10^5 \, \textrm{mK}^{2}$ at $k=0.038$ $h\, \mathrm{cMpc}^{-1}$, while at $z=17.0$, the best limit is $\Delta^{2}_{21}<5.0 \times 10^6 \, \textrm{mK}^{2}$ at $k=0.041$ $h\, \mathrm{cMpc}^{-1}$. These are the strongest constraints on the 21-cm power spectrum at the respective redshifts, with the limit at $z = 20.3$ being deeper by more than an order of magnitude over all previous Cosmic Dawn power spectrum limits. Comparison against simulated exotic 21-cm signals shows that while the $z=20.3$ limits begin to exclude the most extreme models predicting signals stronger than the EDGES detection, an order-of-magnitude improvement would constrain signals compatible with EDGES. A coherence analysis reveals that the excess variance is largely incoherent across nights for the $z=20.3$ redshift bin, suggesting that deeper integrations could yield significantly stronger constraints on the 21-cm signal from the Cosmic Dawn.
\end{abstract}


%%%%%%%%%%%%%%%%%%%%%%%%%%%%%%%%%%%%%%%%%%%%%%%%%%

%%%%%%%%%%%%%%%%% BODY OF PAPER %%%%%%%%%%%%%%%%%%
\defcitealias{munshi2024first}{M24}

\section{Introduction}
The 21-cm line of neutral hydrogen has emerged as one of the most powerful probes of the astrophysics and cosmology of the early Universe. The emerging field of 21-cm cosmology aims to detect the redshifted 21-cm signal from the early Universe, seen in absorption or emission against the background radiation, the Cosmic Microwave Background (CMB). As the first luminous sources formed during the Cosmic Dawn and subsequently reionised the intergalactic medium during the Epoch of Reionisation (EoR), the astrophysical processes during these periods imprinted distinct signatures on the 21-cm brightness temperature distribution. Detection of these signatures thus offers a potential for direct experimental verification of our theories of early star formation and galaxy evolution \citep[e.g.,][]{furlanetto2006cosmology,pritchard201221}. Over the last decade, two complementary approaches to detecting the 21-cm signal from the early universe have gained momentum: global and interferometric 21-cm experiments. 

Global 21-cm experiments measure the sky-averaged 21-cm brightness temperature as a function of redshift. The claimed detection of an absorption trough in the global 21-cm signal during the Cosmic Dawn by EDGES\footnote{Experiment to Detect the Global EoR Signature} \citep{bowman2018absorption} has generated considerable interest in this epoch. The feature detected by EDGES has an unusual shape and is more than twice as deep as the absorption trough that is expected during the Cosmic Dawn based on standard models, which are incompatible with the detection. It thus becomes necessary to invoke exotic non-standard models of the Cosmic Dawn to explain the EDGES detection. These non-standard models fall under two categories. The first class of models proposes additional components to the background temperature, effectively boosting the contrast between the 21-cm signal and the background, which could result in the observed absorption trough \citep[e.g.,][]{feng2018enhanced,ewall2018modeling,dowell2018radio,fialkov2019signature}. Another class of models proposes that alternative mechanisms such as baryon-dark-matter interaction could lead to super-cooling of the gas, lowering the spin temperature and enhancing the contrast against the background radiation, thus producing an unusually deep absorption trough \citep[e.g.,][]{barkana2018possible,fialkov2018constraining,berlin2018severely,munoz2018small,liu2019reviving}. There have been no independent measurements with other instruments that have confirmed the EDGES detection. On the contrary, results from the SARAS\footnote{Shaped Antenna measurement of the background RAdio Spectrum} experiment reject the EDGES detection with 95\% confidence \citep{singh2022detection}. Several other global 21-cm experiments, such as BIGHORNS\footnote{Broadband Instrument for Global HydrOgen ReioNisation Signal} \citep{sokolowski2015bighorns}, PRIZM\footnote{Probing Radio Intensity at High-Z from Marion} \citep{philip2019probing}, REACH\footnote{Radio Experiment for the Analysis of Cosmic Hydrogen} \citep{de2022reach}, MIST\footnote{Mapper of the IGM spin temperature} \citep{monsalve2024mapper}, and RHINO\footnote{Remote HI eNvironment Observer} \citep{bull2024rhino}, aim to settle this debate, with many of them having already been deployed and recording data.

In contrast to the sky-averaged signal probed by global experiments, interferometers probe the three-dimensional spatial fluctuations in the 21-cm signal brightness temperature distribution and can potentially provide us with a wealth of information about the early universe. However, the signal-to-noise ratio (S/N) requirement for the detection of the 21-cm signal with interferometers is even more demanding than global 21-cm experiments, owing to the faint nature of the signal compared to instrumental thermal noise sensitivities reached in reasonable observation hours using the most sensitive currently operating radio interferometers. Thus, the current generation of experiments attempts to make a statistical detection of the 21-cm fluctuations by estimating its power spectrum. The upcoming SKA\footnote{Square Kilometre Array} is expected to have the thermal noise sensitivity to not only detect the high-$z$ 21-cm power spectrum, but also to generate tomographic maps of the EoR in reasonable observing hours \citep{koopmans2015cosmic,bonaldi2025square}. Most interferometric experiments aiming to detect the high-redshift 21-cm signal have focused on the EoR, where they probe the distribution and evolution of ionised bubbles around the first stars and galaxies that eventually reionised the Universe. These efforts have been led by interferometers such as the GMRT\footnote{Giant Metrewave Radio Telescope} \citep{paciga2013simulation}, LOFAR\footnote{Low-Frequency Array} \citep{patil2017upper,mertens2020improved,mertens2025deeper,ceccotti2025first}, PAPER\footnote{Precision Array to Probe EoR} \citep{kolopanis2019simplified}, MWA\footnote{Murchison Widefield Array} \citep{barry2019improving,li2019first,trott2020deep,nunhokee2025limits}, and HERA\footnote{Hydrogen Epoch of Reionization Array} \citep{abdurashidova2022first,adams2023improved}. Though there has not been a detection of the signal yet, these experiments have progressively set stronger upper limits on the 21-cm power spectrum, enabling some scenarios of reionisation to be ruled out \citep{ghara2020constraining,greig2021exploring,abdurashidova2022hera,adams2023improved,ghara2025constraints}.

Detection of the 21-cm signal power spectrum from the Cosmic Dawn ($z>15$) is even more challenging than the EoR, primarily due to the higher system noise of instruments at these low radio frequencies, contributed mainly by the significantly higher sky temperature. The astrophysical foreground emission is much stronger at these low radio frequencies, owing to the power law spectral shape of synchrotron emission, imposing extremely high requirements for the accurate subtraction of foregrounds to reach the instrument's thermal noise sensitivity. Additionally, the ionosphere and wide-field nature of the instruments present stronger data processing challenges compared to higher frequencies. Instruments that have attempted detection of the Cosmic Dawn 21-cm signal power spectrum include the MWA \citep{ewall2016first,yoshiura2021new}, LOFAR \citep{gehlot2019first}, AARTFAAC\footnote{Amsterdam ASTRON Radio Transients Facility And Analysis Center} \citep{gehlot2020aartfaac}, OVRO-LWA\footnote{Owens Valley Radio Observatory - Long Wavelength Array} \citep{eastwood201921,garsden202121}, and NenuFAR\footnote{New Extension in Nan\c cay Upgrading LOFAR} \citep{munshi2024first}. Currently, these power spectrum upper limits are a few orders of magnitude above the levels expected from the standard 21-cm models of the Cosmic Dawn. However, the exotic non-standard models of the Cosmic Dawn that have emerged to explain the EDGES detection also predict the fluctuations of the 21-cm brightness temperature to be much stronger than those expected from standard models \citep{fialkov2018constraining,feng2018enhanced}, effectively boosting the corresponding power spectrum. Still, the strongest power spectrum upper limits on the 21-cm signal during the Cosmic Dawn are at least an order of magnitude beyond that expected from the most extreme exotic models at $z>15$. Deep power spectrum limits on the 21-cm signal during the Cosmic Dawn can thus observationally confirm or rule out scenarios of the exotic physics suggested by the EDGES detection through an independent experimental approach.

Besides the challenging S/N requirements, the strong foregrounds impose steep technical requirements in instrument calibration necessary for a detection. These challenges include contamination from radio frequency interference (RFI), ionospheric scintillation, modelling of far-field sources, and accurate estimation of the instrumental primary beam. In HERA, for example, cross-talk between feeds in the closely packed redundant array layout has proved to be one of the current limiting factors \citep{abdurashidova2022first,adams2023improved}, which is mitigated to some extent using fringe-rate filtering \citep{kern2019mitigating,kern2020mitigating,charles2023use,charles2024mitigating,garsden2024demonstration}. LOFAR, MWA, and NenuFAR upper limits, on the other hand, suffer from excess noise above the theoretical noise sensitivity of the instrument \citep{mertens2020improved,trott2020deep,munshi2024first}. For LOFAR, inaccurate modelling of bright sources \citep{ceccotti2025spectral}, the turbulent ionosphere \citep{brackenhoff2024ionospheric} and incomplete sky models used in calibration \citep{hofer2025impact} have been ruled out as the main contributors to the excess variance. Transient RFI from aeroplanes has also been shown to currently have a negligible contribution to the power spectra estimated using the widefield instrument AARTFAAC \citep{gehlot2024transient}. As the calibration and analysis pipelines of the instruments continue to improve, several recent studies \citep{gan2022statistical,munshi2024first,chokshi2024necessity,brackenhoff2025robust} point to the inaccurate modelling of the instrumental primary beam near nulls and the leakage of power from bright sources via their point spread function (PSF) sidelobes as one of the major limiting factors in reaching the thermal noise limit and exploiting the full sensitivity of instruments.

NenuFAR, a new low-frequency radio interferometer in France, is one of the most sensitive instruments to the Cosmic Dawn 21-cm signal fluctuations at large scales, with the sensitivity to detect the predicted exotic model power spectra in about 100 observing hours. The first analysis of NenuFAR data obtained on the north celestial pole (NCP) field by \cite{munshi2024first}, hereafter \citetalias{munshi2024first}, produced upper limits on the 21-cm signal power spectrum at $z=20.3$, the best limit being $\Delta^{2}_{21} < 2.4 \times 10^7 \, \textrm{mK}^{2}$ at $k=0.041$ $h\, \mathrm{cMpc}^{-1}$. However, these limits were over two orders of magnitude above the thermal noise power spectrum due to excess variance contributed primarily by distant off-axis sky sources and local sources of RFI. Through a combination of analytical modelling and forward simulations, two separate studies by \cite{munshi2025beyond} and \cite{munshi2025near} demonstrated the impact of wide-field sources and local RFI sources, respectively, on the power spectrum estimated with phase tracking instruments and offered strategies to mitigate their impact. 

In this paper, we analyse multiple nights of NenuFAR observations of an optimal target field at two redshift bins centred at $z=17.0$ and $z=20.3$, and implement improvements to the analysis pipeline that decrease the excess variance in the data caused by these two contaminating effects. We report $2\sigma$ upper limits on the 21-cm signal power spectrum at both redshift bins and compare the resulting limits against the power spectra predicted by exotic models of the Cosmic Dawn 21-cm signal. The paper is organised as follows: Section~\ref{sec:ksp} introduces the instrument and the observation strategy. Section~\ref{sec:calibration} describes the preprocessing and calibration-based sky-model subtraction. The power spectrum estimation is described in Section~\ref{sec:ps_estimation}, and the residual foreground removal and robustness tests are described in Section~\ref{sec:gpr}. We present our results in Section~\ref{sec:results} and discuss their implications and future prospects in Section~\ref{sec:discussion}. We summarise the main conclusions from this analysis in Section~\ref{sec:conclusion}. In this paper, we have used a flat $\Lambda$CDM cosmology, with the cosmological parameters ($\mathrm{H}_0=67.7$, $\Omega_m=0.307$) obtained by \cite{planck2016planck}.

\section{NenuFAR Cosmic Dawn observations}\label{sec:ksp}
One of the main drivers of NenuFAR is its Cosmic Dawn Key Science Program (CD KSP), which aims to detect the redshifted 21-cm signal power spectrum from the Cosmic Dawn. NenuFAR operates at very low frequencies ($10-85$ MHz), and the CD KSP focuses on the $40-85$ MHz frequency range corresponding to a redshift range of $z=15-31$ \citep{2021sf2a.conf..211M}.

\subsection{Instrument and observation status}
\begin{figure}
	\includegraphics[width=\columnwidth]{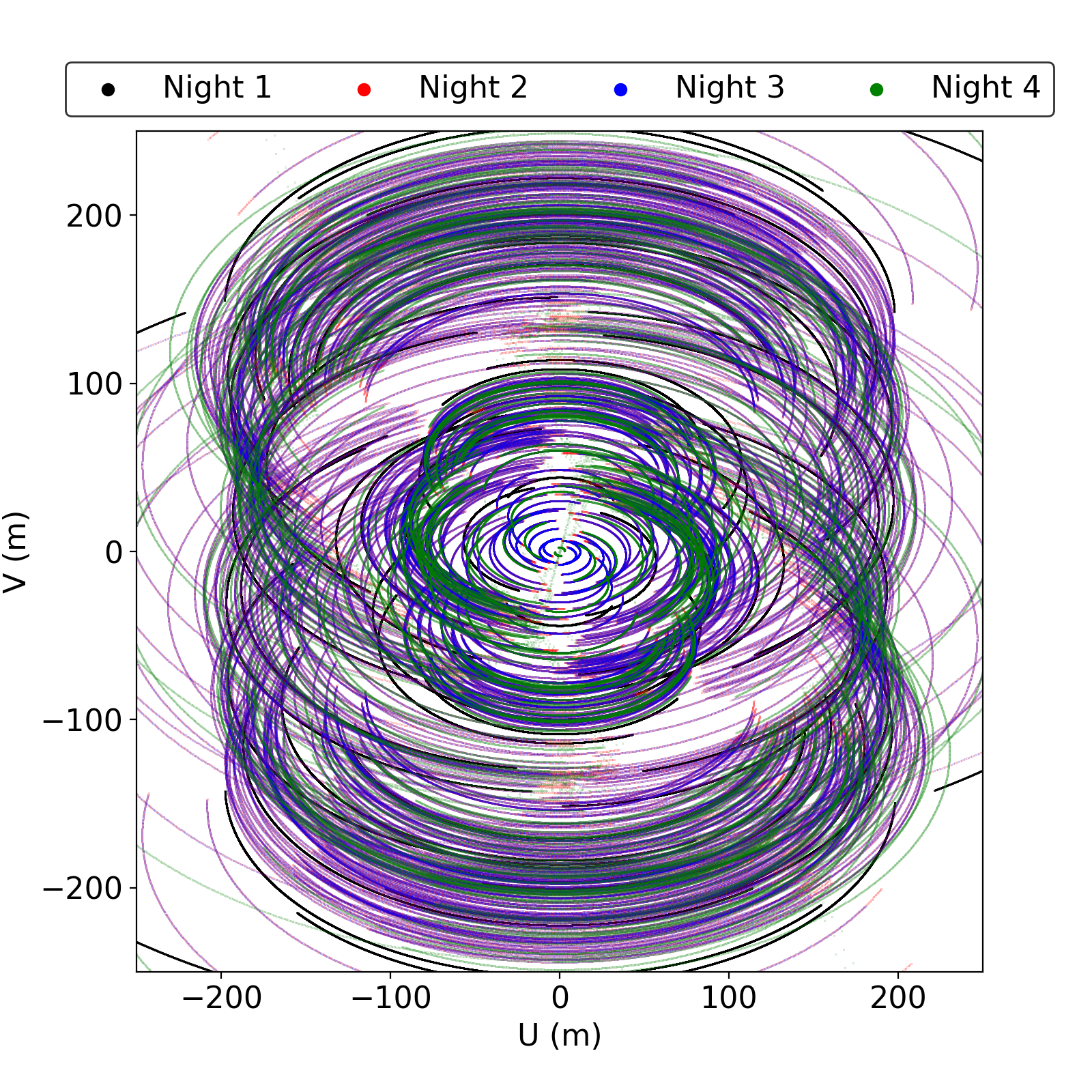}
    \caption{The baseline coverage of observations of the NT04 field. The different colours correspond to the four different nights of observation.}
    \label{fig:uvplot}
\end{figure}

NenuFAR is composed of inverted V-shaped dual-polarised antennas that are grouped into analogue steerable stations composed of 19 antennas in a hexagonal configuration, called Mini-Arrays (MAs hereafter). Most of the MAs are distributed within a $400\,$m diameter dense core, with a few remote MAs located at distances of a few kilometres from the core. We refer the reader to \cite{zarka2012lss,zarka2015nenufar,zarka2020low} for detailed descriptions of the instrument. The observations analysed in this paper used a maximum of 78 core MAs (out of the planned 96) and 4 remote MAs (out of the planned 8). The defining feature of NenuFAR for Cosmic Dawn studies is its extremely dense core, which provides a very high collecting area and filling factor. Fig.~\ref{fig:uvplot} shows the baseline coverage of the four nights of observation used in this analysis and illustrates the dense sampling at small $u\varv$ by the baselines formed by pairs of core stations. Owing to this large collecting area, NenuFAR is one of the most sensitive instruments to the fluctuations in the 21-cm signal at these frequencies and the most sensitive instrument below 50 MHz, where both HERA and the upcoming SKA will not observe. The KSP has already garnered over 2000 hours of observations across multiple target fields, focusing on two primary deep fields: the NCP field and the NT04 (RA = $7\,$h 20 min, Dec = $35^{\circ}$) field.
\begin{table}
\centering
\caption{Observational parameters of the four nights of observation of the NT04 field used in this analysis.}
\label{tab:obs_params}
\begin{tabular}{@{}l ll cc cc cc@{}}
\toprule
\multirow{2}{*}{Id} & \multirow{2}{*}{Start date} & \multicolumn{2}{c}{Calibrator} & \multicolumn{2}{c}{Target} & \multicolumn{2}{c}{RFI \% (target)} \\  
\cmidrule(lr){3-4} \cmidrule(lr){5-6} \cmidrule(lr){7-8}
                          &                      & Start & Span & Start & Span & Z20 & Z17 \\  
                          &                      & (UTC) & (h) & (UTC) & (h) &  &  \\  
\midrule
1                         & 2023-12-08           & 21:00            & 0.5             & 21:30            & 9.5          &   1.6        &     10.7      \\
2                         & 2023-12-29           & 19:00            & 0.5             & 19:30            & 11.5         &   1.7        &     10.7      \\
3                         & 2024-01-29           & 17:00            & 0.5             & 17:30            & 10.5         &   1.6        &      9.9     \\
4                         & 2024-03-02           & 17:00            & 0.5             & 17:30            & 8.5          &   1.6        &    10.0       \\
\bottomrule
\end{tabular}
\end{table}

\begin{figure*}
    \centering
    \begin{minipage}{0.7\textwidth}
        \includegraphics[width=\linewidth]{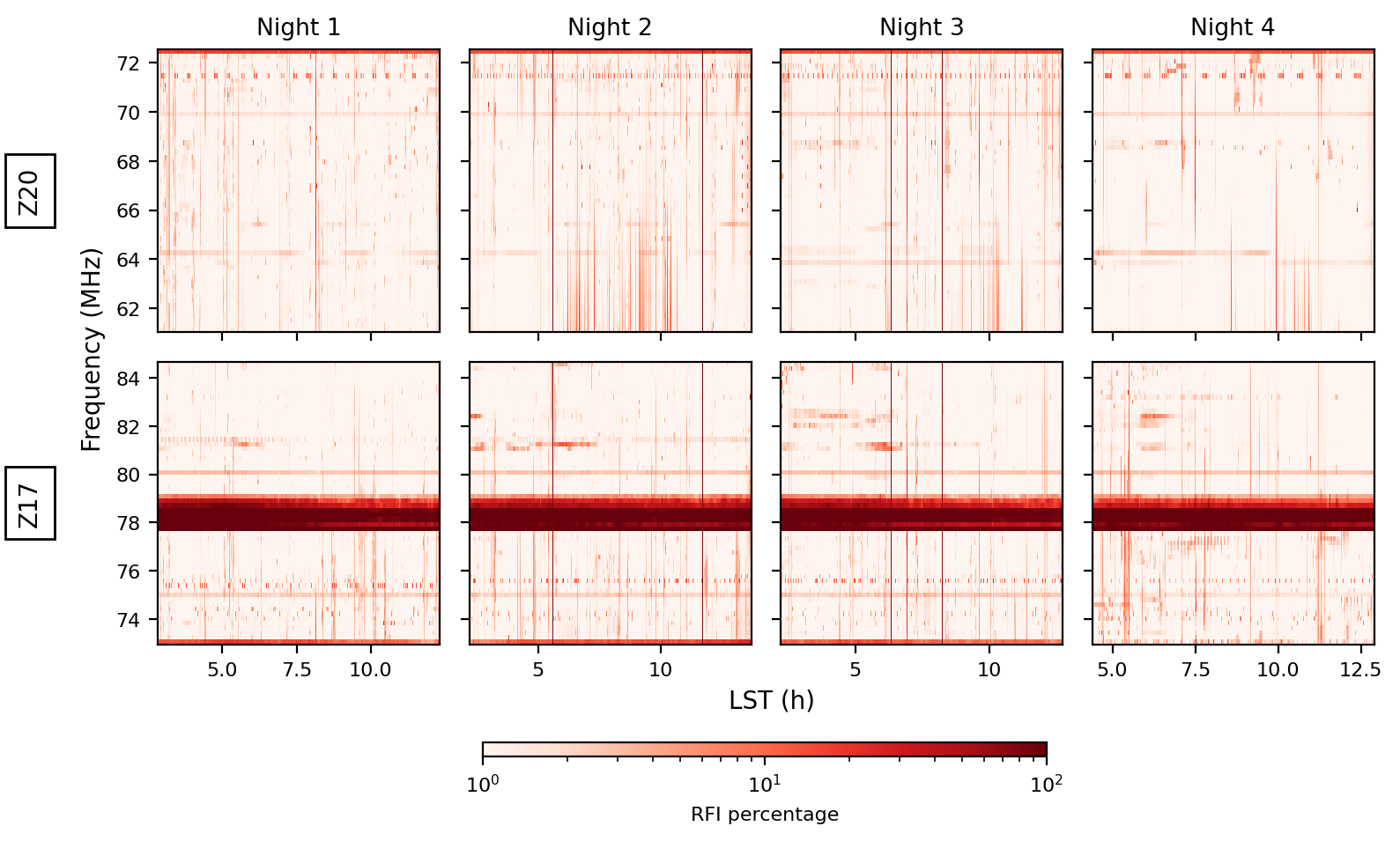}
    \end{minipage}\hfill
    \begin{minipage}{0.3\textwidth}
        \includegraphics[width=\linewidth]{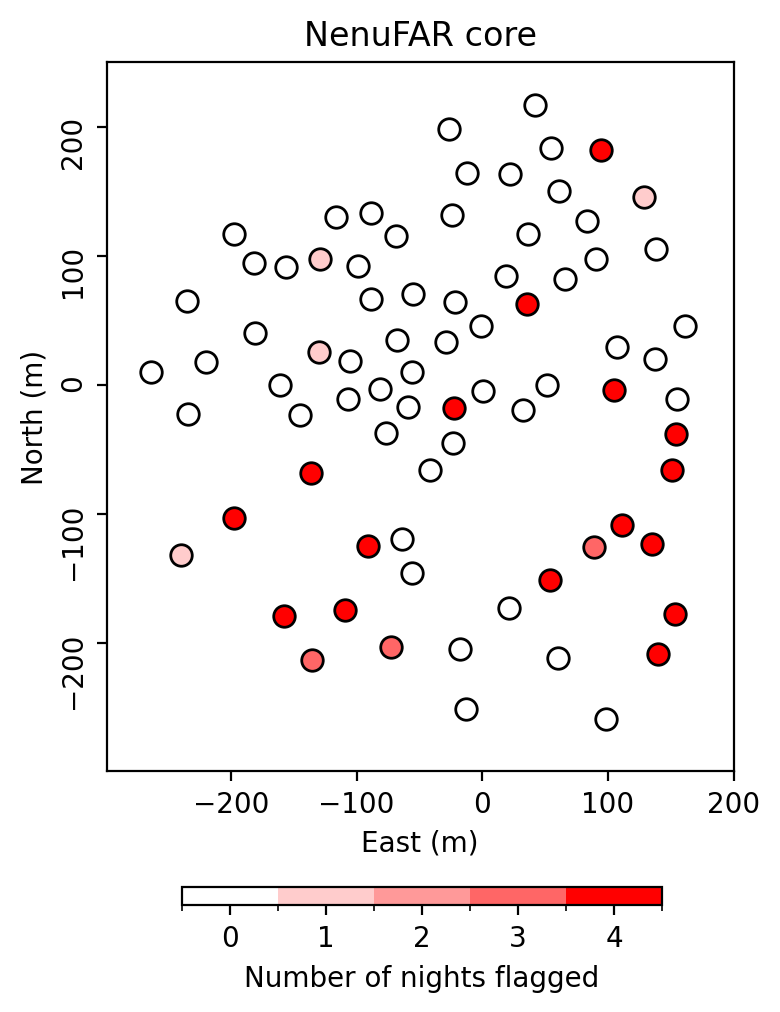}
    \end{minipage}
    \caption{RFI flagging statistics at different stages of processing. Left: The flagged RFI percentage in the highest time and frequency resolution correlated data for the two spectral windows (rows) for the four nights (columns). Right: A scatter plot of NenuFAR MA locations showing the number of nights each MA has been identified and flagged (in Z20) based on bad solutions in the A-team subtraction step.}
    \label{fig:rfi_stats}
\end{figure*}
\subsection{Field and night selection}
\citetalias{munshi2024first} showed that, after foreground subtraction, the residual data from NenuFAR observations of the NCP field are still dominated by a strong excess variance above the thermal noise power. This excess variance is contributed primarily by residuals from strong off-axis sources picked up by NenuFAR's primary beam grating lobes and local sources of RFI near the NenuFAR core. Based on these results, we initiated a survey of six fields with NenuFAR (Mertens et al. in prep.), to identify optimal target fields where foreground contamination is minimised. The selection of the fields was performed based on two sets of criteria. The major criteria constitute the distance from bright A-team sources, the apparent flux from these sources as seen through the primary beam sidelobes, low emission from the Galactic plane, and phase centre transit close to the zenith. The minor selection criteria constitute the presence of a calibrator source within the field and year-round observability. The power spectra estimated after subtraction of the brightest sources were used as a metric to identify the NT04 field as the optimal target field for 21-cm studies with NenuFAR. We discuss the field selection survey in detail in a forthcoming paper by Mertens et al. (in prep). We have now initiated a deep survey of this field with NenuFAR, with over 600 hours of observation approved over Cycles 3 to 6 of NenuFAR observations, out of which more than 500 hours of observation have already been collected. 

In this paper, we select four nights from observations of the NT04 field that were available from Cycle 3 of NenuFAR observations (December 2023 to June 2024) based on superior RFI statistics. The left panel of Fig.~\ref{fig:rfi_stats} shows the RFI statistics of the four selected nights. All four observations of the target field were preceded by a $30\,$min observation of the calibrator source Cassiopeia A (Cas A). In this analysis, we focus on the higher frequencies of $61-85$ MHz, which we divide into two spectral windows: $61-72.5$ MHz ($z=20.3$, Z20 hereafter) and $73-84.5$ MHz ($z=17.0$, Z17 hereafter), which are analysed independently. This is done to minimise the light cone effect caused by the evolution of the signal within the redshift range being analysed \citep{datta2012light,datta2014light}. The 0.5 MHz band between Z20 and Z17 is not used in order to avoid known RFI. The observational details of the four selected nights are summarised in Table~\ref{tab:obs_params}. The baseline coverage of the four nights of observation shown in Fig.~\ref{fig:uvplot} illustrates how combining the four nights not only improves the sensitivity by sampling the same modes multiple times but also improves the sampling of the $u\varv$ plane.

\section{Calibration and sky-model subtraction}\label{sec:calibration}
\begin{figure*}
\includegraphics[width=2\columnwidth]{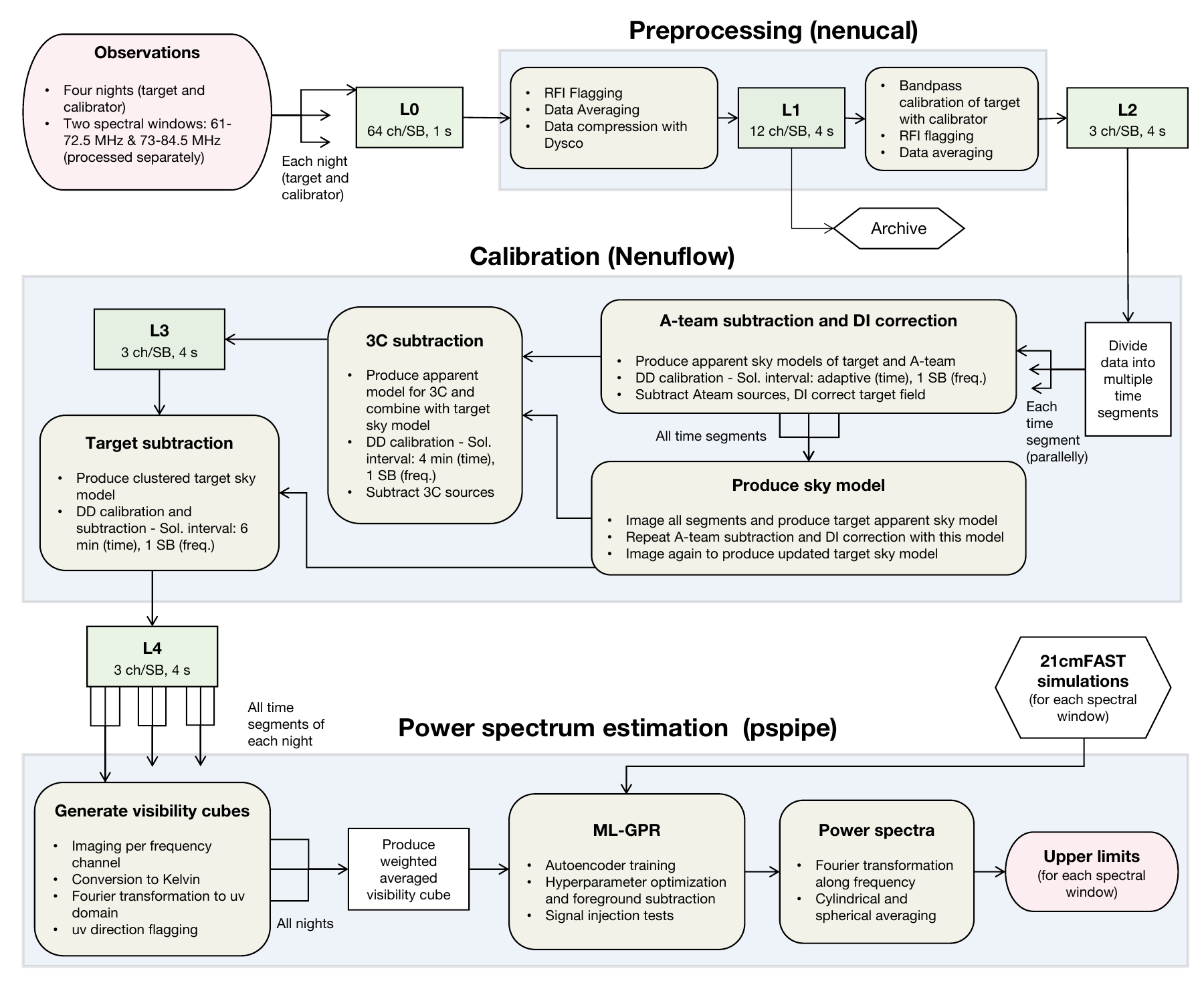}
    \caption{A flowchart describing the NenuFAR 21-cm cosmology processing pipeline. The green rectangles indicate the different data levels, and the blue regions correspond to the three main sections of the pipeline.}
    \label{fig:flowchart}
\end{figure*}

The data processing pipeline employed by the NenuFAR CD KSP can be divided into three main steps: Preprocessing, calibration-based sky model subtraction, and power spectrum estimation (including residual foreground removal). The routine preprocessing step is performed in the \texttt{nancep} computing cluster at the Nan\c cay radio observatory, while the remaining two steps of the processing are performed in the \texttt{DAWN} cluster in Groningen \citep{pandey2020integrated}.  A workflow describing the sequence of steps in the full data processing pipeline is presented in Fig.~\ref{fig:flowchart}. NenuFAR data is calibrated using \texttt{DDECal} \citep{gan2023assessing,gan2022assessing}, and we currently use the \texttt{diagonal} mode in which only the diagonal elements of the Jones matrices are estimated. The calibration parameters used at the different stages of the pipeline are summarised in Table~\ref{tab:calib_params}.

\subsection{Preprocessing}
The correlated data of NenuFAR is available at a time resolution of $1\,$s, divided into sub-bands of width $195.3\,$kHz each, using a polyphase filter bank. Each sub-band is further channelised into 64 channels at the raw data stage, also called the L0 level. The data for both the target and calibrator fields for each night, at the highest time and frequency resolution, are flagged using \texttt{AOFlagger} \citep{offringa2012morphological} to remove intermittent strong RFI. The results of this RFI flagging step are shown in the left panel of Fig.~\ref{fig:rfi_stats}. While Z20 is relatively clean, Z17 suffers from strong known RFI in the middle of the band, which results in nearly $100\%$ flagging over a range of these frequencies. Next, two channels of width 15.3 kHz each are flagged at either edge of each 195.3 kHz sub-band to avoid edge effects of the polyphase filter. The data are then averaged into $4\,$s time resolution and 12 channels per sub-band frequency resolution, and compressed with \texttt{Dysco} \citep{offringa2016compression,chege2024impact} to produce the L1 visibilities. 

The cables connecting the MAs in NenuFAR vary in length between 20$\,$m and 150$\,$m, making it crucial to account for cable reflections using a bandpass calibration step to avoid artefacts at high delay, which can be limiting for 21-cm cosmology studies. The bandpass solutions for each MA are calculated from the calibrator observation with \texttt{DDECal} (Table~\ref{tab:calib_params} lists the calibration parameters), and the solutions are transferred to the target field observations.\footnote{See Appendix B of \citetalias{munshi2024first} for an example of how this accounts for cable reflections and other direction-independent (DI) bandpass effects.} Since the difference in the beam response between the calibrator and the target direction is not corrected here, the data does not have the correct flux scale yet, and this is corrected only after the A-team subtraction and DI correction step (Section~\ref{sec:ateam}). The bandpass calibrated data, after an additional round of RFI flagging with \texttt{AOFlagger} and averaging into 3 channels per sub-band frequency resolution, form the L2 visibilities. The analysis in the rest of the paper is performed on the L2 data.

\subsection{Nenuflow}
The faint nature of the 21-cm signal imposes stringent precision requirements on calibration algorithms that need to be tuned to the specific instrument so that they can account for the most prominent issues it suffers from. The first analysis by \citetalias{munshi2024first} presented such a calibration pipeline for NenuFAR, which consists of a series of steps that iteratively improve the sky model of the target field while removing the contribution of strong foregrounds from the visibilities through direction-dependent (DD) calibration. In this work, we have structured the calibration and sky model subtraction part of the pipeline into a streamlined workflow, called \texttt{Nenuflow}\footnote{\url{https://github.com/kariukic/nenuflow}}. \texttt{Nenuflow} is built on \texttt{Nextflow} \citep{di2017nextflow}, which provides key benefits such as reliability, reproducibility, robust error handling, and streamlined parallelisation. Additionally, an automated pipeline is very beneficial for systematically processing data from a large number of nights, which is expected to be needed in 21-cm cosmology analyses to detect the signal power spectrum. While the preprocessing and power spectrum estimation sections of the pipeline could be incorporated in \texttt{Nenuflow}, in this work, it is used only for calibration-based sky model subtraction since this is where the robust structure and parallelisation offer the most benefits. The preprocessed L2 data of the target field for each night and spectral window is taken as input by \texttt{Nenuflow}.

\subsubsection{Bright A-team source subtraction}\label{sec:ateam}
The regular hexagonal layout of the 19 antennas making up the MAs in NenuFAR gives rise to strong grating lobes in their primary beams.\footnote{The different NenuFAR MAs are rotated by multiples of $10^{\circ}$ with respect to each other, resulting in six different non-redundant sets of primary beam imprints due to the hexagonal symmetry.} The residual power from bright off-axis radio sources passing through the grating lobes was determined by \citetalias{munshi2024first} to be one of the primary causes of the strong excess power in NenuFAR data. So, a major portion of our efforts to improve the calibration pipeline must go into the bright source subtraction step. The strongest sources in the radio sky (e.g. Cygnus A, Cas A, Taurus A) are collectively known as the A-team. In the case of NenuFAR, due to its wide-field nature and the very high flux densities of these sources, the contaminating effects of the A-team on the data are particularly strong. The PSF sidelobes of these sources can dominate the target field, making it necessary to remove their contribution from the visibilities before the gain of the instrument in the target field direction can be corrected. We achieve this through a DD calibration step using \texttt{DDECal}, which provides calibration solutions for each MA separately in the direction of each A-team source and the target field (as a single direction), as a function of time and frequency. Currently, the NenuFAR beam model is not integrated in \texttt{EveryBeam}\footnote{\url{https://git.astron.nl/RD/EveryBeam}}, which is used by \texttt{DDECal}. Thus, the model visibilities, against which the calibration is performed, do not have an imprint of the NenuFAR primary beam. As a result, the calculated gains primarily capture the effect of the NenuFAR primary beam for each MA. For each A-team source, the predicted visibility contribution, corrupted by the corresponding calculated gains, is subtracted from the visibilities. This is followed by correction of the residual visibilities using the gains in the target direction. This essentially inverts and applies the station-based gain solutions in a single direction (towards the target field) to the visibilities, in a DI gain correction step.

\begin{figure}
    \includegraphics[width=\columnwidth]{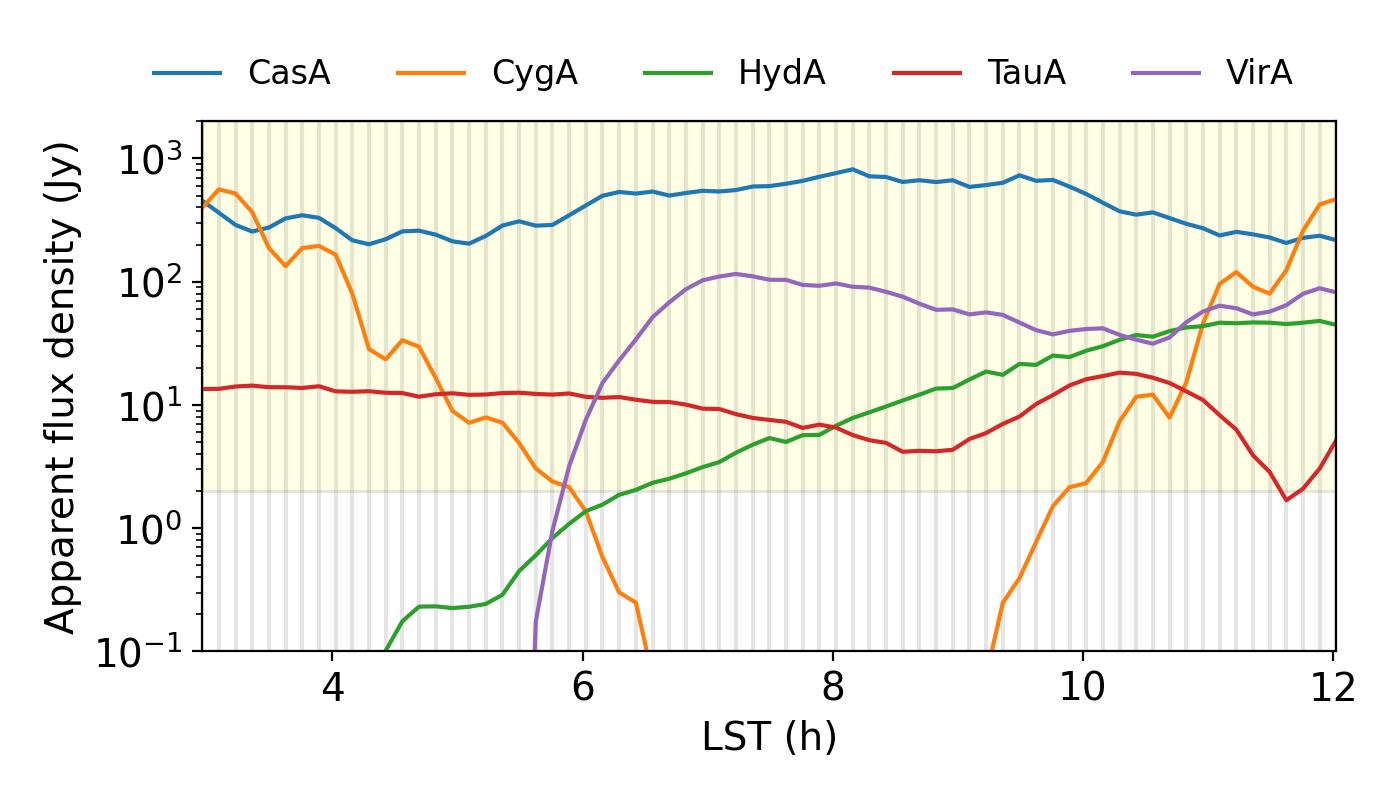}
    \includegraphics[width=\columnwidth]{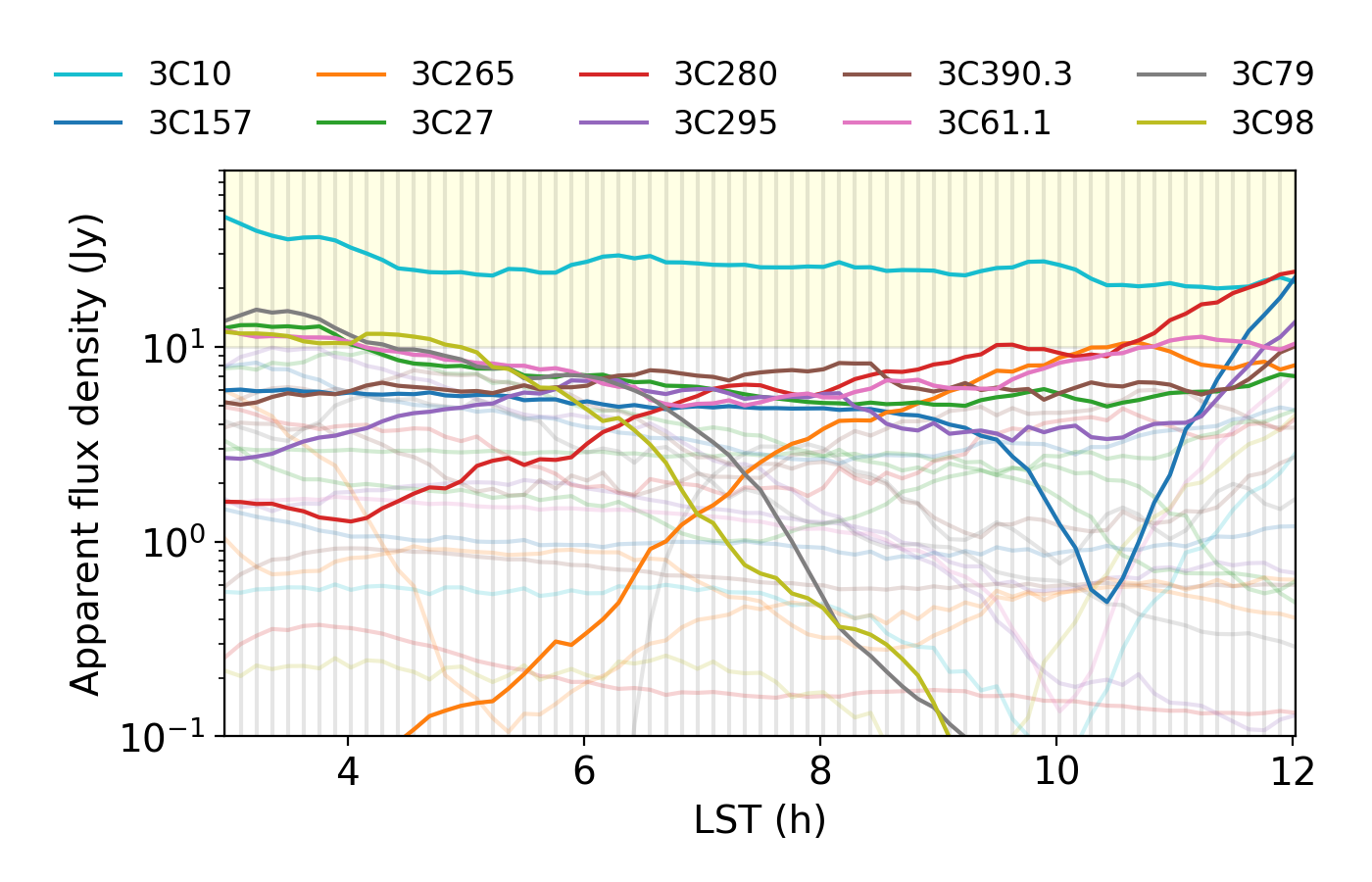}
    \caption{Apparent fluxes of A-team and 3C sources as a function of LST for Z20, during Night~1, estimated using a simulated NenuFAR primary beam model. The top and bottom panels show the apparent fluxes for the A-team and 3C sources, respectively. The yellow shaded regions indicate the range of apparent fluxes above the flux threshold used to select sources to include in the DD calibration step for each segment. The vertical lines demarcate the different time segments.}
    \label{fig:apparent_flux}
\end{figure}

\begin{table}
\centering
\caption{Calibration parameters used at the different stages of processing.}
\begin{threeparttable}
\label{tab:calib_params}
\begin{tabular}{ccccc}
    \hline
    \addlinespace[2pt]
    \multirow{2}{*}{\makecell{Stage}}  & \multirow{2}{*}{Direction(s)} & \multicolumn{2}{c}{Solution interval} & \multirow{2}{*}{Smoothness}\\
    \cmidrule(lr){3-4}
                             &                           & Time & Frequency &  \\
    \addlinespace[2pt]    
    \hline\hline
    \addlinespace[3pt]
    \makecell{Bandpass\\calibration} & Cas A & 4 s & $15.3\,$kHz & None\\
    \addlinespace[3pt]
    \makecell{A-team\\subtraction} & A-team, NT04 & \makecell{$4\,$s$ - 2\,$min\\(adaptive)} & $195.3\,$kHz & $2\,$MHz\\
    \addlinespace[3pt]
    \makecell{3C\\subtraction} & 3C, NT04 & $4\,$min & $195.3\,$kHz & $4\,$MHz\\
    \addlinespace[3pt]
    \makecell{Target\\subtraction} & NT04 clusters & $6\,$min\tnote{a} & $195.3\,$kHz & $6\,$MHz\\
    \hline
\end{tabular}
\begin{tablenotes}
    \item[a] For the first night, this value is 8 min, matching the duration of the time segments.
\end{tablenotes}
\end{threeparttable}
\end{table}
The analysis by \citetalias{munshi2024first} divided the data into $52\,$min segments in order to parallelise calibration. Here, we refine this approach and divide the data into $12\,$min segments\footnote{Only for Night~1 of Z20, which was processed first, we used 8 min segments instead.}, such that the solution time intervals for all calibration steps are divisors of the time segment duration, avoiding residual narrower solution time intervals at the end of each segment. Additionally, the NenuFAR analogue beamformer repoints every 6 min, which is naturally accommodated by the 12 min calibration solution intervals. Apart from increasing the number of threads in parallelisation, increasing the number of segments allows us to calibrate against an apparent sky model for each segment, where the simulated NenuFAR primary beam, calculated using \texttt{nenupy}\footnote{Nenupy (\url{https://github.com/AlanLoh/nenupy}) uses a dipole beam model obtained from electromagnetic simulations to compute the full NenuFAR primary beam model using the array factor of each MA.} \citep{alan_loh_2023_7994526} and averaged over time, frequency and MAs, is applied to the intrinsic sky model before calibration. This provides a reasonable starting point for the calibration solutions, which need to capture the beam on timescales shorter than a 12 min segment and across the 11.5 MHz frequency bandwidth. The top panel in Fig.~\ref{fig:apparent_flux} shows the A-team's apparent fluxes calculated using the simulated NenuFAR primary beam as a function of time for Z20 of Night~1. The ripples in the apparent fluxes with time are produced by the sources passing through the primary beam sidelobes. We impose a flux cut of 2 Jy such that a source is included in the calibration run for a segment only if it has an apparent flux above this threshold (yellow shaded region). This is slightly different to the approach adopted by \citetalias{munshi2024first}, where all A-team sources located above the horizon were included. The flux cut is an input parameter to \texttt{Nenuflow}, allowing multiple calibration runs to ensure calibration solutions have sufficient S/N and describe the primary beam well.
\begin{figure}
    \includegraphics[width=\columnwidth]{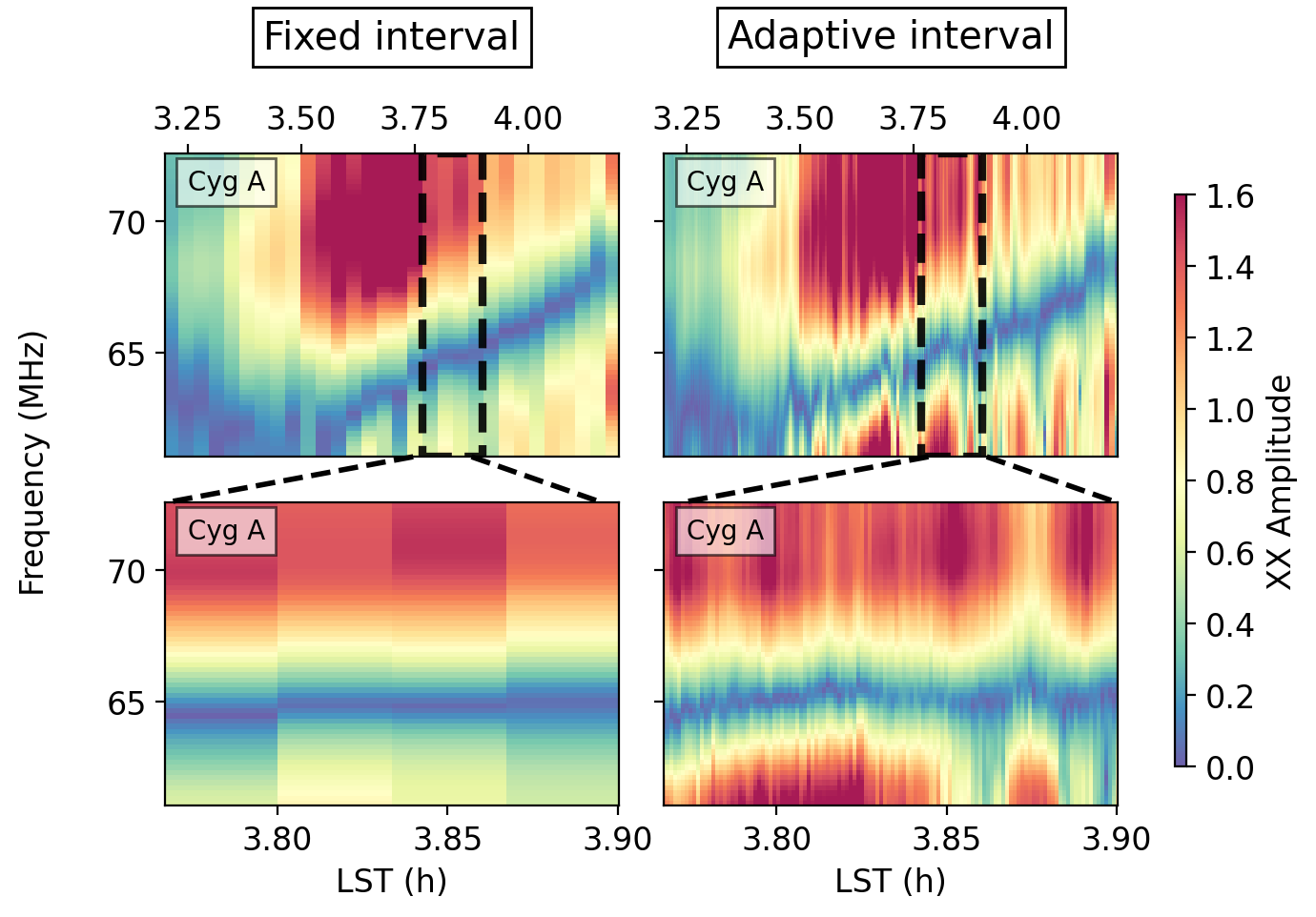}
    \includegraphics[width=\columnwidth]{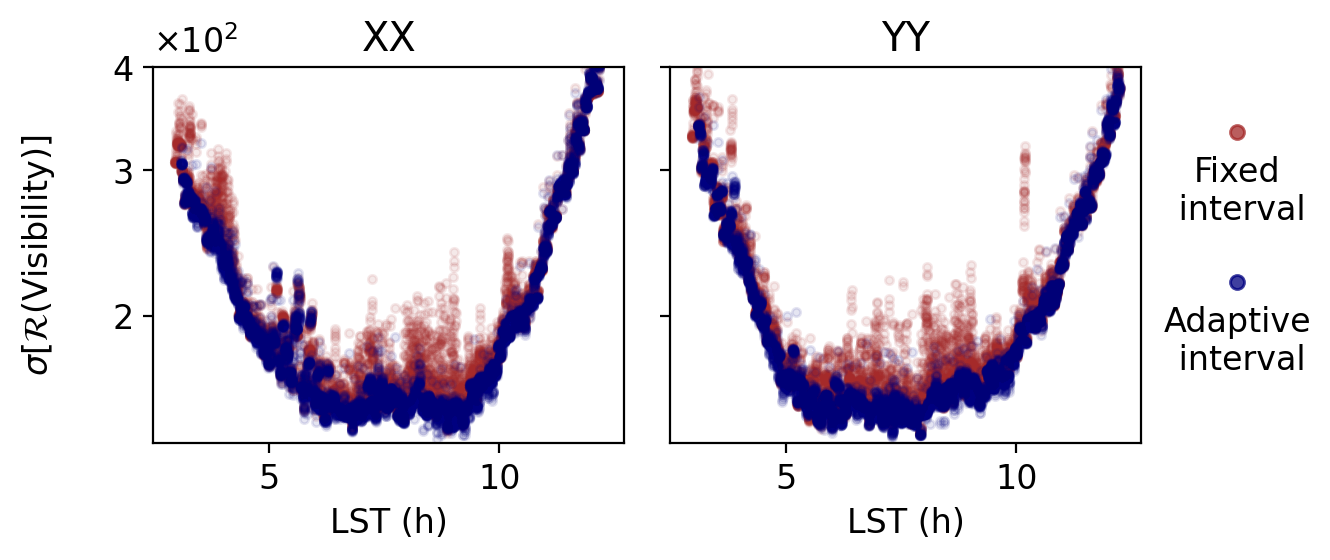}
    \caption{Comparison between calibration with a fixed and adaptive solution time interval. The top row shows the XX gain amplitudes for an example station, in the direction of Cyg A, during a 1 h portion of Night~1. The middle row shows a zoomed-in view of a single 8 min time segment, corresponding to the dashed rectangular region of the top row. The bottom row shows the standard deviation of the real part of the visibilities for the full night, after A-team subtraction and DI correction, for both XX and YY correlations.}
    \label{fig:solperdir}
\end{figure}

Another improvement we make to the A-team source subtraction compared to \citetalias{munshi2024first} is the implementation of adaptive solution time intervals. In the DD calibration run for any given segment, the apparent flux for the different directions can be very different, as seen in Fig.~\ref{fig:apparent_flux}. A direction with a lower apparent flux needs a longer solution interval to garner enough S/N. However, not accounting for the primary beam on such timescales could lead to a large residual flux in a direction with high apparent flux. We address this by adapting the solution time interval as a function of direction. The number of sub-solutions needed per direction is calculated internally by \texttt{Nenuflow}, based on the estimated apparent fluxes for each segment. The directions with higher flux are allocated a higher number of sub-solutions per interval, and the higher total flux ensures that there is enough S/N within a sub-solution interval to obtain reliable solutions. The different A-team directions are allocated solution intervals between 4 s (the time resolution of L2 data) and 2 min, depending on their apparent flux in a 12 min time segment, as opposed to a fixed solution interval of 2 min used by \citetalias{munshi2024first}. The target direction is solved with the full 2 min time interval, since the beam variations in the target direction are expected to be small on such timescales. The top and middle rows of Fig.~\ref{fig:solperdir} illustrate the effect of using adaptive solution intervals for an example MA by comparing the gain amplitude solutions in the direction of Cygnus A (Cyg A) for the XX visibility. The top panels show the solutions for a one-hour duration during Night~1, while the middle panels give a zoomed-in view of a single 8 min time segment. We find that Cyg A passes through multiple primary beam sidelobes and nulls, which are captured considerably better through the adaptive solution interval approach. Within the highlighted time segment, Cyg A direction is allocated the smallest possible (4 s) solution interval, which can better capture the nulls in both time (around 3.86 h) and frequency (around 65 MHz). The bottom row of Fig.~\ref{fig:solperdir} compares the standard deviation of the residual visibilities of Night~1 after A-team source subtraction and DI correction using the two approaches. Using an adaptive interval reduces the residuals significantly, especially during the middle of the observation when the phase centre is closer to the zenith and the sensitivity is maximum.

\begin{figure*}
    \centering
        \includegraphics[width=\columnwidth]{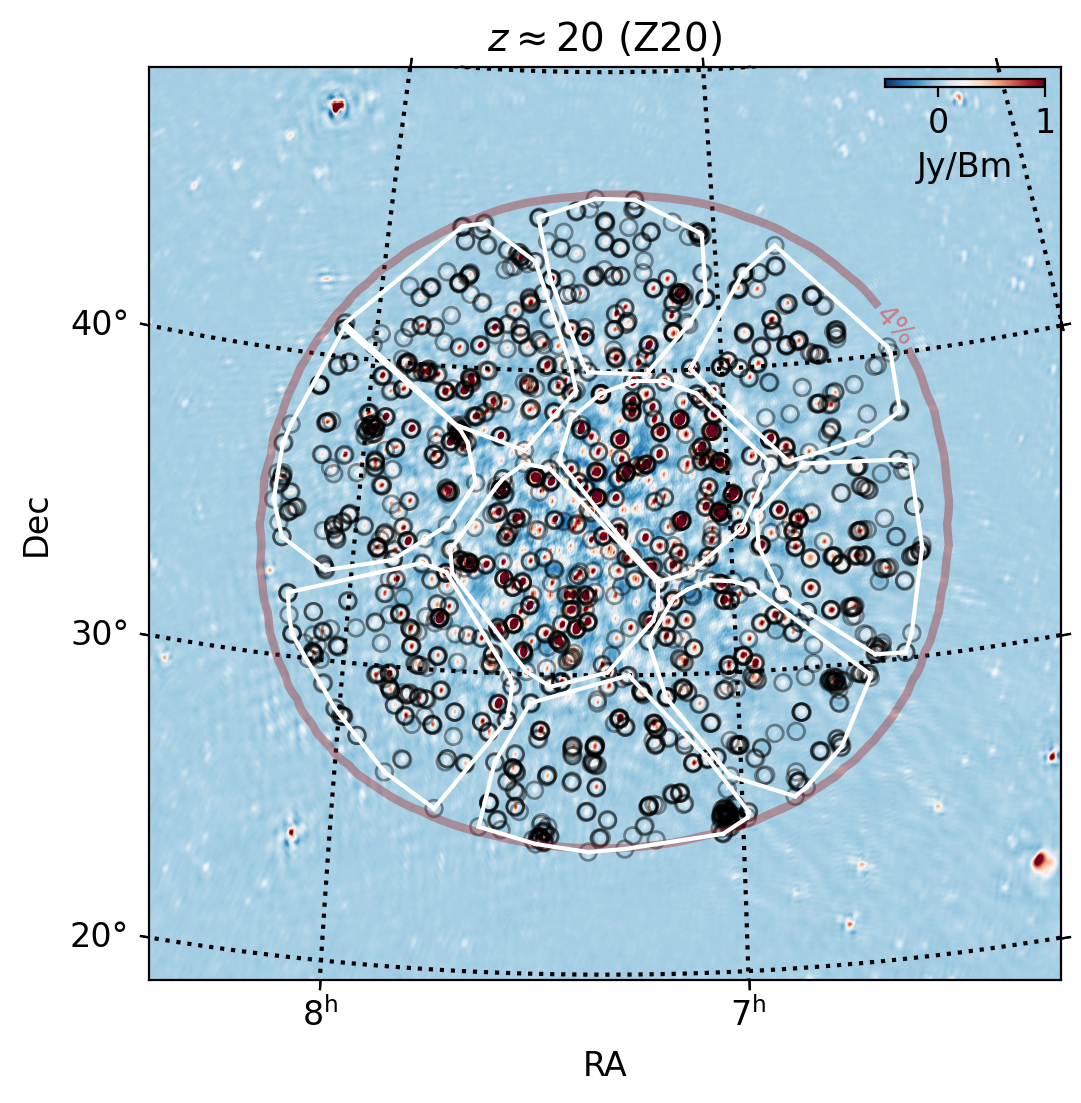}
        \includegraphics[width=\columnwidth]{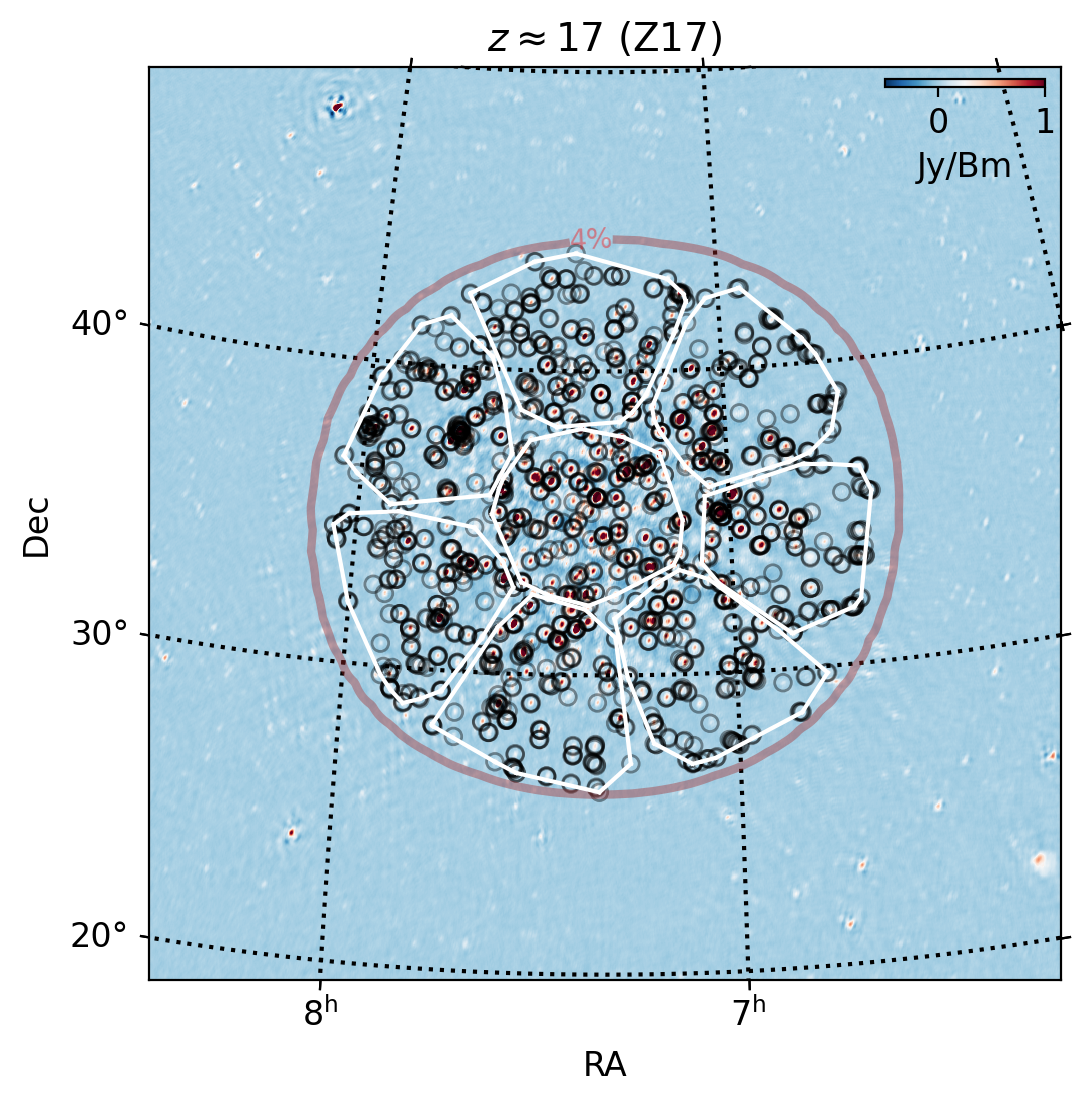}
    \caption{Cleaned images of the target field after DI correction, with the sky models from Night~1 overplotted. The two panels correspond to the two spectral windows. The overplotted white curves indicate the clusters of sources which are used as separate directions when the target field subtraction is performed in a DD calibration and subtraction step.}
    \label{fig:narrow_images}
\end{figure*}

\begin{figure*}
    \centering
    \includegraphics[width=\columnwidth]{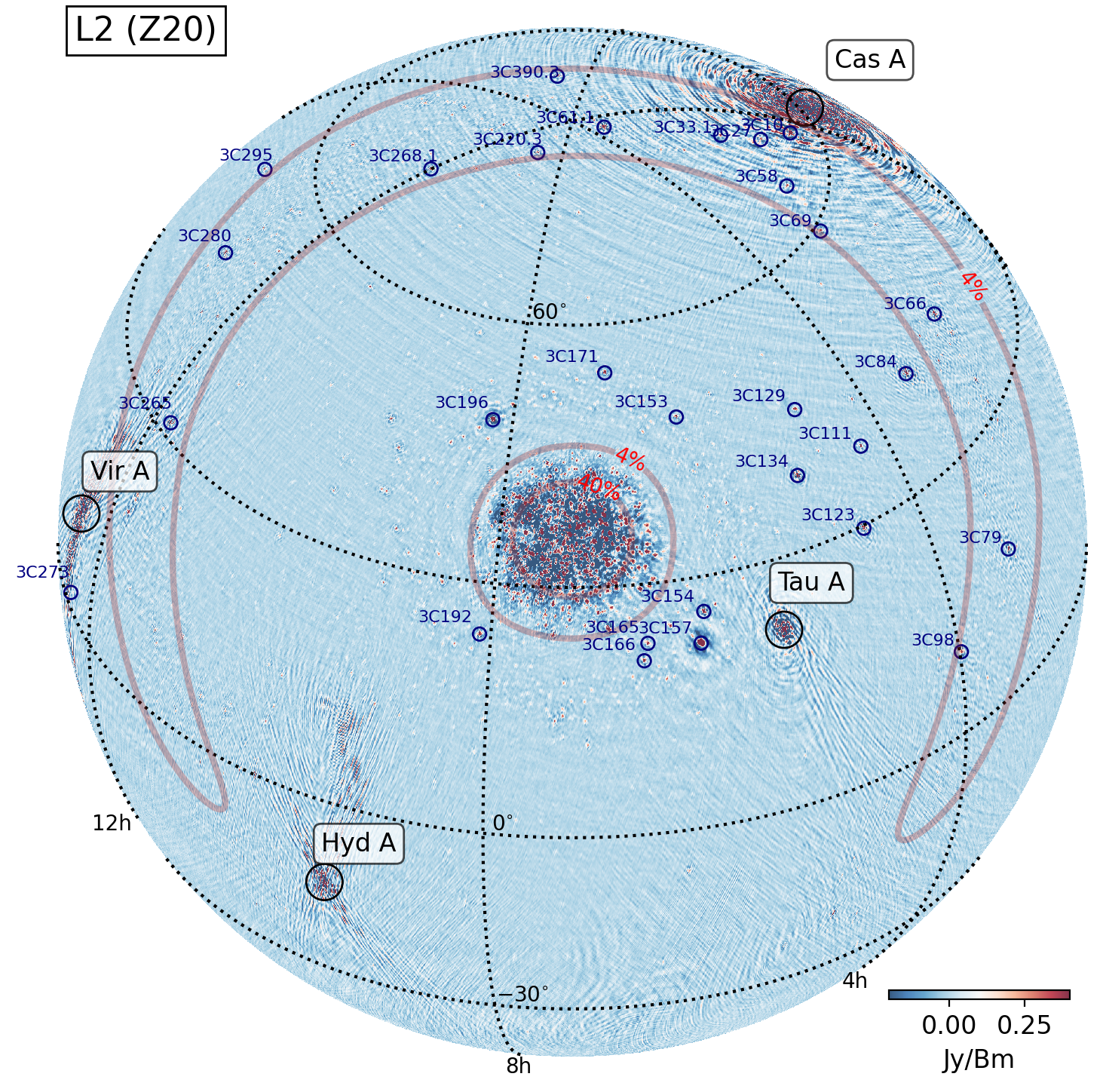}
    \includegraphics[width=\columnwidth]{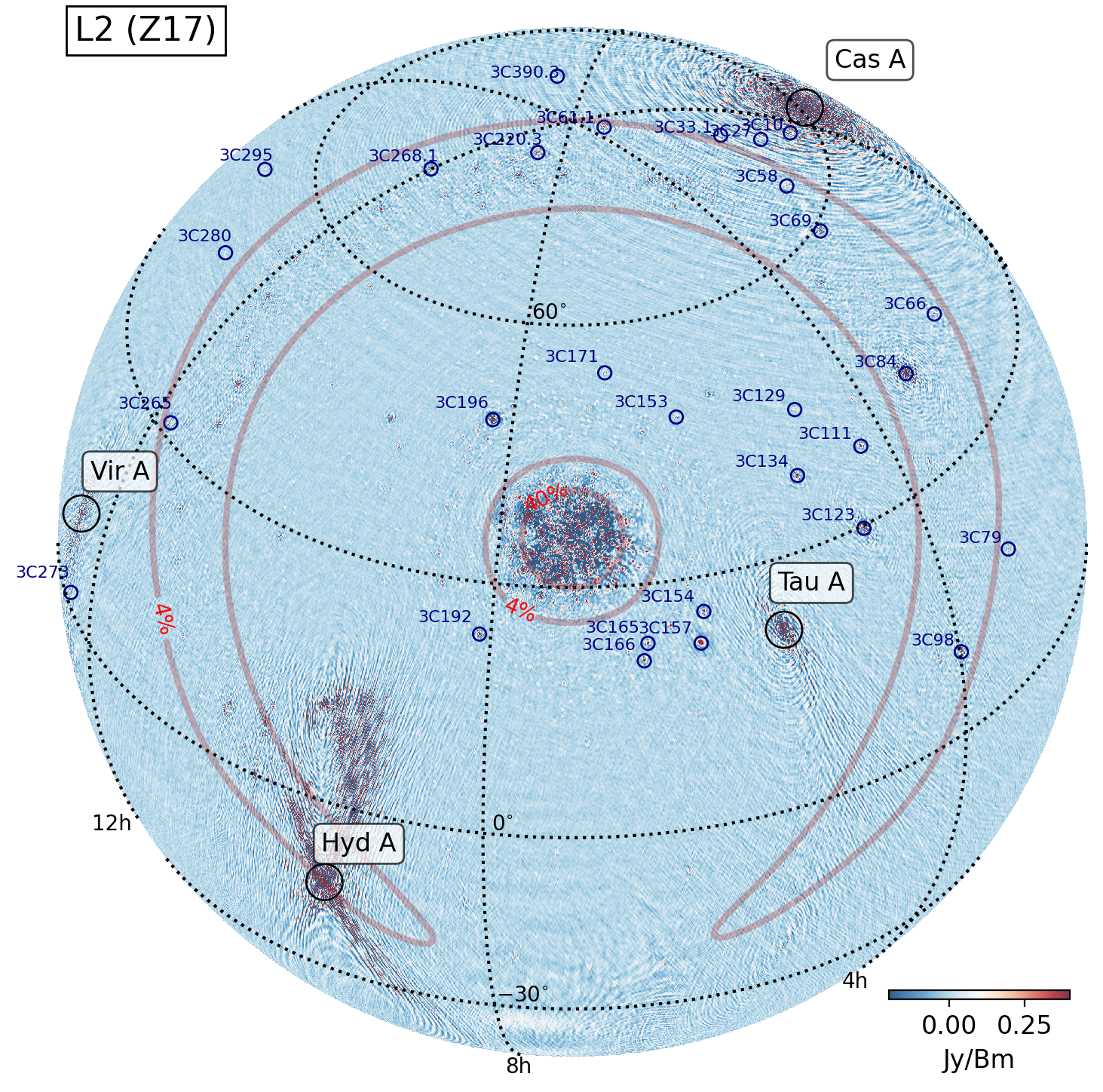}
    \includegraphics[width=\columnwidth]{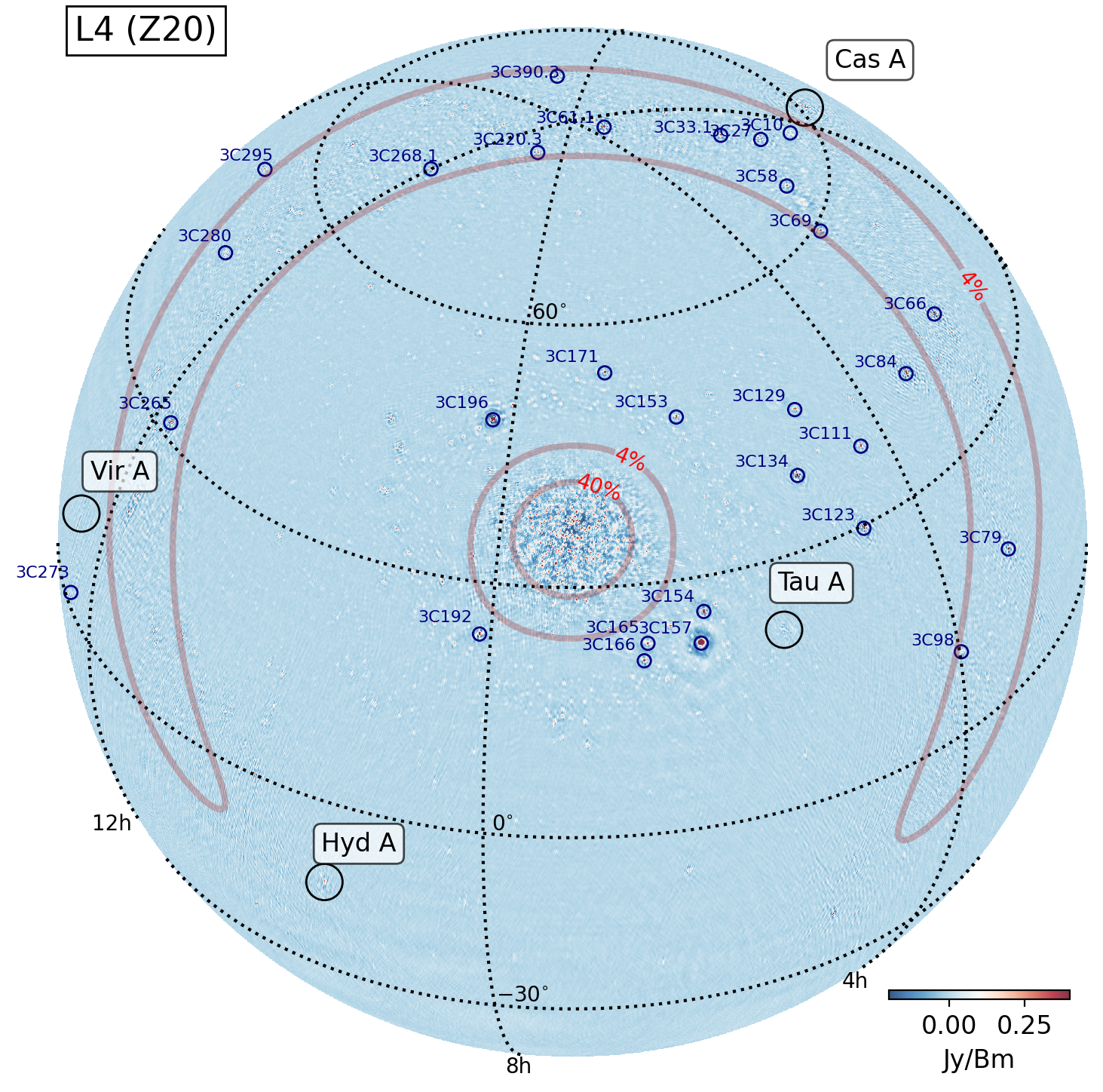}
    \includegraphics[width=\columnwidth]{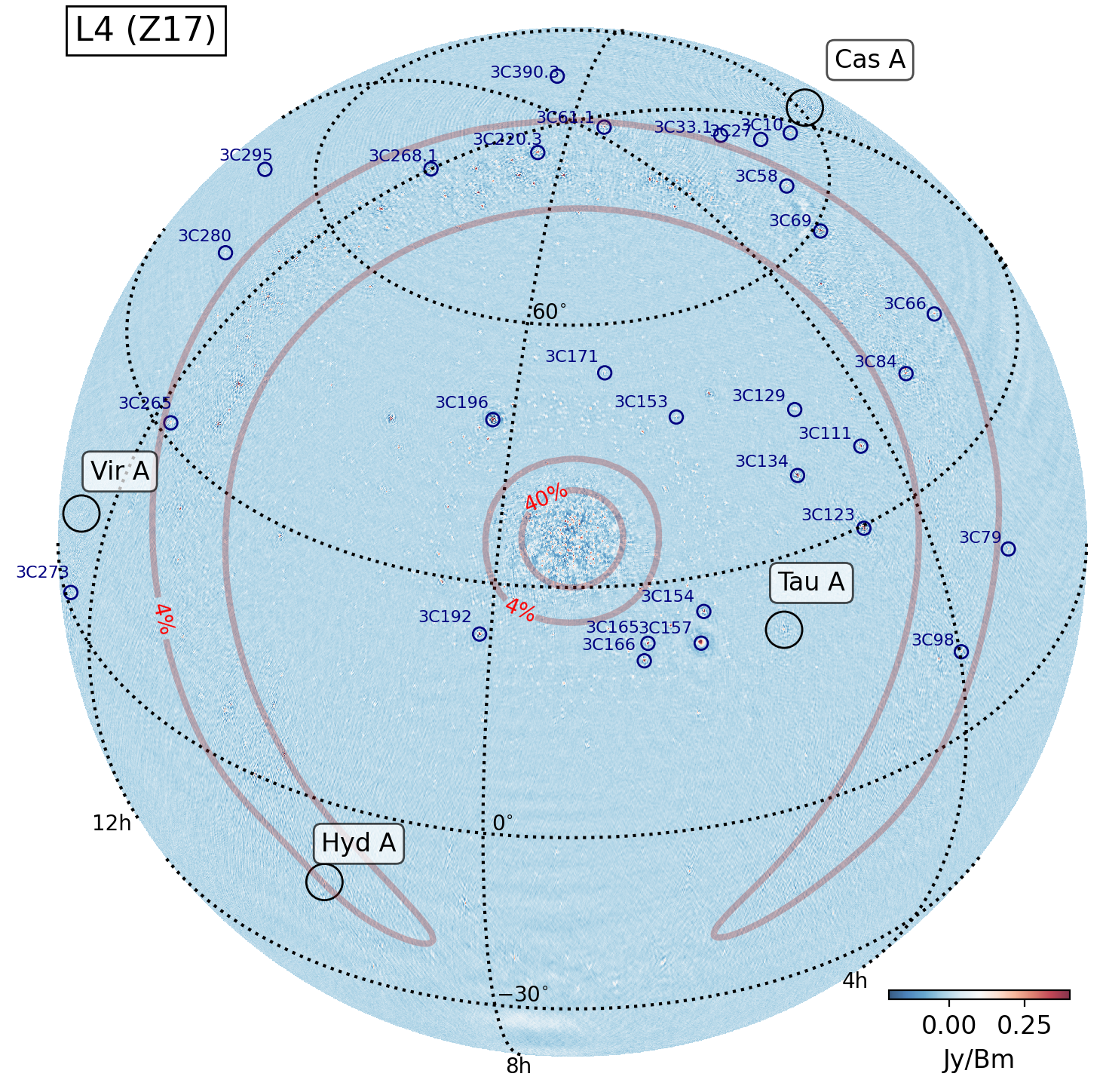}
    \caption{Wide-field dirty images before and after calibration-based sky model subtraction. The two columns correspond to the two spectral windows. The top and bottom rows show the images before and after sky model subtraction, respectively. The NenuFAR primary beam is simulated at a time resolution of 1 h for all four nights and averaged over time, frequency and MAs for both spectral windows. The red contours indicate 4\% and 40\% levels of the primary beam.}
    \label{fig:wide_images}
\end{figure*}
\begin{figure*}
    \centering
    \includegraphics[width=2\columnwidth]{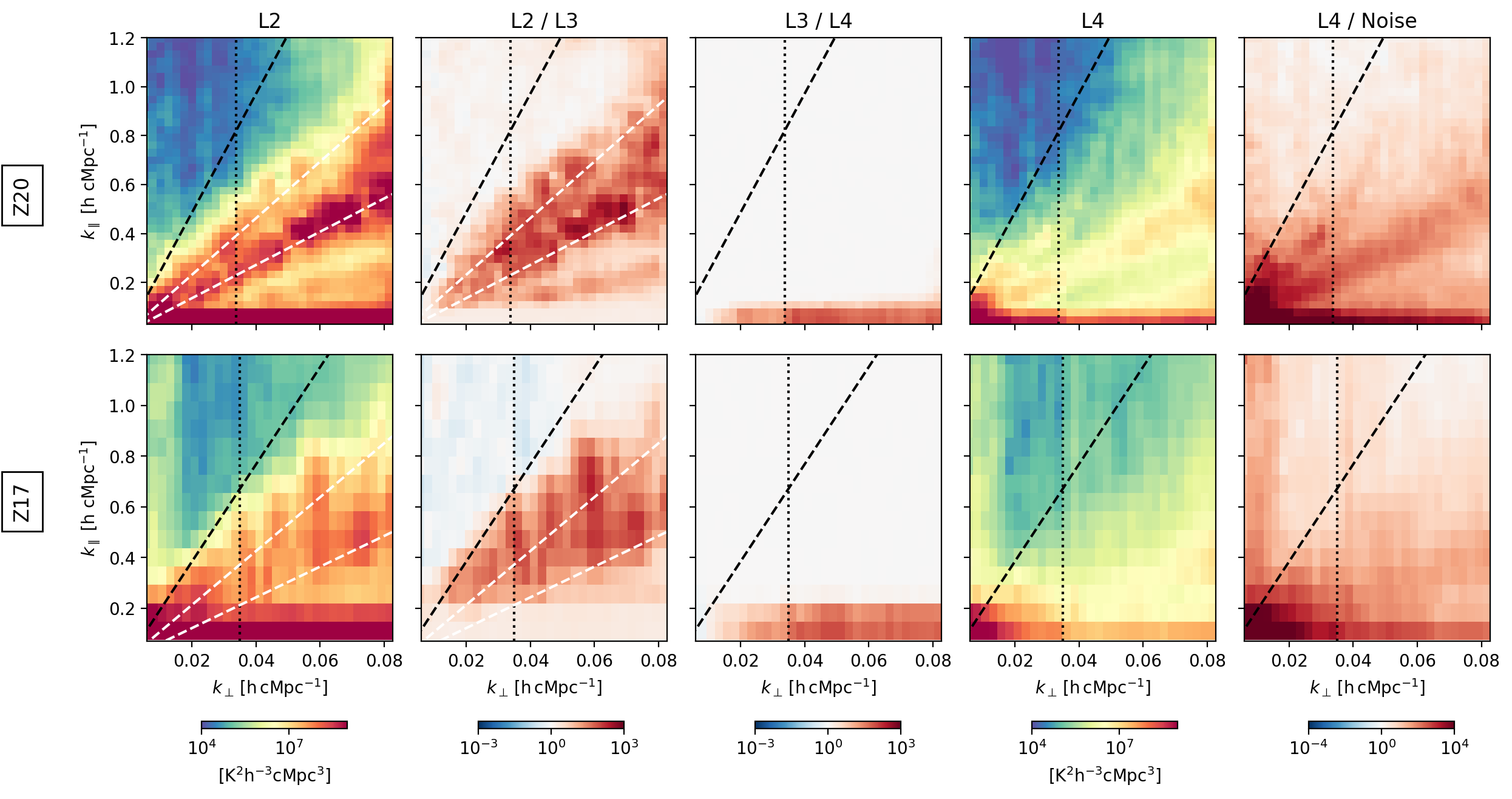}
    \caption{Power spectra estimated at different data levels. The two rows correspond to the two spectral windows. The first column corresponds to the L2 data. The second and third columns show the level of power subtracted in the off-axis source subtraction and target subtraction, respectively. The fourth column shows the power spectrum after sky-model subtraction, and the fifth column shows its ratio with the thermal noise power spectrum. The black dashed lines indicate the full-sky horizon limit, and the white dashed lines indicate the source wedge of Cas A. The vertical dotted lines indicate the baseline cut. We note that the power spectra presented in this figure are estimated before the steps described in Sections \ref{sec:rfi_uv} and \ref{sec:optimal_elevation} are applied to the data.}
    \label{fig:ps_stages}
\end{figure*}
In total, we include five A-team sources, Cas A, Taurus A (Tau A), Virgo A (Vir A), Hydra A (Hyd A), and Cyg A, in the A-team subtraction step. The sky model for the target field is initially an intrinsic catalogue model obtained from the global sky model\footnote{\url{https://lcs165.lofar.eu/}} in a 15-degree radius around the NT04 field phase centre. This model is composed of compact sources modelled using delta functions and Gaussians, and was created using data from The GMRT Sky Survey \citep[TGSS;][]{2017A&A...598A..78I}, NRAO VLA Sky Survey \citep[NVSS;][]{condon1998nrao}, WEsterbork Northern Sky Survey \citep[WENSS;][]{1997A&AS..124..259R}, and VLA Low-frequency Sky Survey \citep[VLSS;][]{cohen2007vla}. \citetalias{munshi2024first} used a 20$\lambda$ baseline cut in calibration and showed, through signal injection tests, that this induces signal suppression of $\approx$3\% in the A-team subtraction step due to overfitting in the baselines used in calibration. Here, we avoid this suppression by performing the calibration and the final power spectrum estimation with two mutually exclusive sets of baselines \citep{patil2016systematic,patil2017upper,mouri2019quantifying,mevius2022numerical}. This is possible due to the presence of an extra remote MA compared to \citetalias{munshi2024first}, which forms a large number of additional baselines with the core stations. Only baselines longer than $40\lambda$ are used in calibration, and the resulting station-based gains are applied to the predicted models and subtracted from all visibilities. We note that this baseline cut, selected by \citetalias{munshi2024first}, is smaller than the $250\lambda$ cut used by LOFAR, since the baselines in NenuFAR are shorter than LOFAR, and a higher baseline cut would drastically reduce the number of baselines available during calibration. After A-team subtraction and DI correction of the target field, an apparent sky model (effectively beam-averaged over the full night) of the target field is constructed by performing multi-frequency synthesis imaging and deconvolution of the visibilities from all time segments within a night using \texttt{WSClean} \citep{offringa2014wsclean}. The spectral information in the model is captured by fitting a second-order polynomial along frequency for each clean component during deconvolution. The sources outside a fixed radius ($11^{\circ}$ for Z20 and $9^{\circ}$ for Z17) from the phase centre are filtered out to only keep genuine sources within the main lobe of the primary beam. Fig.~\ref{fig:narrow_images} shows the extent of the sky model after this filtering, which remains confined within the first null of the primary beam, as indicated by the contours corresponding to 4\% of the peak primary beam amplitude. The A-team subtraction, DI correction, and imaging are then repeated using this improved sky model on the original (L2) visibilities, following the approach taken by \citetalias{munshi2024first}, to produce an updated model of the target field. Fig.~\ref{fig:narrow_images} shows the cleaned images constructed from all four nights, with the black circles indicating the location of the clean components of the updated sky model for Night~1.

The A-team source subtraction and DI correction are performed in a strongly parallelised fashion, over 15 nodes of the \texttt{DAWN} cluster, with up to 5 calibration runs performed in parallel on each node. This enables multiple runs that are necessary to identify and flag stations with bad calibration solutions over a series of iterations\footnote{Typically, fewer than three iterations are sufficient to identify and flag such unstable stations.}. The stations that are completely flagged in this step for each night are illustrated in the right panel of Fig.~\ref{fig:rfi_stats}. This calibration solution-based station flagging results in the removal of 15 stations from all nights, and these were identified to have electronic malfunctions\footnote{This has since been fixed.}. A few more stations are flagged on individual nights based on artefacts in the calibration solutions. Fig.~\ref{fig:wide_images} shows wide-field dirty images constructed before and after sky-model subtraction for both spectral windows. Cyg A is more than 90 degrees from the phase centre, so its contribution cannot be distinguished in these images. Cas A is the strongest source in both spectral windows, and for Z17, Hyd A is picked up more strongly by the grating lobes, which are closer to the phase centre compared to Z20. We find that the A-team sources are subtracted well, and their PSF sidelobe ripples that run across the entire sky are significantly reduced after the A-team source subtraction step.

\subsubsection{3C source subtraction}
NenuFAR's grating lobes pick up a large number of bright radio sources (top row of Fig.~\ref{fig:wide_images}). \citetalias{munshi2024first} adopted a manual approach of identifying the sources that have the most significant contribution in the $u\varv$ plane and performing a DD calibration and subtraction. Here, we use a more systematic approach, similar to what is used for the A-team subtraction. A large number of 3C sources \citep{bennett1962preparation} visible near the primary beam grating lobes in the full-sky images are included in an intrinsic 3C source model. This is given as an input to \texttt{Nenuflow}, which applies the simulated primary beam for each segment and includes those sources above a minimum flux limit. The bottom panel of Fig.~\ref{fig:apparent_flux} shows the simulated apparent fluxes of the selected 3C sources for Z20 during Night~1. The sources that do not make the flux threshold for any segment are shown with faded colours. The effective S/N within a calibration solution interval for a certain direction depends primarily on the apparent flux received from this direction. However, if the DD calibration is performed for a large number of directions, the number of free parameters being solved for increases, effectively increasing the solver noise and raising the S/N requirement. We find that including all sources above a 2 Jy flux limit causes such an overfitting where, even though the contribution of the 3C sources is subtracted on the baselines used in calibration ($> 40\lambda$), the shorter baselines used for power spectrum estimation exhibit an excess variance. We thus adopt a more conservative approach of imposing a much higher flux limit of 10 Jy, such that fewer than six directions are included for any time segment. Thus, the source contributions are still subtracted when they have a very high apparent flux, and the subtraction is accurate and avoids overfitting. The choice of the flux cut, and hence the maximum number of directions to include in the 3C subtraction step depends on the trade-off between the fraction of the flux that can be reliably subtracted, and the excess variance caused on the shorter baselines due to overfitting. The adaptive solution interval is not used in the 3C subtraction since the apparent flux of the strongest 3C sources is much lower than that of the strongest A-team sources, so the errors made within the solution time interval of 4 min are less significant. Comparing the top row of Fig.~\ref{fig:wide_images} with the bottom row, we see that while the 3C sources are not completely subtracted in this more conservative approach, the peak fluxes of the brightest sources decrease in the process.

\subsubsection{Target field subtraction}
The apparent sky model of the target field, after filtering out sources outside the main lobe from the phase centre, is divided into multiple clusters (ten for Z20, eight for Z17) using a spatial $k$-means clustering algorithm \citep{lloyd1982least,macqueen1967some}. The choice of the number of clusters ensures sufficient S/N for the faintest cluster within the solution interval. The clusters for Night~1 for both spectral windows are shown in white convex hulls in Fig.~\ref{fig:narrow_images}. Each cluster forms a separate direction in DD calibration and subtraction, which is performed on the L3 visibilities in the same manner as \citetalias{munshi2024first}, using a 6 MHz frequency smoothness constraint (see Table~\ref{tab:calib_params}). This results in the subtraction of sources till the confusion noise limit. The resulting full-sky dirty images are shown in the bottom row of Fig.~\ref{fig:wide_images}. The maximum flux densities within the main lobe of the primary beam calculated from the residual dirty images are 0.74 Jy (for Z20) and 0.51 Jy (for Z17), which are close to the expected confusion noise levels of 0.70 Jy/PSF (for Z20) and 0.47 Jy/PSF (for Z17) estimated from equation~(6) of \cite{van2013lofar}. Comparing the same measured band in \citetalias{munshi2024first} to Z20, the residual peak flux for Z20 is lower than the value of 1.05 Jy reached by \citetalias{munshi2024first} due to a higher maximum baseline length of 3.3 km as opposed to that of 1.4 km for \citetalias{munshi2024first}, resulting in a narrower PSF and thus higher angular resolution and lower confusion noise. The residual visibilities after target field subtraction are used to estimate the power spectrum after a final foreground removal step using Gaussian process regression (Section~\ref{sec:gpr}).

\section{Power spectrum estimation}\label{sec:ps_estimation}
We estimate power spectra using \texttt{pspipe}\footnote{\url{https://gitlab.com/flomertens/pspipe}} \citep{mertens2020improved,mertens2025deeper}. The visibility data is composed of multiple measurement sets for a large number of time segments for each night and spectral window. Gridded data cubes for each night and spectral window are produced by constructing dirty image cubes using \texttt{WSClean} \citep{offringa2014wsclean,offringa2019precision}, converting into Kelvin units, applying a spatial Hann window, and spatial Fourier transforming into the $u\varv\nu$ domain. The combined data cube for all four nights is obtained by computing a weighted average of the visibility cubes of the individual nights using the corresponding weight cubes obtained from the naturally weighted PSFs.

Though the final power spectrum estimation for setting the upper limits on the 21-cm signal is performed after residual foreground removal using GPR, we also estimate the power spectrum from the visibility data at each stage of calibration and foreground subtraction for diagnostic purposes. The visibility cubes are Fourier transformed along the frequency direction with a Blackman-Harris window, modulus squared, and averaged in spherical and cylindrical shells to estimate the spherical and cylindrical power spectra, respectively. We refer the reader to \citetalias{munshi2024first} for more details on the power spectrum estimation. For Z17, a large number of frequency channels in the middle of the frequency band are flagged (left panel of Fig.~\ref{fig:rfi_stats}). Thus, as long as there is significant power in the lowest $k_{\parallel}$ modes, the Fourier transformation along the frequency axis causes strong artefacts in the cylindrical power spectrum. Thus, the power spectra before GPR-based foreground removal for Z17 are computed only in the frequency range 79 to 85 MHz, for illustration purposes.

\subsection{Residual power spectra}
Before constructing image cubes and power spectra, we perform an additional flagging step based on the standard deviation of the residual visibilities per time segment, which is discussed in Appendix~\ref{sec:std_residuals}. We then compute the cylindrical power spectra from the Stokes~\textit{I} visibilities at the different data levels L2, L3 and L4. The thermal noise power spectrum is estimated from the time-differenced Stokes~\textit{V} visibilities. Fig.~\ref{fig:ps_stages} shows the power spectra at different data levels for both spectral windows, and illustrates the $k_{\perp},k_{\parallel}$ modes at which the power is subtracted by computing the ratio of the power spectra from the combined data cube at successive data levels. The black dashed lines indicate the full-sky horizon limit of the foreground wedge \citep{munshi2025beyond} for the four nights. The white lines indicate the source wedge of Cas A, the region where the peak power of Cas A is expected to dominate \citep{munshi2025beyond}. The vertical dotted line indicates the $40\lambda$ baseline cut. Only $k_{\perp}$ modes below this threshold are used to set power spectrum upper limits, as they are unaffected by signal suppression during calibration. For both spectral windows, we find that off-axis A-team and 3C source subtraction (L2/L3) removes over two orders of magnitude of power, particularly in the $k_{\perp},k_{\parallel}$ modes occupied by Cas A. We note that the slight increase in power in the EoR window between L2 and L3 data for Z17 is caused by time steps near the start and end of observations for each night, with higher RFI contamination (Fig. \ref{fig:std_residuals}). These time steps are eventually flagged in the next two sections (Sections \ref{sec:rfi_uv} and \ref{sec:optimal_elevation}). The target field subtraction (L3/L4), as expected, removes power at the lowest $k_{\parallel}$ modes corresponding to the region near the phase centre. After sky model subtraction, the L4 data is still more than four orders of magnitude beyond the thermal noise power at the lowest $k_{\parallel}$ modes due to confusion-limited foreground sources. The high power at the lowest $k_{\perp},k_{\parallel}$ modes is possibly caused by diffuse Galactic foreground emission. At high $k_{\parallel}$, the residual power approaches the thermal noise power for both spectral windows. However, the excess power above the thermal noise is significantly higher for Z17, possibly because of higher RFI contamination. For Z20, we find a region of high residual power at low $k_{\perp}$ and high $k_{\parallel}$ modes. Z17 also exhibits high residual power at the lowest $k_{\perp}$ modes across the entire $k_{\parallel}$ range.

\subsection{RFI and bright source flagging in the $u\varv$ plane}\label{sec:rfi_uv}
\begin{figure}
    \centering
    \includegraphics[width=\columnwidth]{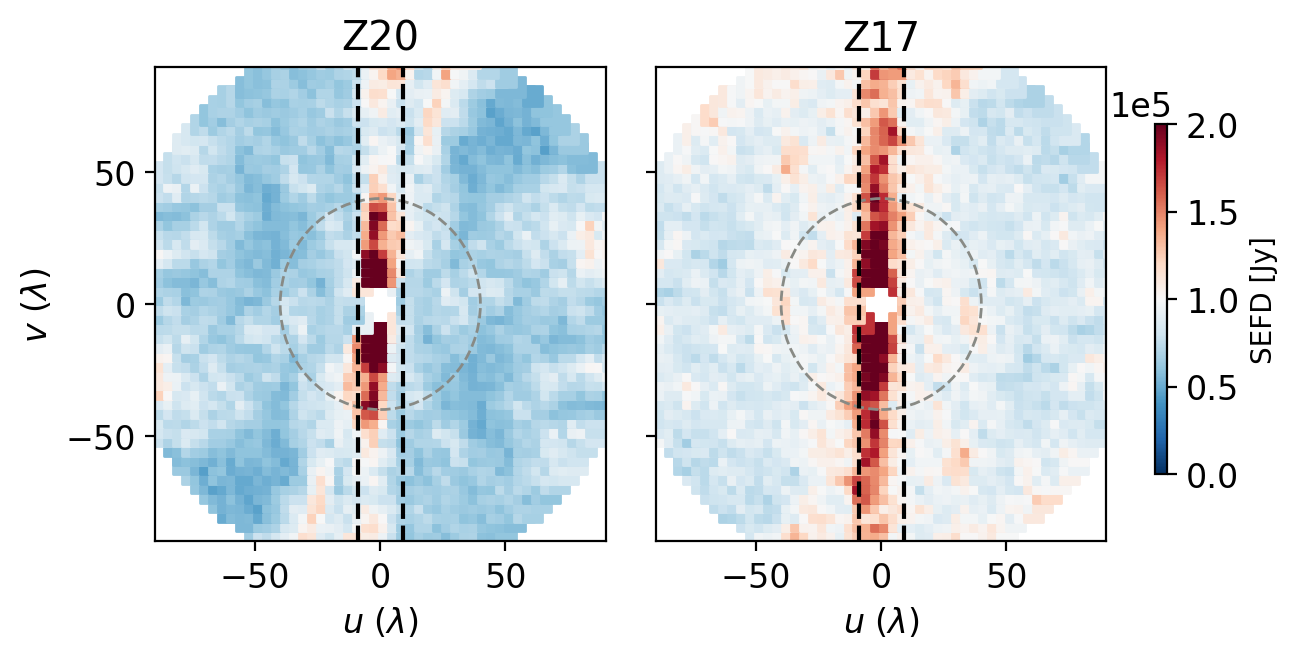}
    \includegraphics[width=\columnwidth]{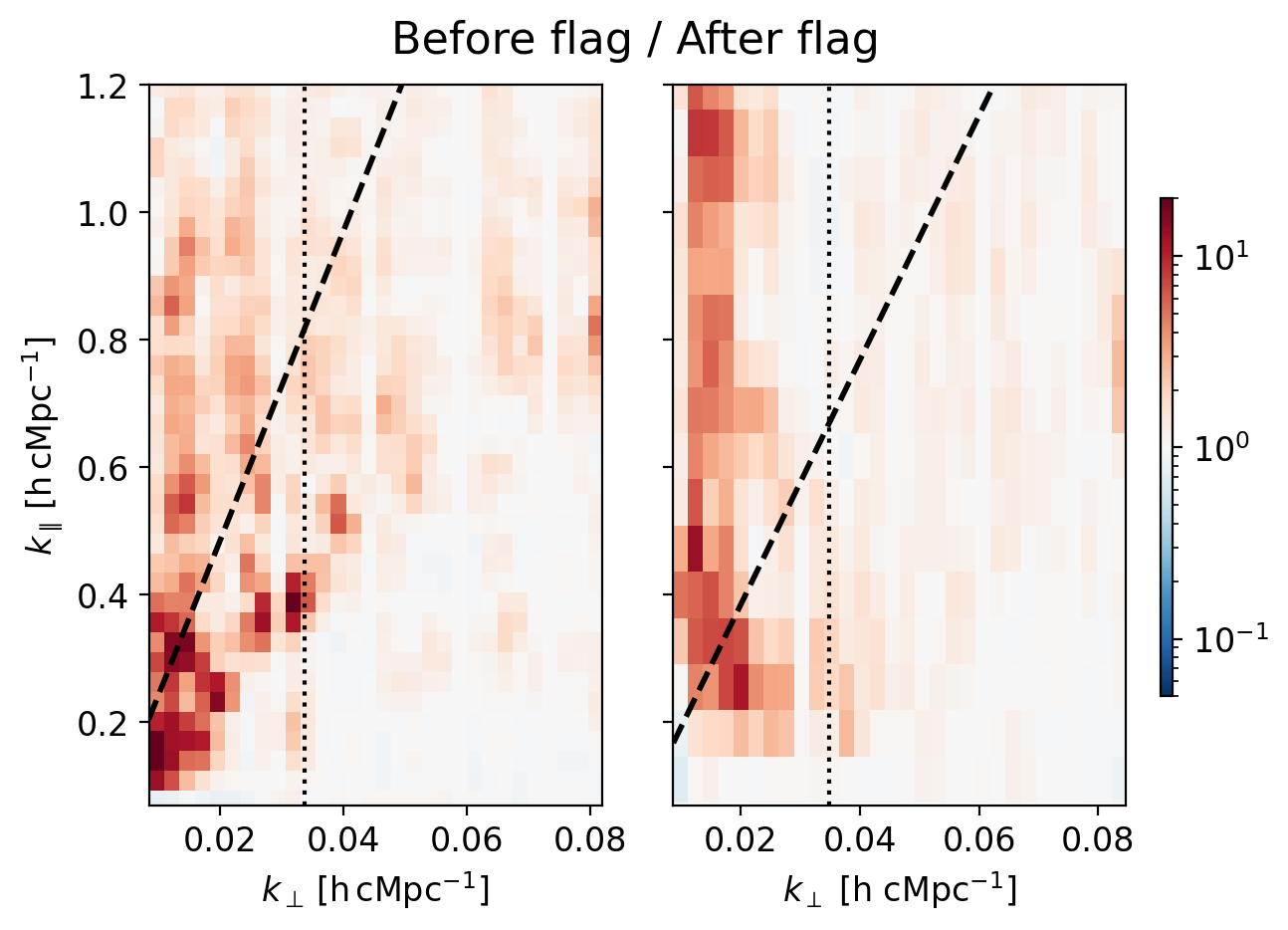}
    \caption{The signature of RFI in the data. The two columns correspond to the two spectral windows. The top row shows the frequency channel differenced noise in the $u\varv$ plane. The grey dashed circles indicate the baseline cut. The bottom row shows the ratio of the power spectrum estimated before and after five $u\varv$ cells in the vicinity of the $u=0$ line (indicated by vertical dashed lines in the top row) are flagged. In the bottom row, the black dashed lines indicate the full-sky horizon limit, and the vertical dotted lines indicate the baseline cut.}
    \label{fig:ps_rfi}
\end{figure}

We estimate the frequency difference noise in the gridded data cube after sky model subtraction for each $u\varv$ cell, to identify residual contributions to the visibilities coming from different directions. The noise in the $u\varv$ plane, in units of system equivalent flux density (SEFD), is shown in the top row of Fig.~\ref{fig:ps_rfi}. For both spectral windows, we find a region of higher residual rms along the $u=0$ line. Such a signature can be caused by stationary RFI with respect to the array and has been demonstrated through forward simulations of local RFI by \cite{munshi2025near}. Since the celestial poles are the only points in the sky whose delays are fixed with respect to the array during a synthesis observation, stationary RFI adds up coherently at the NCP. For phase centres other than the NCP, this resembles an RFI induced source toward the north of the image, which results in a region of higher power along the $u=0$ line due to sampling in $u\varv$ tracks and gridding in the $u\varv$ plane \citep{munshi2025beyond}\footnote{A slight deviation from the $u=0$ direction seen in Fig.~\ref{fig:ps_rfi} could be attributed to the $u\varv$ tracks not being perfect circles (see appendix B of \citealt{munshi2025beyond}).}. We see that not only is the peak RFI variance higher for Z17 than for Z20 due to higher RFI contamination, but the variance across the rest of the $u\varv$ plane is also increased. We flag five $u\varv$ cells centred on the $u=0$ line for all $\varv$ values before constructing the power spectrum (indicated by the black dashed lines in the top panels of Fig.~\ref{fig:ps_rfi}). The bottom row of Fig.~\ref{fig:ps_rfi} shows the ratio of the power spectra estimated before and after such a flagging step. We find that the higher power at the small $k_{\perp}$ modes is reduced by more than an order of magnitude through this flagging in both the spectral windows. We note that such a reduction of RFI power by flagging a region of the $u\varv$ plane is only possible for phase centres away from the NCP field, where the RFI power is spread throughout the $u\varv$ plane \citep{munshi2025near}. As a result, to decrease the contamination from RFI in the NCP field, \citetalias{munshi2024first} identified and flagged baselines strongly affected by RFI by inspecting their delay power spectra. A second (fainter) line of higher variance in the top panels of Fig.~\ref{fig:ps_rfi}, along the southwest-northeast direction, is likely caused by the residuals of Cas A. Five $u\varv$ cells in the vicinity of this direction are also flagged, to suppress the contribution from Cas A residuals.

\subsection{Optimal elevation range}\label{sec:optimal_elevation}
Over the course of each night, the elevation of the NT04 field changes with time. The foreground power in the cylindrical power spectrum is confined to a region known as the foreground wedge. The boundary of the foreground wedge, called the horizon line, has conventionally been assumed to be independent of the observation parameters. However, accounting for curved sky effects, \cite{munshi2025beyond} showed that the slope of the horizon line depends on the elevation of the phase centre. These full-sky horizon line equations suggest that the horizon line is steepest when the phase centre is at the lowest elevation. Thus, selecting only a subset of the LST range for power spectrum estimation could significantly reduce the foreground contamination, albeit at the cost of a loss of sensitivity due to a smaller integration time. 
\begin{figure}
    \centering
    \includegraphics[width=\columnwidth]{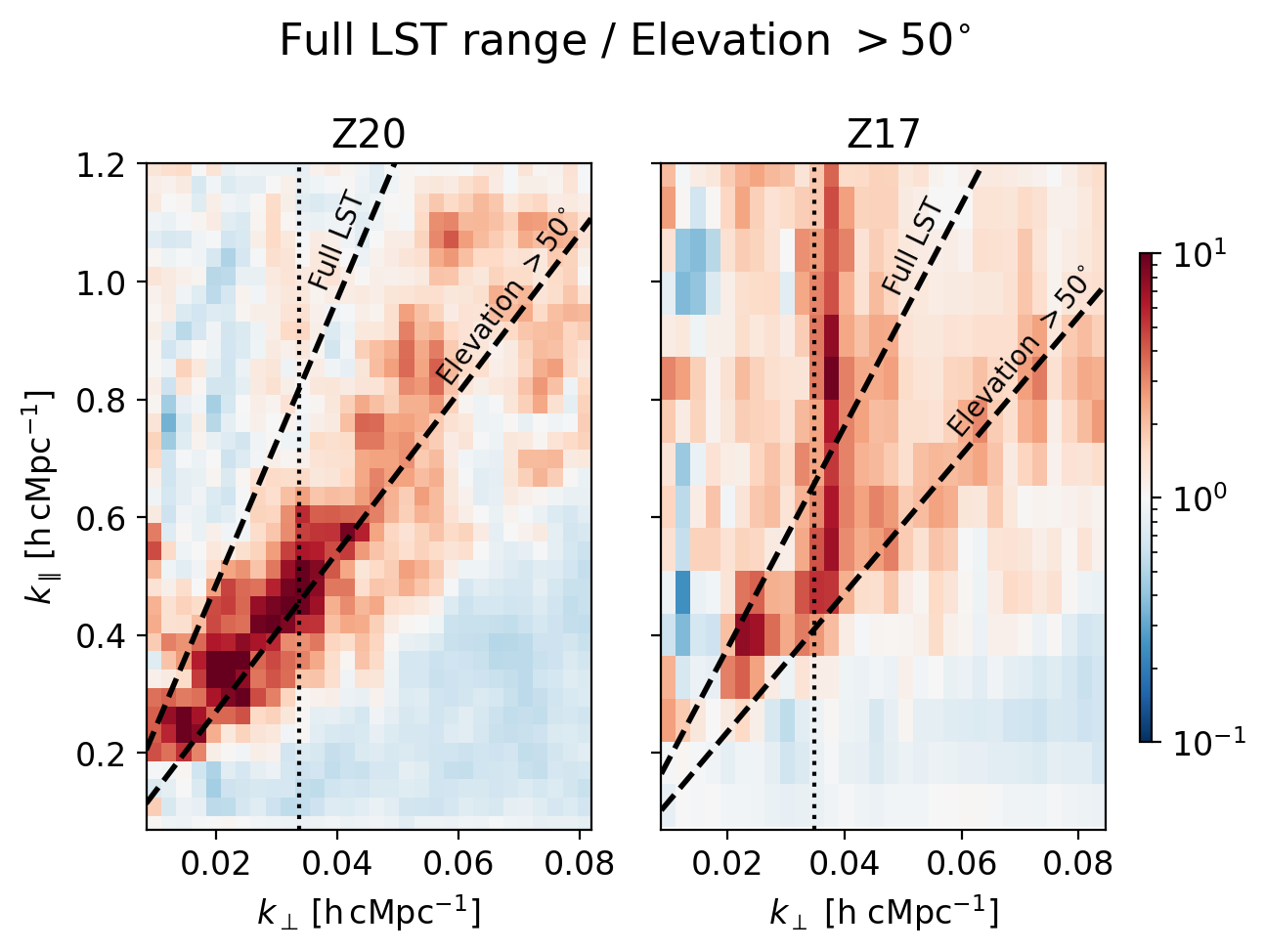}
    \caption{The ratio of power spectra estimated from the full data to that estimated from the LST range where the phase centre elevation is above $50^{\circ}$. The two columns correspond to the two spectral windows. The black dashed lines indicate the full-sky horizon limit, corresponding to the two cases. The vertical dotted lines indicate the baseline cut.}
    \label{fig:ps_alt}
\end{figure}
To reduce the foreground contamination at high $k_{\parallel}$, we estimate the power spectrum including only the LST range where the elevation of the NT04 field is higher than $50^{\circ}$ for all four nights. This is compared against the power spectrum estimated from the data for the full LST range in Fig.~\ref{fig:ps_alt}. The horizon delay line for both cases is plotted in black dashed lines. We find that confining the elevation range results in a reduction of power in the $k_{\perp},k_{\parallel}$ modes between the two horizon lines for both spectral windows, with the reduction for Z20 being over an order of magnitude in some modes.\footnote{Since the horizon delay lines indicate the maximum extent of the wedge, this extra power need not be confined completely between the horizon lines. Specifically, this power corresponds to the modes occupied by the foreground sources when the target field is below $50^{\circ}$ elevation.} We find that for Z17, this also reduces power above the full LST range horizon line. This power could possibly be caused by RFI sources with a non-smooth spectrum, during the LST range when the phase center is below $50^{\circ}$ elevation. The exact origin of the higher contamination on baselines with lengths close to the baseline cut is currently unknown. As expected, we find an increase in power in other modes caused by increased thermal noise power due to the smaller integration time. Still, the significant reduction in foreground power at high $k_{\parallel}$ modes far outweighs this minor reduction of sensitivity, especially since high $k_{\parallel}$ modes are the hardest to clean in later steps. We note that the NCP is at a fixed elevation, and the horizon line is fixed throughout the observation duration, making it impossible to perform such an optimisation of the LST range to confine the foreground contamination to low $k_{\parallel}$.

\section{Residual foreground removal}\label{sec:gpr}
After the calibration-based sky-model subtraction, RFI flagging in the $u\varv$ plane, and elevation-based data selection, the data are still dominated by unmodelled foreground emission that is several orders of magnitude stronger than the thermal noise level. In NenuFAR data, this is constituted primarily by foreground emission within the main lobe of the primary beam below the confusion noise limit and residuals of bright off-axis sources, which still contaminate the target field through their PSF sidelobes. These foregrounds can be disentangled from the 21-cm signal based on their spectral smoothness, compared to the intrinsic spectral fluctuations of the 21-cm signal along the line-of-sight. We perform Gaussian process regression \citep[GPR: ][]{mertens2018statistical} on the combined gridded visibility cube in the $u\varv\nu$ space to model and subtract such residual foreground contributions from the data.

\subsection{Gaussian process regression}
Following \citetalias{munshi2024first}, we use machine-learning enhanced GPR \citep[ML-GPR:][]{mertens2024retrieving,acharya202421} for residual foreground removal. In the context of foreground removal in 21-cm cosmology, GPR models the gridded data in the $u\varv\nu$ space ($\mathbf{d}$) as a sum of Gaussian processes along frequency that describe foregrounds ($\mathbf{f}_{\mathrm{fg}}$), the 21-cm signal ($\mathbf{f}_{21}$), and thermal noise ($\mathbf{n}$). Often, an additional excess variance component ($\mathbf{f}_{\mathrm{ex}}$) is needed to account for power at small frequency coherence scales that cannot be described by the foreground or the 21-cm signal components. The GPR model can then be written as
\begin{equation}
\mathbf{d} = \mathbf{f}_{\mathrm{fg}}(\nu) + \mathbf{f}_{21}(\nu) + \mathbf{f}_{\mathrm{ex}}(\nu) + \mathbf{n}(\nu).
\end{equation}
Each component is characterised by a frequency-frequency covariance function that describes its spectral behaviour. The total covariance kernel of the data ($\mathbf{K}$) is then a sum of the covariance kernels of the foreground ($\mathbf{K}_{\mathrm{fg}}$), 21-cm signal ($\mathbf{K}_{21}$), noise ($\mathbf{K}_{n}$) and excess ($\mathbf{K}_{\mathrm{ex}}$) components:
\begin{equation}
\mathbf{K} = \mathbf{K}_{\mathrm{fg}} + \mathbf{K}_{21} + \mathbf{K}_{\mathrm{ex}} + \mathbf{K}_{n}.
\end{equation}
The covariance kernels for the foreground and excess components are generated from analytical covariance functions that are parametrised by a set of hyperparameters, such as the frequency coherence scale and variance of the components. We use covariance functions from the Matern class, defined by
\begin{equation}\label{eq:matern}
K_{\mathrm{Matern}}(\Delta\nu) = \sigma^{2}\dfrac{2^{1-\eta}}{\Gamma(\eta)}\left(\dfrac{\sqrt{2\eta}\Delta\nu}{l}\right)^{\eta}K_{\eta}\left(\dfrac{\sqrt{2\eta}\Delta\nu}{l}\right).
\end{equation}
Here, $\Gamma$ is the Gamma function, $K_{\eta}$ is the modified Bessel function of the second kind, $\eta$ defines the nature of the Matern class function, $\Delta\nu$ is the absolute frequency separation, and $\sigma^2$ and $l$ are hyperparameters that describe the variance and frequency coherence scale of the component, respectively. For the 21-cm signal covariance kernel, however, instead of an analytical covariance function, a variational auto-encoder (VAE) kernel trained on simulations of the 21-cm signal is used in order to describe the signal of interest using physically motivated models. Table~\ref{tab:mlgpr} summarises the GPR covariance model. All $\sigma^2$ values are specified as fractions of the data variance in log space, and the $l$ values are in units of MHz. Below, we describe the covariance kernels that are used for the different components in ML-GPR.
\begin{table}
\caption{The covariance model used in ML-GPR for both spectral windows. The prior columns indicate the prior bounds in log space as a fraction of data variance for all $\sigma^2$ parameters and in linear space in units of MHz for all $l$ parameters. In the posterior column, parameters that are unconstrained within the prior range are indicated with a `-', and for those that reach the lower prior boundary, only the upper limit is listed.}
\label{tab:mlgpr}
\centering
\renewcommand{\arraystretch}{1.5}
\begin{tabular}{@{}lccccc@{}}
\toprule
\multirow{2}{*}{Kernel} & \multirow{2}{*}{Parm.} & \multicolumn{2}{c}{Z20} & \multicolumn{2}{c}{Z17} \\ 
\cmidrule(lr){3-4} \cmidrule(lr){5-6}
& & Prior & Posterior & Prior & Posterior \\ \midrule\midrule
\multirow{2}{*}{\parbox{0.5cm}{$\mathbf{K}_{\mathrm{int}}$ (RBF)}}   & $\sigma^{2}$ & $-0.5,0.5$  & $0.104^{+0.014}_{-0.016}$   & $-0.5,0.5$  & $0.046^{+0.011}_{-0.013}$  \\
                        & $l$          & $20,60$ & $27.171^{+0.563}_{-0.681}$    & $20,60$ & $30.705^{+0.857}_{-0.721}$    \\ \midrule
\multirow{2}{*}{\parbox{0.5cm}{$\mathbf{K}_{\mathrm{mix}}$ (RBF)}} & $\sigma^{2}$ & $-2,-1$  & $-1.657^{+0.003}_{-0.004}$  & $-2,-1$  & $-1.157^{+0.008}_{-0.009}$  \\
                        & $l$          & $0.1,1$ & $0.503^{+0.001}_{-0.001}$ & $0.1,1$ & $0.582^{+0.003}_{-0.003}$ \\ \midrule
\multirow{3}{*}{\parbox{0.5cm}{$\mathbf{K}_{\mathrm{21}}$ (VAE)}} & $x_1$              & $-4,4$  & -                 & $-4,4$  & -                 \\
                        & $x_2$              & $-4,4$  & -                 & $-4,4$  & -                 \\
                        & $\sigma^{2}$  & $-6,-1$ & $<-3.208$         & $-6,-1$ & $< -2.749$         \\ \midrule
\multirow{2}{*}{\parbox{0.5cm}{$\mathbf{K}_{\mathrm{ex1}}$ (Matern 3/2)}}            & $\sigma^{2}$  & $-5,-2$  & $-4.586^{+0.048}_{-0.062}$  & $-5,-2$  & $-3.163^{+0.012}_{-0.014}$  \\
                        & $l$           & $0.02,0.1$   & $0.045^{+0.004}_{-0.003}$   & $0.02,0.1$   & $0.057^{+0.002}_{-0.002}$   \\ \midrule
\multirow{2}{*}{\parbox{0.5cm}{$\mathbf{K}_{\mathrm{ex2}}$ (Matern 3/2)}}            & $\sigma^{2}$  & $-5,-2$  & $-3.812^{+0.067}_{-0.050}$  & $-4,-1$  &$-2.159^{+0.088}_{-0.092}$\\
                        & $l$           & $0.1,1$   &$0.251^{+0.025}_{-0.020}$   & $0.1,1$   & $0.561^{+0.045}_{-0.048}$ \\ \bottomrule
\end{tabular}
\end{table}

\begin{figure}
    \centering
    \includegraphics[width=\columnwidth]{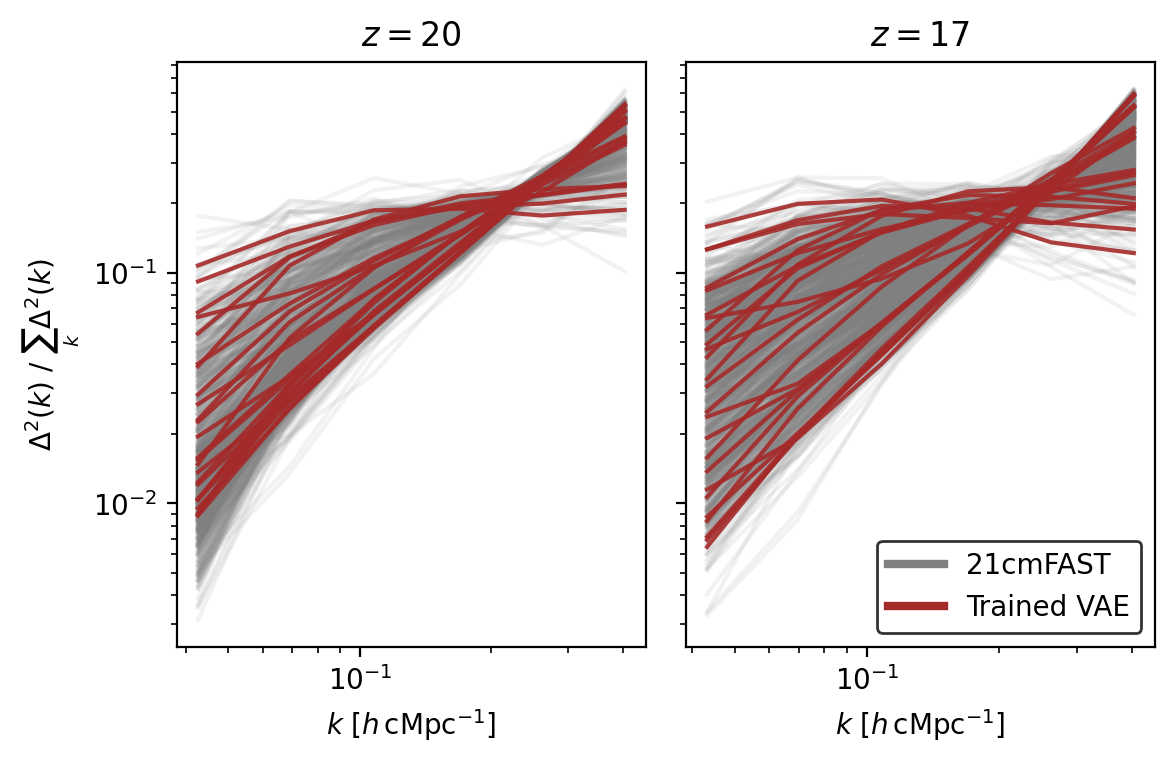}
    \caption{Power spectrum shapes captured by the 21-cm covariance kernel in ML-GPR. The normalised power spectra of the 1000 \texttt{21cmFAST} simulations for both redshifts are shown in grey curves. The normalised power spectra generated by the trained VAEs at a set of uniformly spaced points in the latent space are overplotted in brown.}
    \label{fig:gpr_train}
\end{figure}

\paragraph*{Foregrounds:} We divide the foreground component further into an intrinsic component $\mathbf{f}_{\mathrm{int}}$ with covariance $\mathbf{K}_{\mathrm{int}}$ and a mode-mixing component $\mathbf{f}_{\mathrm{mix}}$ with covariance $\mathbf{K}_{\mathrm{mix}}$. The intrinsic component describes the power at the lowest $k_{\parallel}$, which is constituted by residual foregrounds within the field of view and has extremely large frequency coherence scales. The mode-mixing component captures the off-axis foreground sources that occupy the rest of the foreground wedge created by the chromatic and non-ideal $u\varv$ coverage. For both the intrinsic and mode-mixing foreground components, we use radial basis function (RBF) covariance kernels, which are obtained by setting $\eta=\infty$ in equation~(\ref{eq:matern}). RBF covariances have a rapid decay in power along $k_{\parallel}$ and thus give rise to frequency-smooth functions that are ideal for modelling foregrounds, enabling easier separation from the 21-cm signal and other components with spectral fluctuations. We distinguish between the intrinsic and mode-mixing foregrounds based on the priors on $l$, with the intrinsic component having priors at much higher $l$ values. In future, this will be replaced by a physically motivated foreground covariance kernel that naturally produces the mode-mixing covariance using the primary beam models and foreground source distribution functions.
\paragraph*{21-cm signal:} The 21-cm covariance kernels are VAEs trained on simulations of the 21-cm signal. The code \texttt{21cmFAST} \citep{mesinger201121cmfast,murray202021cmfast} is used to perform simulations of the 21-cm signal at $z=17$ and 20 by varying a set of astrophysical parameters. We reuse the trained kernel that was used by \citetalias{munshi2024first} for $z=20$. For $z=17$, we perform an additional 1000 simulations of the 21-cm signal, estimate the power spectra, and train a VAE kernel (see \citetalias{munshi2024first} for details about the simulations and training). We note that the VAE is trained to capture only the shape of the 21-cm power spectrum, described by the two latent space dimensions $x_1$ and $x_2$, while the variance is left as a free parameter optimised during GPR. This enables the trained VAE to describe not only the standard 21-cm signal models but also extreme and exotic models of the 21-cm signal that can have a significantly boosted power spectrum\footnote{While this naturally accounts for excess radio backgrounds, if the power spectrum boost arises from effects that alter the 21-cm power spectrum shapes, the framework assumes that such shapes can still be described by the variety of shapes accommodated by the trained VAE.}. Fig.~\ref{fig:gpr_train} shows the normalised power spectra obtained from the \texttt{21cmFAST} simulations at both redshifts (in grey) and the power spectra generated from the trained VAEs at 25 uniformly spaced points in the two-dimensional (2D) latent space (in brown). We see that the VAE is able to capture the variety of power spectrum shapes at both redshifts.

\paragraph*{Excess:} To account for residual power in the data at smaller frequency coherence scales than the foregrounds, we need to include an excess component. The excess component is described by a Matern covariance kernel with $\eta=3/2$, which gives it a covariance structure that decays much slower than foregrounds along the $k_{\parallel}$ direction, enabling it to capture rapidly fluctuating spectral structures that are still correlated in frequency. Here, we divide the excess further into two components: $\mathbf{f}_{\mathrm{ex1}}$ (with covariance $\mathbf{K}_{\mathrm{ex1}}$), that captures very small coherence scale structures in the data and accounts for almost noise-like excess variance above the estimated thermal noise, and $\mathbf{f}_{\mathrm{ex2}}$ (with covariance $\mathbf{K}_{\mathrm{ex2}}$), which has a larger coherence scale, and accounts for residual power in the data possibly caused by low level systematics.  The priors on $l$ are chosen in such a way as to reflect the corresponding frequency coherence scales.
\paragraph*{Thermal noise:} The thermal noise covariance is a diagonal matrix, with its variance estimated from the time-differenced Stokes~\textit{V} visibility cube.

\begin{figure}
    \centering
    \includegraphics[width=\columnwidth]{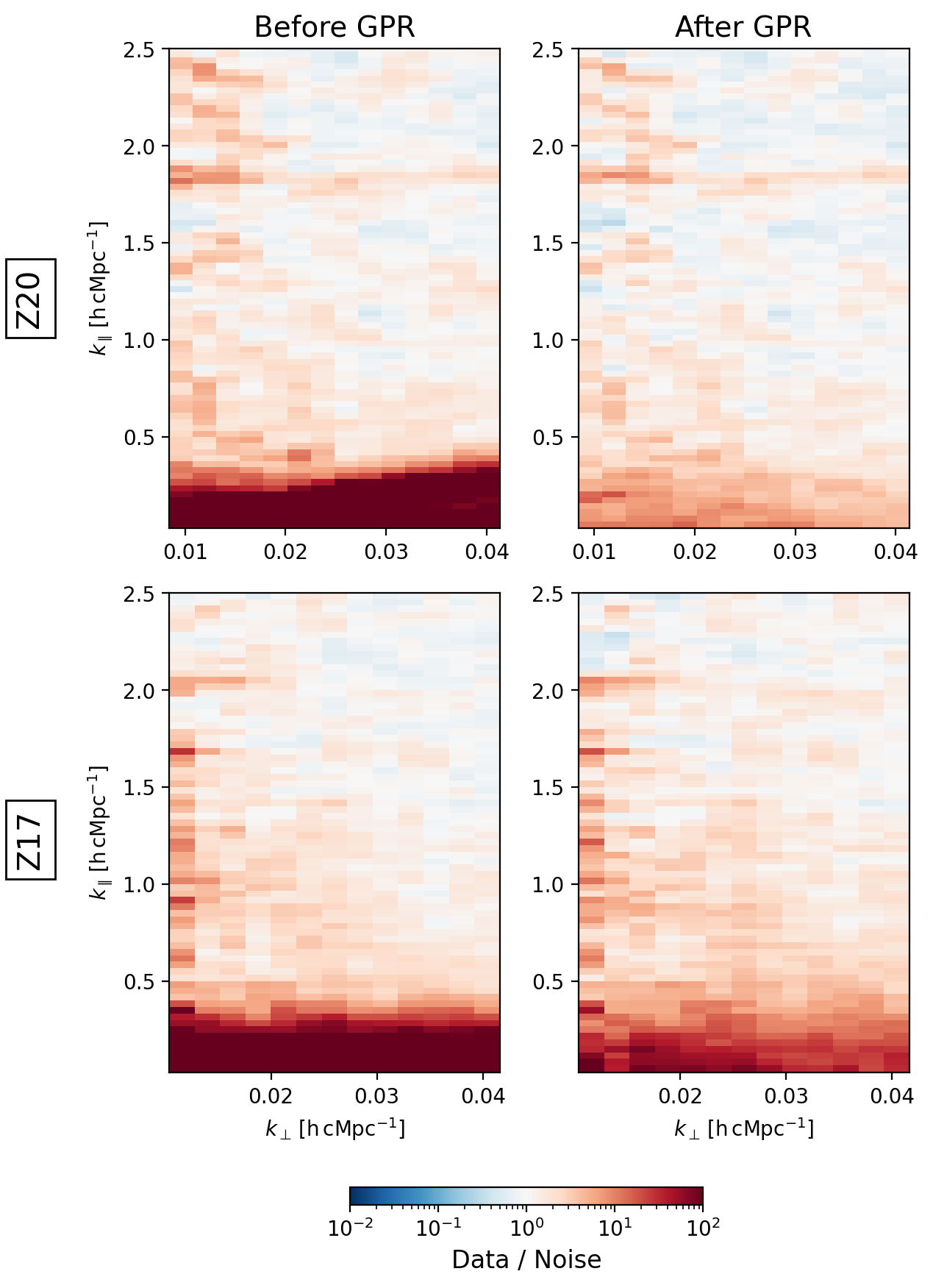}
    \caption{Cylindrical power spectra, expressed as a ratio over the thermal noise power spectrum, before and after GPR-based foreground subtraction. The two rows correspond to the two spectral windows.}
    \label{fig:gpr_res}
\end{figure}

\begin{figure*}
    \centering
    \includegraphics[width=0.75\columnwidth]{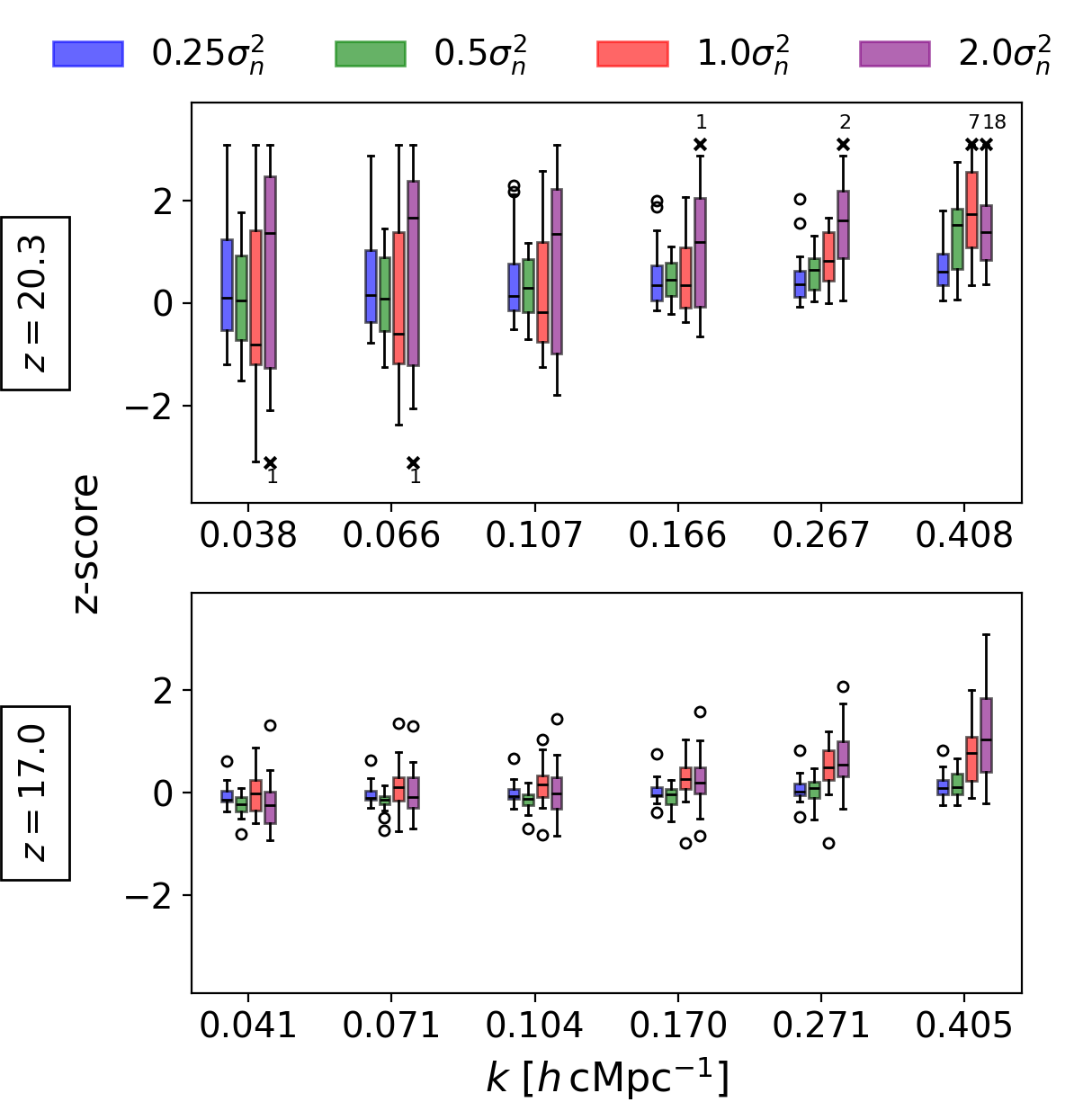}
    \includegraphics[width=1.25\columnwidth]{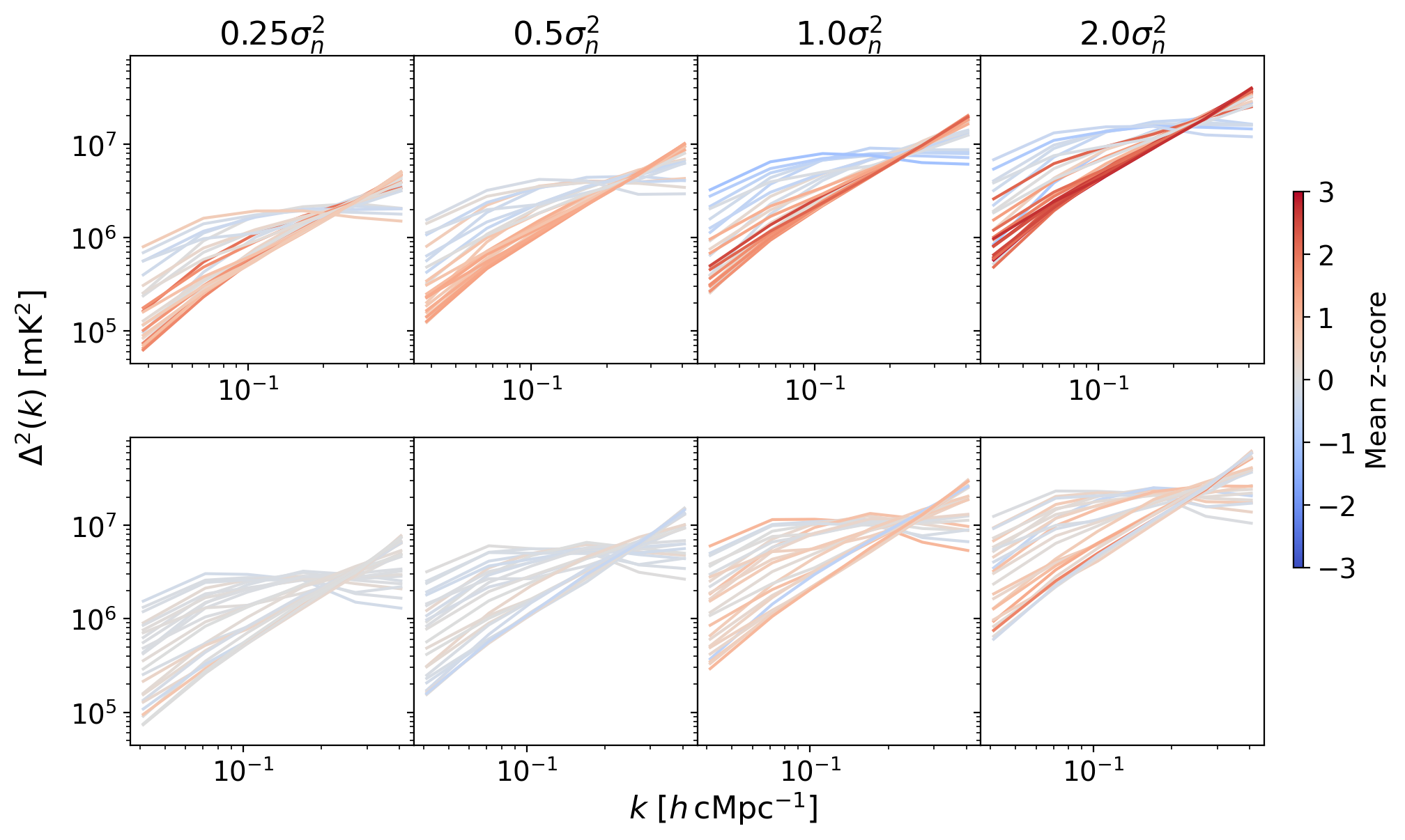}
    \caption{Results of the signal injection tests performed on GPR for each spectral window. The two rows correspond to the two redshift bins. The left panels show the distribution of the z-scores as a function of $k$ values and injected signal variances. The black crosses with the numbers alongside them indicate the number of signals for which the recovery is beyond the 3$\sigma$ limit. The right panel shows the mean z-score as a function of the shape and strength of the injected signal power spectrum.}
    \label{fig:z_scores}
\end{figure*}

\vspace{0.5cm}
We apply GPR separately to the visibility cubes of the two spectral windows. The posterior distribution of the hyperparameters of the Gaussian process covariance model is estimated with Nested sampling using \texttt{dynesty} \citep{speagle2020dynesty}, and is shown in Fig.~\ref{fig:corner} of Appendix \ref{sec:gpr_post}. We find that all parameters are constrained within compact regions in the parameter space within the prior bounds, except the 21-cm component. Here, the parameters controlling the 21-cm power spectrum shape, $x_1$ and $x_2$, remain unconstrained, and the variance $\sigma^2$ reaches the lower prior boundary, indicating that the 21-cm signal is not detected. The lengthscale of $\mathbf{K}_\mathrm{int}$ converges to values of 27.2 MHz and 30.7 MHz for Z20 and Z17, respectively, and thus describes extremely frequency smooth components, with lengthscales well above the bandwidth of 11.5 MHz. Small differences in the converged intrinsic foreground lengthscales between the two spectral windows are thus expected, firstly because they describe different portions of the foreground source spectrum, and secondly because the primary beams are different. $\mathbf{K}_\mathrm{mix}$ and $\mathbf{K}_\mathrm{ex1}$ also converge to similar lengthscales for both Z17 and Z20. We note, however, that $\mathbf{K}_\mathrm{ex2}$ converges to significantly different lengthscales for Z20 and Z17, with its variance being a higher fraction of the data variance for Z17 compared to Z20 (Table~\ref{tab:mlgpr}). This is likely because $\mathbf{K}_\mathrm{ex2}$ describes the excess variance due to the RFI contamination, which can have different spectral signatures in Z17 and Z20. The higher excess variance in Z17 is also expected due to the significantly more severe RFI contamination in Z17 compared to Z20 (Fig.~\ref{fig:rfi_stats}) and was evident in the cylindrical power spectra before residual foreground removal (Fig.~\ref{fig:ps_stages}).

For each sample in the posterior distribution, the predictive mean and covariance of the foreground component can now be estimated
\begin{align}
\mathcal{E}(\mathbf{f}_{\mathrm{fg}}) &= \mathbf{K}_{\mathrm{fg}}\mathbf{K}^{-1}\mathbf{d},\nonumber\\
\mathrm{cov}(\mathbf{f}_{\mathrm{fg}}) &= \mathbf{K}_{\mathrm{fg}}-\mathbf{K}_{\mathrm{fg}}\mathbf{K}^{-1}\mathbf{K}_{\mathrm{fg}}
\end{align}
A realisation of the foreground component is then subtracted from the visibility cube to yield the residual data cube. Following the approach described in section 3.8.1 of \cite{mertens2025deeper}, an ensemble of power spectra estimated from the residual data cubes corresponding to 1000 hyperparameters sampled from their posterior distribution is then used to estimate the median residual power spectrum with uncertainties. The cylindrical power spectra estimated before and after GPR for the two spectral windows are shown in Fig.~\ref{fig:gpr_res}, as ratios over the corresponding thermal noise power spectra. To minimise the artefacts in the cylindrical power spectra before GPR caused by the flagged channels in Z17, we interpolate the gridded data values at the flagged channels using the converged data covariance ($\mathbf{K}$) following the description in section~3.8.2 of \cite{mertens2025deeper}. This interpolation is performed only for visualisation and is not used anywhere else, as post-GPR power spectra inherently suppress such artefacts due to the reduced dynamic range along $k_{\parallel}$. Additionally, here the Fourier transformation is performed using least squares spectral analysis \citep{vanivcek1969approximate} to further suppress these artefacts. We find that the power at low $k_{\parallel}$ is reduced significantly through GPR, leaving excess power above the thermal noise level, which is much stronger in Z17 than Z20, likely caused by stronger RFI contamination, as discussed earlier. The residual power near $k_{\parallel}=1.8\ h\,\mathrm{cMpc}^{-1}$ (for Z20) and $k_{\parallel}=2.0\ h\,\mathrm{cMpc}^{-1}$ (for Z17) are artefacts caused by the polyphase filterbank used in channelisation into 195.3 kHz sub-bands.

\subsection{Signal injection tests}
To verify the robustness of the foreground removal with GPR, we perform a suite of signal injection tests. The test consists of injecting the visibilities corresponding to the 21-cm signal into the gridded data cube before GPR, running GPR in the same manner as done on the data, and comparing the recovered 21-cm signal against the injected signal to determine the level of bias. For a detailed description of how the signal injection tests are performed, we refer the reader to \citetalias{munshi2024first}. For each spectral window, the injection test is repeated for 25 different shapes of the signal sampled uniformly within the range $x_1,x_2\in[-2,2]$ from the 2D latent space of the VAE. For each shape, we repeat the test for four different strengths of the input 21-cm signal corresponding to 0.25, 0.5, 1 and 2 times the thermal noise variance. For each of these 100 injected signals, we use the z-score as a metric for the level of bias that is incurred from GPR. The recovered power spectrum, corresponding to 1000 different points from the posterior distribution of the GPR hyperparameters, is estimated for each injected signal, and the z-score at each $k$ bin is obtained by calculating the inverse CDF of the p-value under the assumption of a Gaussian error distribution. The z-score represents the number of standard deviations by which the recovered signal over- or underestimates the injected signal, with values below -2 indicating a 21-cm signal absorption beyond 2$\sigma$ error bars. 

The results from all 100 injection tests for each spectral window are summarized in Fig.~\ref{fig:z_scores}. The left panel shows the z-scores at each $k$ bin, separately for the different strengths of the injected signal. The spread in the z-score values is caused by the results from the different shapes of the injected signals. For $z=20.3$, we find that there is a positive bias at high $k$ bins, and here, several recovered signals are more than 3$\sigma$ above the injected signal. At the lower $k$ bins, the z-score distributions are relatively well-centred around zero, as we would expect if the recovery is unbiased. There are a few $k$ bins for which the z-score goes below -2, but these outliers constitute 1.33$\%$ out of all computed z-score values. While there is a positive bias also at large $k$ bins for $z=17.0$, the z-scores are closer to zero compared to $z=20.3$, indicating that the injected signals are better constrained within the recovered signal uncertainties. This is due to the larger error bars and the stronger excess component in $z=17.0$, which can more easily absorb a portion of the 21-cm signal when the VAE kernel fails to do so. The right panel of Fig.~\ref{fig:z_scores} illustrates the dependence of the average z-score on the shape of the injected signal for each of the four different injected signal strengths. Flatter signal power spectrum shapes have a lower z-score on average compared to steeper power spectra, which are better recovered with GPR, while often incurring a positive bias. We also find that such a positive bias is stronger in injected signals with more power (also seen in the left panel for higher $k$ values). This occurs because stronger signals are recovered with greater precision, resulting in a smaller standard deviation and, consequently, a higher estimated z-score. Nonetheless, since our analysis focuses on setting upper limits on the 21-cm power spectrum, this positive bias is not a limiting factor.

\section{Results}\label{sec:results}
In this section, we discuss the main results from this analysis and set upper limits on the 21-cm signal power spectrum. We then investigate the possible origin of the excess variance seen in the residual data.
\subsection{Comparison with the NCP field}\label{sec:comparison_ncp}
\begin{figure}
    \centering
    \includegraphics[width=\columnwidth]{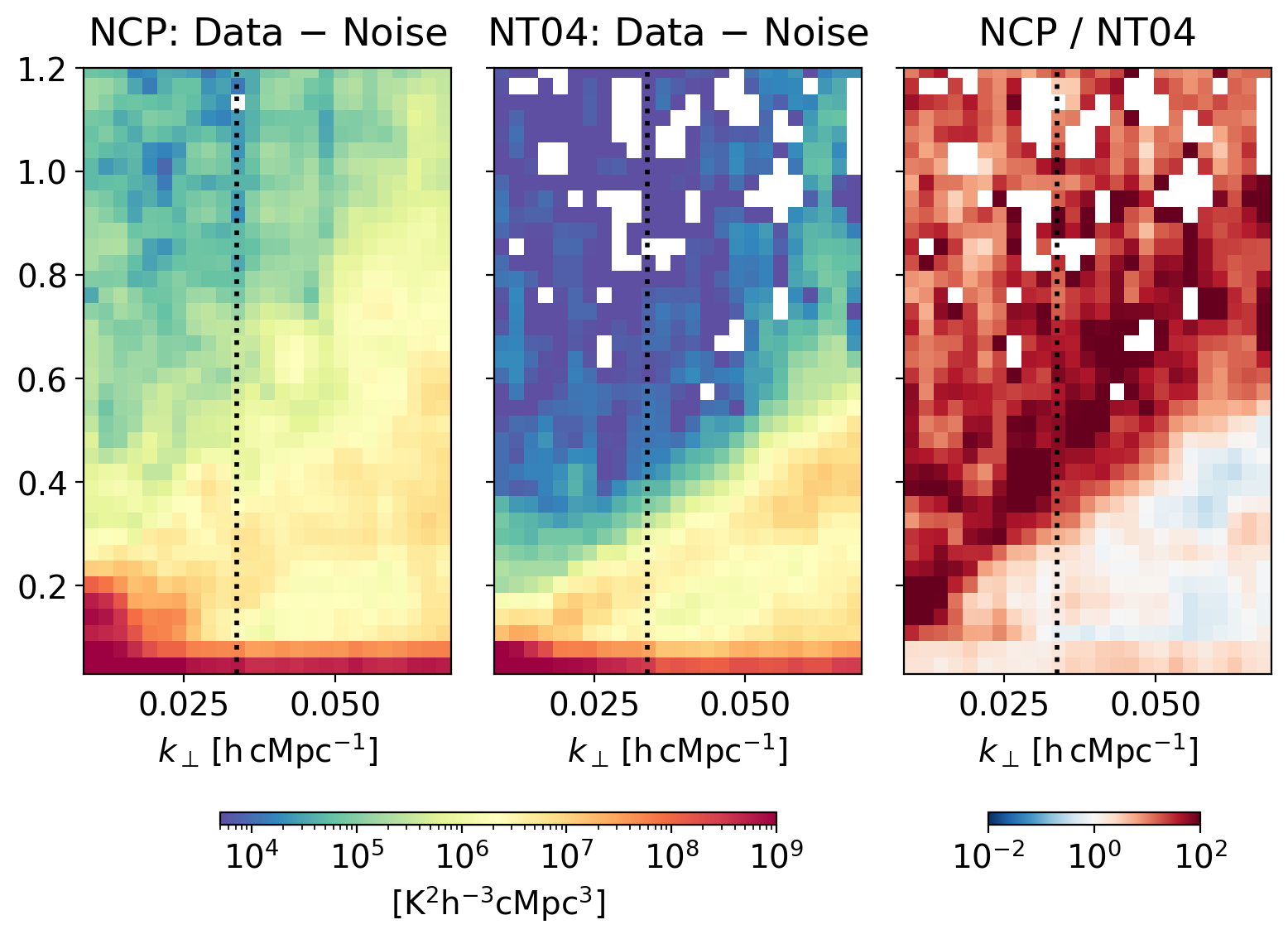}
    \caption{Power spectrum of the Z20 data before GPR, compared against the results obtained from the same spectral window from observations of the NCP field by \citetalias{munshi2024first} at the same stage of analysis. The three panels show the noise-bias-subtracted cylindrical power spectra, corresponding to this (NT04) analysis, the NCP analysis, and the ratio of the NCP and NT04 power spectra.}
    \label{fig:ps_ncp}
\end{figure}

A key distinction in our analysis is the shift from the NCP field used by \citetalias{munshi2024first} to the NT04 field, selected through a targeted survey of multiple candidate fields (Mertens et al., in prep.), and the improvements in reducing the excess variance that this enabled (Section~\ref{sec:ps_estimation}). In Fig.~\ref{fig:ps_ncp}, we compare the cylindrical power spectrum from Z20 after calibration and sky model subtraction to that of the NCP field at the same stage. Since the NT04 field integration is $\sim 2.3$ times longer than the NCP observation, to leave the difference in noise power out of the comparison, we subtract the corresponding thermal noise power spectrum estimated from time-differenced Stokes~\textit{V} data from both power spectra before plotting. The modes with negative power correspond to the noise-dominated regions within the EoR window and are indicated in white. The foreground power seems to be much better confined to low $k_{\parallel}$ modes in the NT04 field analysis, and the residual wedge-like power in the NT04 data is possibly caused by the large number of radio sources in the grating lobes of the NenuFAR primary beam. In the right-most panel of Fig.~\ref{fig:ps_ncp}, the ratio of the noise-bias-subtracted power spectra for the NCP and the NT04 analyses is shown. We find a reduction in power by at least an order of magnitude in the entire EoR window, and by over two orders of magnitude for some $k_{\perp},k_{\parallel}$ modes. This suggests that the improved choice of the field, combined with improvements in data processing methods and the ability to flag RFI contributions and confine the foreground wedge, significantly reduces the excess variance that was observed in the analysis by \citetalias{munshi2024first}. The power in the foreground wedge remains at a roughly similar level for the two fields. The reduction in power at the lowest $k_{\parallel}$ modes in the NT04 power spectrum could be attributed to the better target subtraction using more clusters in this analysis, enabled by the higher angular resolution due to an additional remote MA being used during the NT04 observation. The factor of $\approx 2$ reduction in the power in these modes matches the value expected from the confusion noise values reported in Section~\ref{sec:calibration}.

\begin{table}
\caption{$2\sigma$ upper limits on the 21-cm signal power spectrum derived from this analysis.}
\label{tab:upper_limits}
\centering
\begin{tabular}{@{}cccccc@{}}
\toprule
\multicolumn{3}{c}{$z = 20.3$} & \multicolumn{3}{c}{$z = 17.0$} \\
\cmidrule(lr){1-3} \cmidrule(lr){4-6}
$k$ & $\Delta^{2}_{21}(k)$ & $\Delta^{2}_{\mathrm{ul}}(k)$ &
$k$ & $\Delta^{2}_{21}(k)$ & $\Delta^{2}_{\mathrm{ul}}(k)$ \\
$[h/\mathrm{cMpc}]$ & $[\mathrm{mK}^{2}]$ & $[\mathrm{mK}^{2}]$ &
$[h/\mathrm{cMpc}]$& $[\mathrm{mK}^{2}]$ & $[\mathrm{mK}^{2}]$ \\
\midrule
0.038 & $2.1 \times 10^{5}$ & $4.6 \times 10^{5}$ & 0.041 & $2.5 \times 10^{6}$ & $5.0 \times 10^{6}$ \\
0.066 & $9.0 \times 10^{5}$ & $1.8 \times 10^{6}$ & 0.071 & $1.2 \times 10^{7}$ & $2.3 \times 10^{7}$ \\
0.107 & $3.2 \times 10^{6}$ & $5.6 \times 10^{6}$ & 0.104 & $3.0 \times 10^{7}$ & $5.8 \times 10^{7}$ \\
0.166 & $9.3 \times 10^{6}$ & $1.5 \times 10^{7}$ & 0.170 & $8.3 \times 10^{7}$ & $1.4 \times 10^{8}$ \\
0.267 & $2.6 \times 10^{7}$ & $3.7 \times 10^{7}$ & 0.271 & $1.4 \times 10^{8}$ & $2.1 \times 10^{8}$ \\
0.408 & $4.0 \times 10^{7}$ & $4.9 \times 10^{7}$ & 0.405 & $1.8 \times 10^{8}$ & $2.5 \times 10^{8}$ \\
\bottomrule
\end{tabular}
\end{table}

\begin{figure}
    \centering
    \includegraphics[width=\columnwidth]{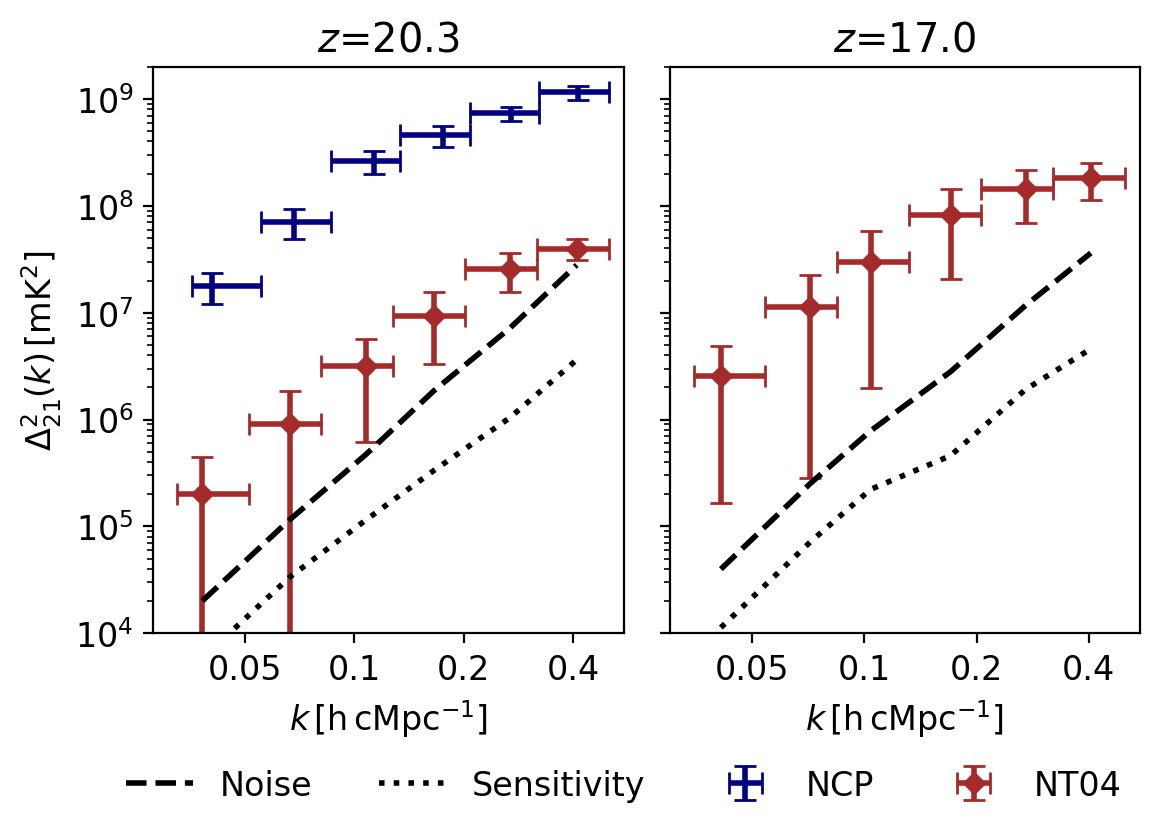}
    \caption{Upper limits on the 21-cm signal power spectrum. The results from this analysis on the NT04 field are shown in brown, while those from the NCP analysis by \citetalias{munshi2024first} are shown in blue. The two panels show the two redshift bins. The black dashed and dotted lines indicate the thermal noise power spectrum and $2\sigma$ sensitivity, respectively.}
    \label{fig:upper_limits}
\end{figure}
\subsection{Power spectrum upper limits}
After GPR-based foreground removal, the region outside the $u\varv$ range of $10\lambda-40\lambda$ is filtered out from the visibility cube. The spherical power spectrum and uncertainties are computed in logarithmically spaced bins in $k$ space, ranging from the smallest accessible $k$ value to a maximum value of $k=0.5\,h\, \textrm{cMpc}^{-1}$. Next, the thermal noise power spectrum is subtracted to yield the noise-bias-subtracted residual power spectrum ($\Delta^{2}_{21}$). We set upper limits on the 21-cm signal at a 2$\sigma$ level above the residual power spectrum at both redshifts ($\Delta^{2}_{\mathrm{ul}}$). Unlike \citetalias{munshi2024first}, a calibration bias correction step is not required here since the baselines used in calibration ($>40\lambda$) are not used in estimating the upper limits. Fig.~\ref{fig:upper_limits} shows the resulting noise-bias-subtracted power spectra. For $z=20.3$, the corresponding limits obtained by \citetalias{munshi2024first} are also shown. The vertical error bars indicate the $2\sigma$ errors on $\Delta^{2}_{21}$, and the horizontal error bars correspond to the $k$ bin range. The dashed and dotted lines are the estimated thermal noise power spectrum and the $2\sigma$ uncertainties on the thermal noise, respectively, the latter describing the sensitivity of the instrument and indicating the minimum value that the upper limits can have. We find that for $z=20.3$, the data power spectrum is within an order of magnitude of the thermal noise power at all $k$ bins, while the upper limits are still up to two orders of magnitude above the sensitivity. The upper limits derived from this analysis are an improvement to those obtained by \citetalias{munshi2024first} at the same redshift bin by over a factor of 50. For $z=17.0$, the data power spectrum is still more than an order of magnitude above the thermal noise power at all but the largest $k$ modes.

Table~\ref{tab:upper_limits} summarises the upper limits on the 21-cm signal obtained from this analysis. At $z=17.0$, the best 2$\sigma$ limit on the 21-cm signal power spectrum is $\Delta^{2}_{21} < 5.0 \times 10^6 \, \textrm{mK}^{2}$ at $k=0.041$ $h\, \mathrm{cMpc}^{-1}$. This limit suffers from excess variance, primarily caused by RFI contamination of NenuFAR observations in this spectral window. Still, this limit is considerably deeper than all Cosmic Dawn power spectrum limits set by other instruments. At $z=20.3$, the best 2$\sigma$ upper limit on the 21-cm signal power spectrum is $\Delta^{2}_{21} < 4.6 \times 10^5 \, \textrm{mK}^{2}$ at $k=0.038$ $h\, \mathrm{cMpc}^{-1}$. This limit is deeper than all previous Cosmic Dawn power spectrum upper limits by more than an order of magnitude \citep{ewall2016first,gehlot2019first,eastwood201921,gehlot2020aartfaac,yoshiura2021new,garsden202121,munshi2024first}.

\begin{figure}
    \centering
    \includegraphics[width=\columnwidth]{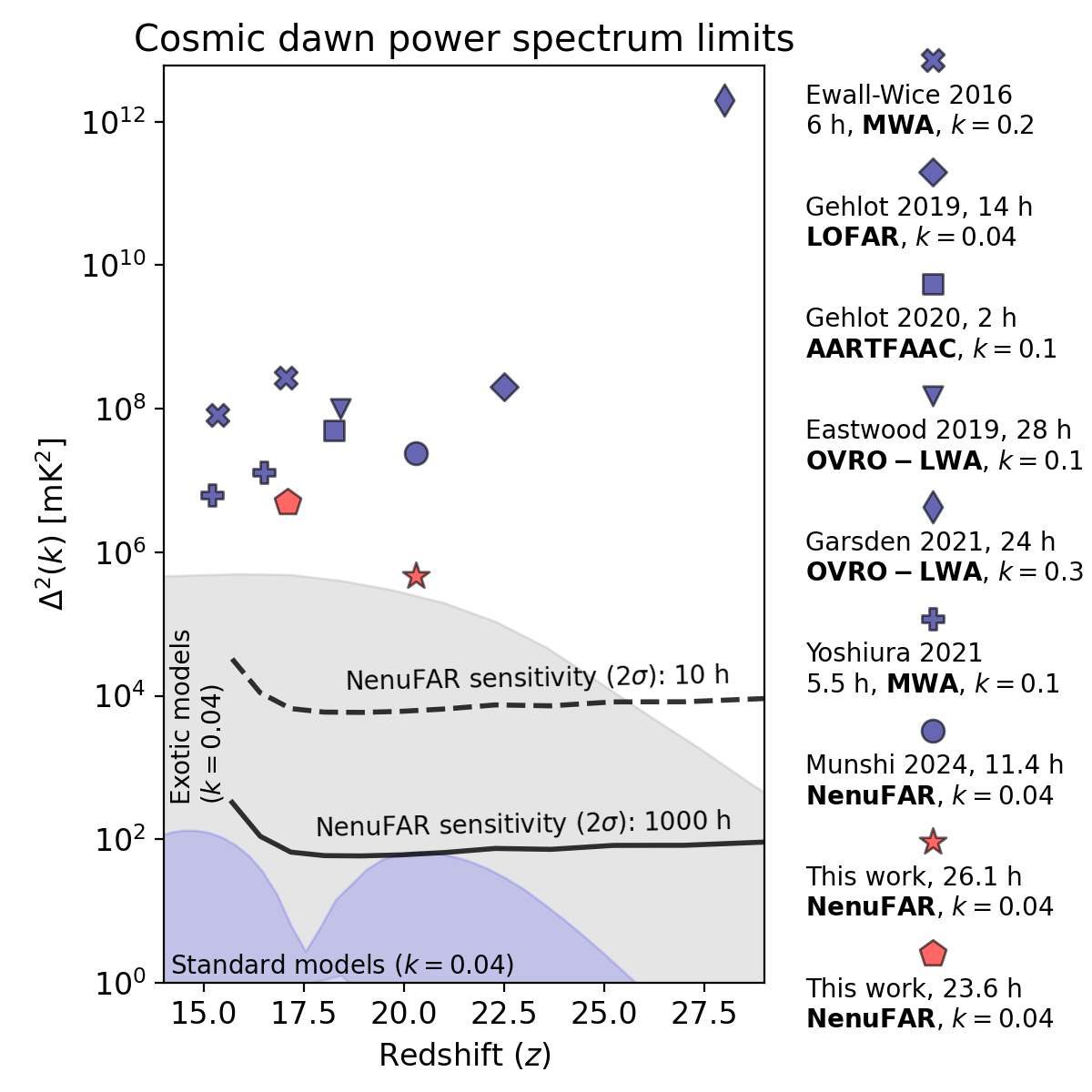}
    \caption{Cosmic dawn power spectrum upper limits from different interferometers. The range of power spectra expected from standard models (blue shaded region) and an approximate range of power spectra expected from exotic models (grey shaded region) are indicated. The $k$ values mentioned in this figure are in units of $h\,\mathrm{cMpc}^{-1}$.}
    \label{fig:cd_limits}
\end{figure}
Fig.~\ref{fig:cd_limits} presents a compilation of all 21-cm power spectrum upper limits from the Cosmic Dawn ($z>15$). The $2\sigma$ sensitivities of the full NenuFAR array configuration at $k=0.04\,h\,\mathrm{cMpc}^{-1}$ for 10 h and 1000 h of integration are simulated and indicated in the figure. The predicted range of power spectra of standard models is obtained from the Evolution of 21 cm Structure (EOS) simulation suite \citep{mesinger2016evolution} and shown in blue. An approximate range for exotic models is also shown in grey. A more comprehensive exploration of the exotic model power spectra and their comparison against the upper limits derived in this analysis is presented in Section~\ref{sec:interpretation}.

\subsection{Coherence of the excess variance}
In both spectral windows, we find excess variance that limits our ability to reach the theoretical sensitivity, with Z17 having a much stronger excess than Z20 due to RFI contamination. It is important to understand the nature of this excess variance in order to remove it or mitigate its impact and assess the future prospects of longer integrations of hundreds of hours of NenuFAR observations at both spectral windows. For this purpose, we use \texttt{pspipe} to compute gridded visibility cubes in brightness temperature units from the sky-model subtracted residual data of each 12 min time segment for all four nights. Next, these visibility cubes are combined by computing a weighted average using the corresponding weight cubes, taking an increasing number of cubes starting from the first cube of night 1, and the corresponding cylindrical power spectra are estimated. To understand the nature of the excess variance as a function of regions in the $k_{\perp},k_{\parallel}$ space, we define three regions:
\begin{enumerate}[left=0pt]
\item Main lobe: The region within an angular distance of $15^{\circ}$ from the phase centre, which contains the main lobe of the NenuFAR primary beam. This is given by the region in the $k_{\perp},k_{\parallel}$ space below the $15^{\circ}$ delay line derived under a flat sky approximation \citep{morales2012four}.\footnote{Though this delay line equation is not strictly correct for phase centres far from zenith \citep{munshi2025beyond}, they are still valid for such small angular distances where the flat-sky approximation holds ($\sin\frac{\pi}{12}\approx\frac{\pi}{12}$).}
\item Sidelobes: The region between the primary beam main lobe and the horizon. In the $k_{\perp},k_{\parallel}$ space, this corresponds to the region between the primary beam delay line and the steepest full sky horizon delay line across the four nights.
\item EoR window: The region in the $k_{\perp},k_{\parallel}$ space beyond the full-sky horizon delay line.
\end{enumerate}

\begin{figure*}
    \centering
    \includegraphics[width=2\columnwidth]{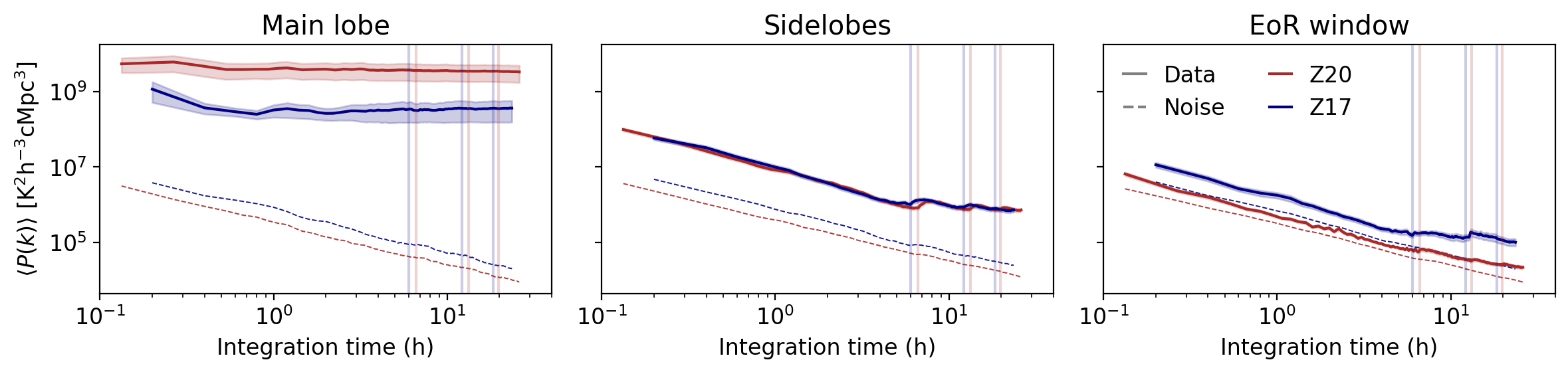}\vspace{0.5cm}
    \includegraphics[width=0.66\columnwidth]{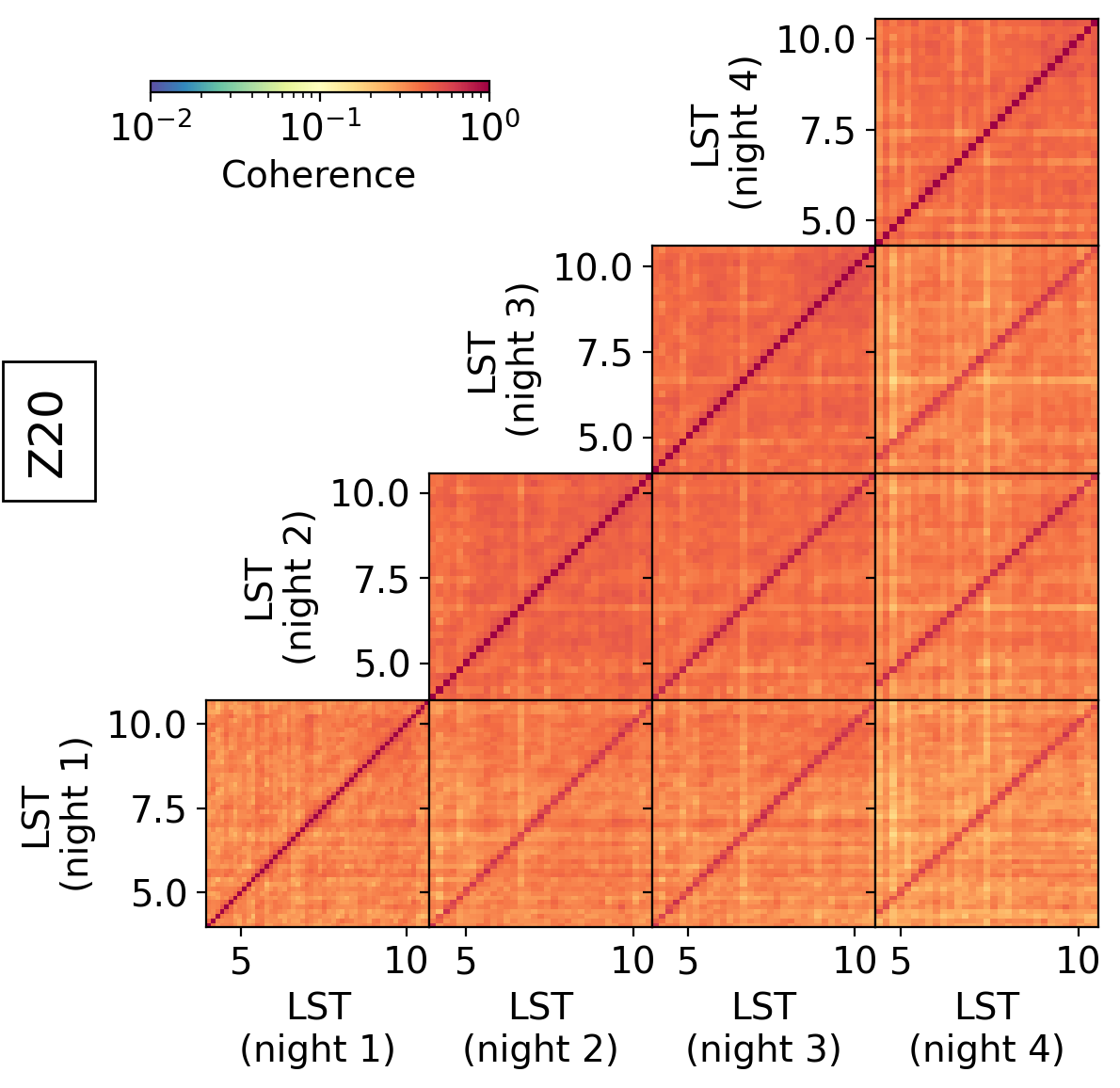}
    \includegraphics[width=0.66\columnwidth]{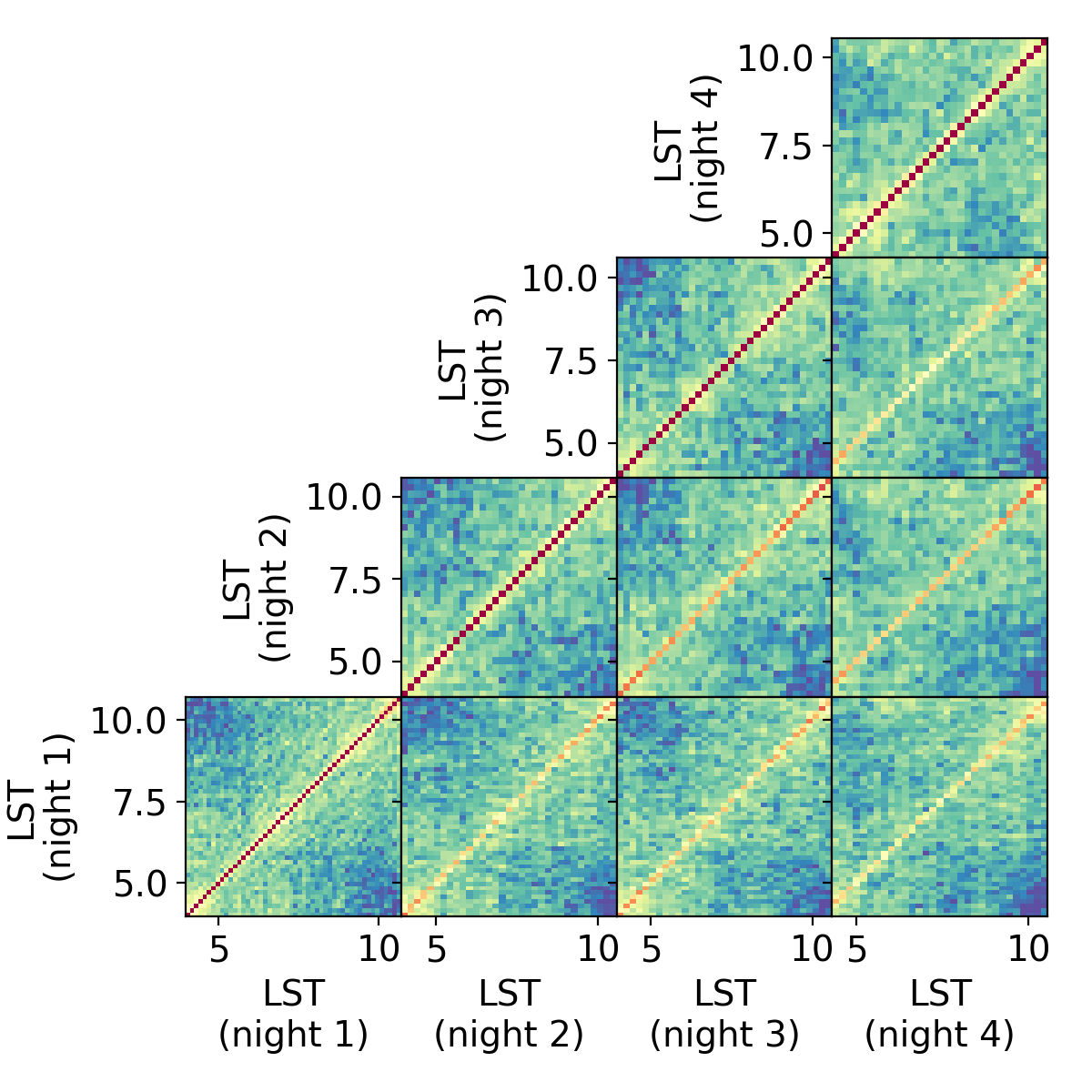}
    \includegraphics[width=0.66\columnwidth]{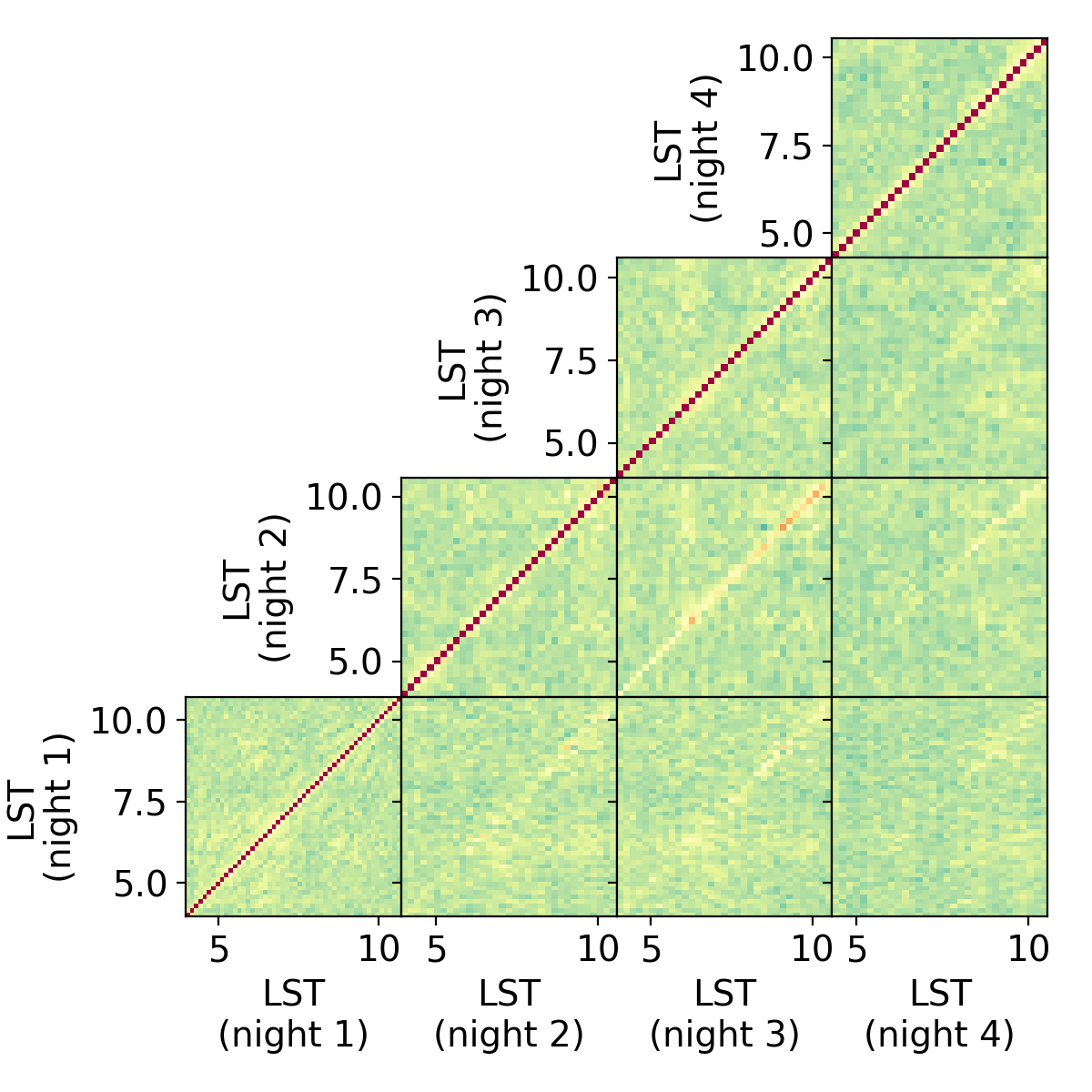}
    \includegraphics[width=0.66\columnwidth]{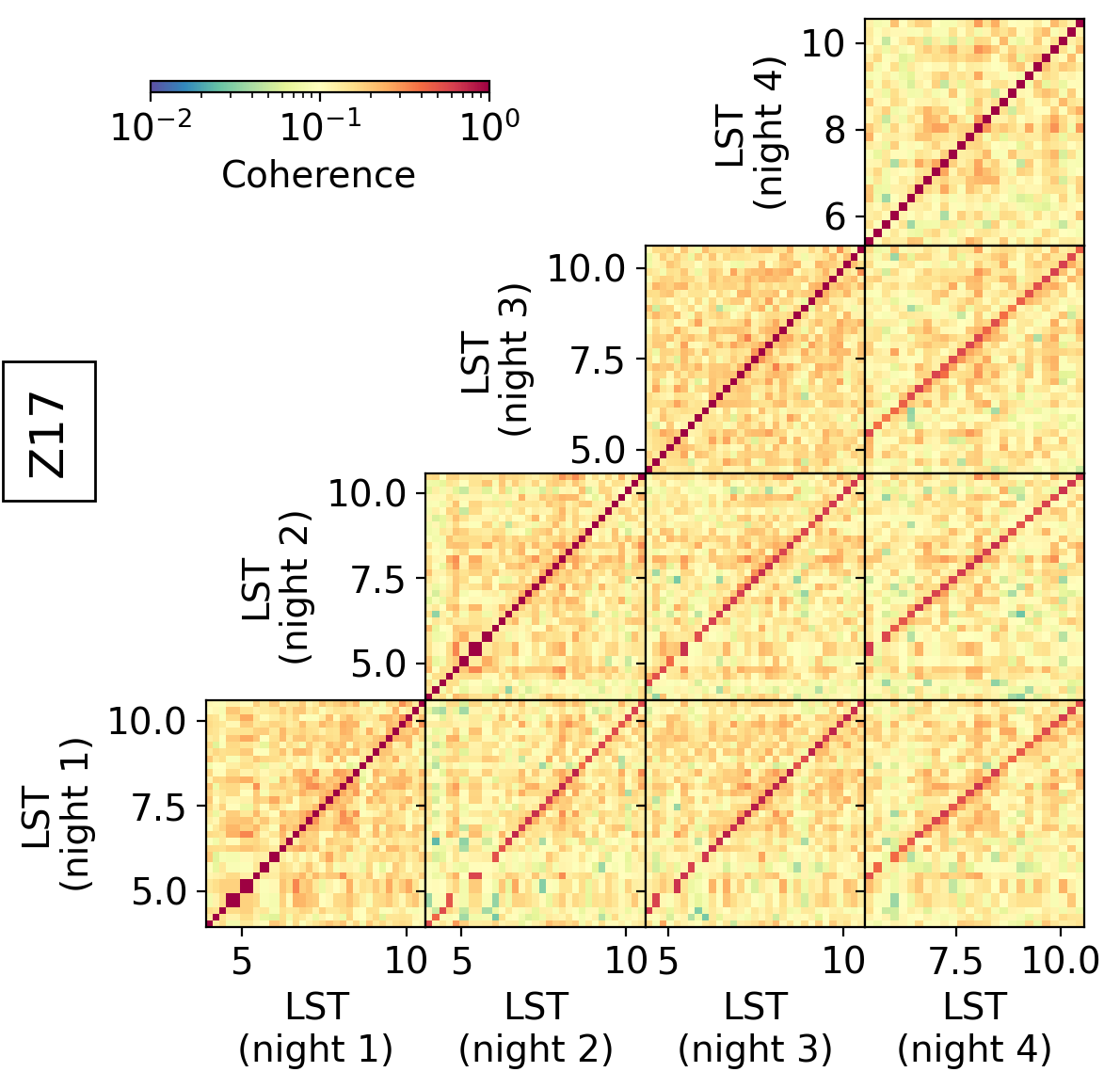}
    \includegraphics[width=0.66\columnwidth]{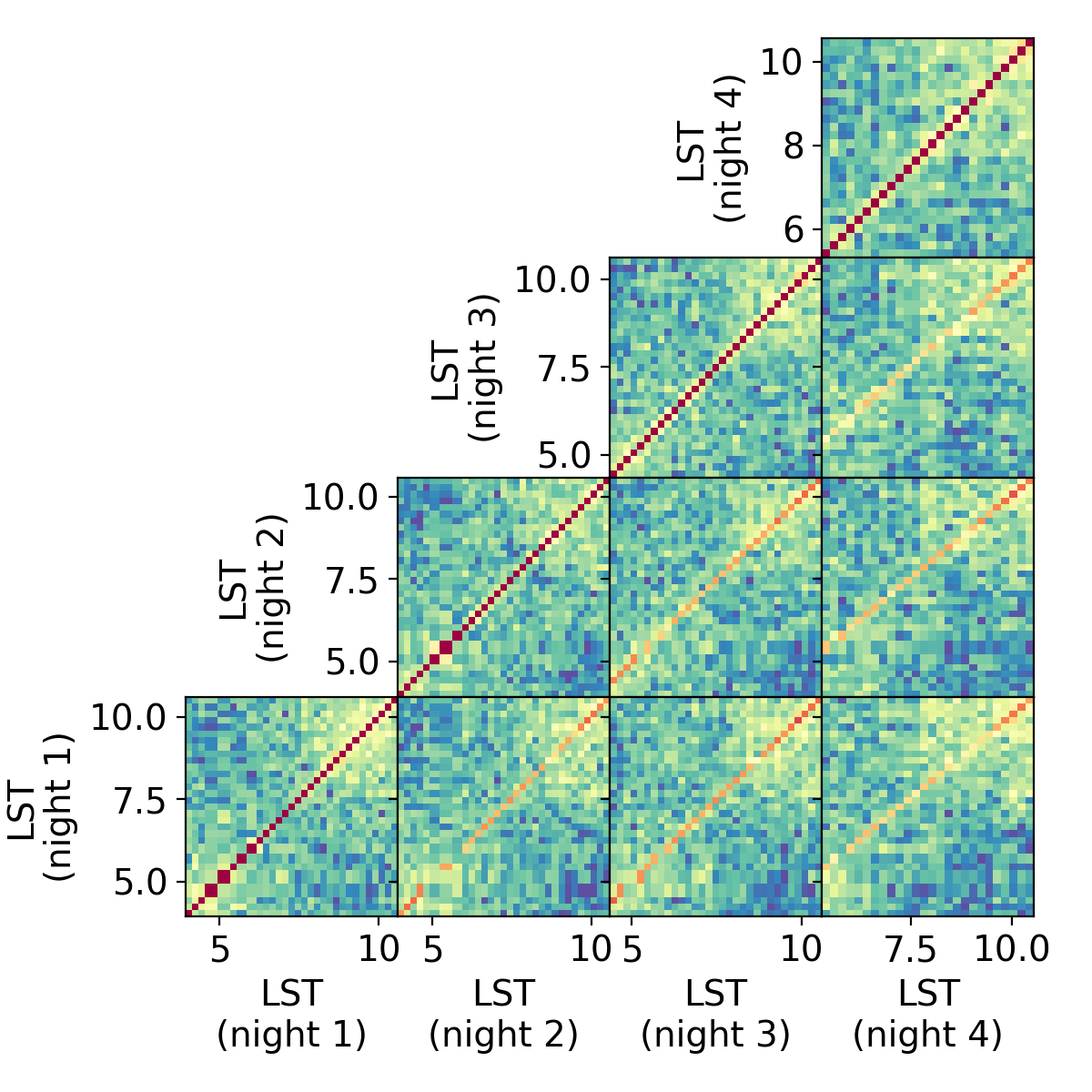}
    \includegraphics[width=0.66\columnwidth]{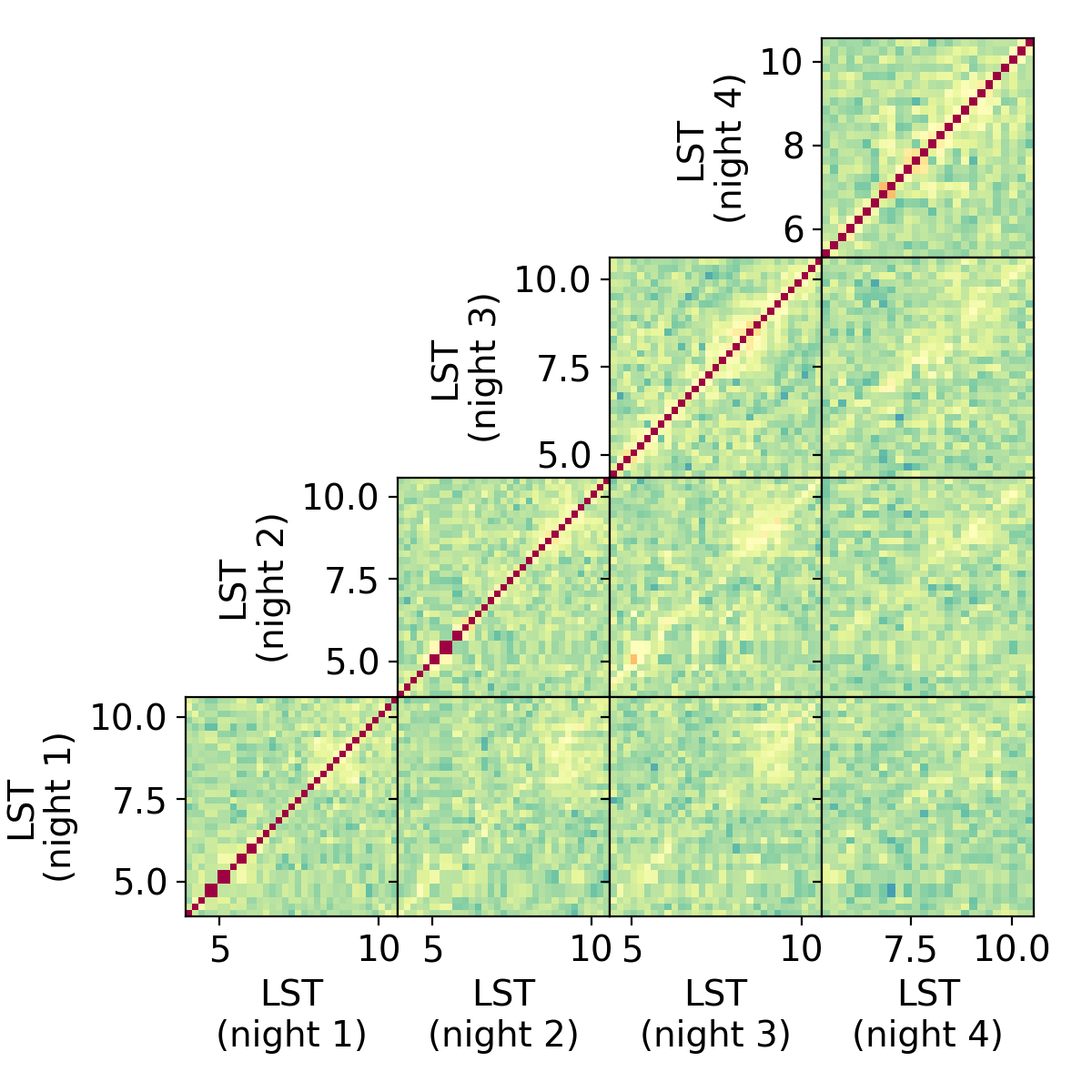}
    \caption{Coherence of the excess variance across LSTs and nights in different regions of the $k_{\perp},k_{\parallel}$ space. The different columns correspond to three regions: the main lobe, the sidelobes, and the EoR window. The top row shows the average data and noise power in the cylindrical power spectrum as a function of integration time, with the vertical solid lines demarcating the four nights. The middle and bottom rows show the cross-coherence values calculated for the two spectral windows.}
    \label{fig:coherence}
\end{figure*}

We note that the NenuFAR primary beam main lobe extends up to $\approx11^{\circ}$ from the phase centre for Z20 and $\approx9^{\circ}$ for Z17. Thus, the region within $15^{\circ}$ from the phase centre that is chosen here also includes the first nulls and sidelobe. However, the finite bandwidth imposes a minimum $k_{\parallel}$ value that we can measure, preventing the use of smaller angular distances where enough modes can be averaged below the delay line. For the sidelobe and wedge regions, we impose a minimum $k_{\parallel}$ value of 0.2 $h \, \mathrm{cMpc}^{-1}$ as a buffer, to better isolate the behavior within the wedge and the EoR window from spectral leakage from the primary beam main lobe. For all regions, a maximum $k_{\parallel}$ cutoff at 1 $h \, \mathrm{cMpc}^{-1}$ is imposed to avoid the artefact caused by the polyphase filter.

The estimated powers in all $k_{\perp},k_{\parallel}$ bins within each region are averaged separately. The entire analysis is repeated for the time-differenced Stokes~\textit{V} data cubes, which give us an estimate of the thermal noise power in each region. The results are shown in the top row of Fig.~\ref{fig:coherence}. The noise power is consistent across the three regions and exhibits the expected inverse dependence on the integration time. In the main lobe, the reduction in data power with integration time in the beginning could be caused by the non-axisymmetric primary beam main lobe picking up different sources as it rotates against the sky. The confusion noise dominated power then adds up coherently with increasing integration time. In the sidelobes, we find that the power adds up relatively incoherently within a night, which was also observed by \citetalias{munshi2024first}. This is a combination of two factors: noise dominating in the first few segments and the grating lobes picking up different groups of sources as the sky rotates. For the subsequent nights, the overall sidelobe data power stops decreasing further since the power from off-axis sources in the grating lobes at the same LST for the different nights still adds up coherently. The slight increase in the data power at the beginning of each night is possibly caused by the stronger foreground contamination in this specified region of the $k_{\perp},k_{\parallel}$ space due to steeper horizon lines corresponding to the low elevation of the phase centre. 

In the EoR window, we find that for Z20, the power decreases monotonically with integration time, though there is an excess above the thermal noise power. However, for Z17, after the first night of integration, the data power starts to average down significantly slower than incoherent noise, indicating that a portion of the excess variance is coherent and starts to dominate the noise power. Residual power in the EoR window is a limiting factor for 21-cm cosmology analyses since the foreground subtraction algorithms cannot easily remove this power, without also suppressing the 21-cm signal. Thus, unless improvements are made to the processing strategies that can mitigate this coherent excess in Z17, possibly caused by RFI contamination, deeper integrations are unlikely to improve significantly upon the current upper limits. However, Z20 looks promising since the excess power seems to integrate down incoherently, roughly like thermal noise, at least within these four nights of observation.

We can investigate the nature of the excess variance in more detail through a cross-coherence analysis. The cylindrically averaged cross coherence between two data cubes $T_i$ and $T_j$ corresponding to the $i$-th and $j$-th segments is defined as
\begin{equation}
C_{i,j}(k_{\perp}, k_{\|}) \equiv \frac{\left\langle | T_i^*(\mathbfit{k}) T_j(\mathbfit{k})|\right\rangle^2}{\left\langle | T_i(\mathbfit{k})|^2\right\rangle\left\langle |T_j(\mathbfit{k})|^2\right\rangle},
\end{equation}
where $\langle\cdots\rangle$ represents a weighted average over all $\mathbfit{k}$ values lying within the $k_{\perp},k_{\parallel}$ bin. The cross-coherence is computed for all pairs of 12 min gridded data cubes across the four nights. 

The computed cross coherence in all $k_{\perp},k_{\parallel}$ bins within each region is averaged separately. The results for both spectral windows are shown in the middle and bottom rows of Fig.~\ref{fig:coherence}, where each pixel corresponds to one such average cross-coherence value. We note that the values of the diagonals in the cross-coherence between the same nights are unity by definition. For the cross-coherence between different nights, the higher values near the diagonal indicate that similar LSTs are more coherent across nights due to the apparent sky being identical and the PSF being similar across nights for the same LST values. The values of cross-coherence at the same LST for different nights are highest within the main lobe and decrease as we go to the sidelobes, ultimately reducing to less than 10\% in the EoR window. Away from the diagonal, the coherence decreases while still remaining above 30\% within the main lobe for Z20. The coherence in the main lobe is lower for Z17, possibly because of the narrower primary beam, resulting in a higher decoherence due to thermal noise within the chosen $15^{\circ}$ angular radius. Additionally, only the frequency range above 79 MHz is used for Z17 to avoid artefacts due to frequency gaps. This reduces the resolution along $k_{\parallel}$ and, consequently, the number of $k_{\parallel}$ bins that lie below the delay line for which the computed coherence values are averaged, increasing the sampling variance in the computed coherence. In the sidelobes, away from the diagonals, there is a higher coherence at the beginning and the end of the observations for both spectral windows. This is possibly caused by the fact that the horizon line is steeper during these periods because the phase centre is at a lower elevation. Thus, the power from off-axis sources can add up more coherently in this region of the $k_{\perp},k_{\parallel}$ space. The residual coherence in the EoR window for both spectral windows can be attributed to leakage from foregrounds or low-level RFI that is partially coherent at the same LST across nights. This coherence is lower for nights that are more separated in time, as we move towards the bottom right square in each panel. We note that the residual coherence away from the diagonal in the EoR window is higher for Z17 for several pairs of nights. A possible origin could be the contaminating effects of RFI, which also result in the much stronger excess in Z17. The coherence analysis supports the insights from the analysis of how the power integrates down with time, in that Z20 is significantly cleaner, with the excess variance being more incoherent with time compared to Z17.

\section{Discussion}\label{sec:discussion}
In this section, we discuss the astrophysical relevance of the upper limits derived in this analysis by comparing them against simulated models. We also highlight the main limitations of the current analysis and discuss prospects for future improvement.

\subsection{Astrophysical interpretation}\label{sec:interpretation}

\begin{figure*}
    \includegraphics[width=\textwidth]{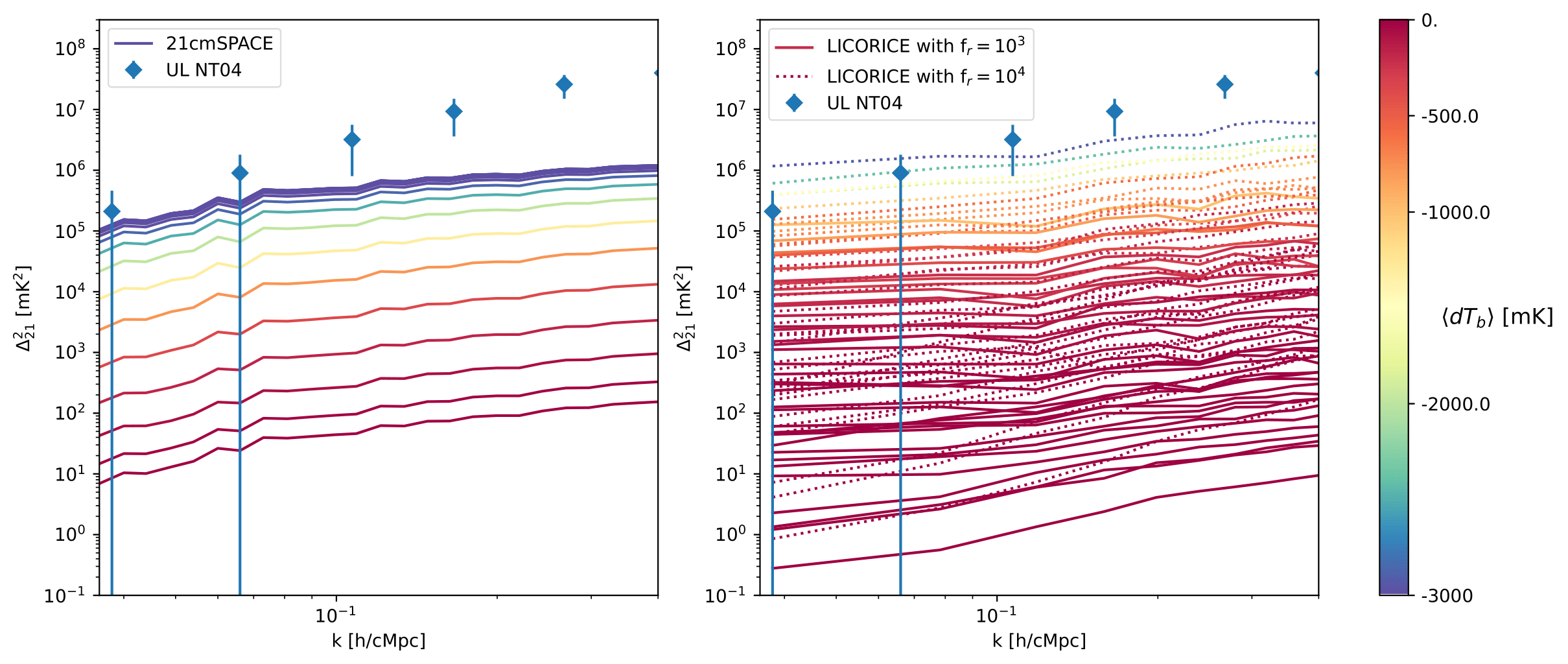}
    \caption{Comparison of the upper limits derived in this analysis at $z=20.3$ against simulated exotic 21-cm signals. The left panel shows the power spectra of signals simulated using \texttt{21cmSPACE} and the right panel shows the power spectra obtained from the \texttt{LICORICE} simulations. The colorbar indicates the corresponding global 21-cm signal depth.}
    \label{fig:exo_models}
\end{figure*}

We focus only on the upper limits at $z=20.3$, which are the more stringent limits from this analysis. Since these upper limits remain above the predictions of most models, including exotic ones, we do not perform a full inference analysis. Nonetheless, it is instructive to illustrate how much these limits must be reduced in order to place meaningful constraints on theoretical models. This is what is shown in Fig.~\ref{fig:exo_models}.

The left panel shows models computed with the simulation code 21-cm Semi-numerical Predictions Across Cosmological Epochs \citep[\texttt{21cmSPACE}, e.g.][]{visbal12,fialkov12,fialkov13,fialkov14}\footnote{\url{https://www.cosmicdawnlab.com/21cmSPACE}}, that include the contribution from radio-emitting high-redshift galaxies. Such models \citep{feng18, mirocha19, Reis2020, sikder2024} were originally inspired by the tentative detection of the sky-averaged (global) 21-cm signal from cosmic dawn \citep{bowman2018absorption} by the EDGES experiment. Here, we use the latest version of the code as described by \citet{liu25}.\footnote{We note that the cosmology assumed in \texttt{21cmSPACE} is Planck 2013 \citep{2014A&A...571A..16P}, while the rest of the work uses Planck 2015 parameters. Although important for precision inference, at the current level of the limits and given that the errors on semi-numerical simulation approaches of $\sim20\%$ \citep{2011MNRAS.414..727Z,2018MNRAS.477.1549H}, this should have a negligible impact on the conclusions.} Since the astrophysical parameters of high-redshift galaxies are highly uncertain, we fix the star-formation efficiency of Pop~II stars to $f_\mathrm{\star,II}=0.5$ \citep[not unlikely for bursty or dense, compact star-forming galaxies at high-$z$, e.g.][]{2025arXiv250321687D,2025arXiv250418618Y}, and Pop~III stars to $f_\mathrm{\star,III}=0.05$ with a fast recovery period after Pop~III supernovae \citep{magg22}. In the plot, only the parameter to which the 21-cm signal at cosmic dawn is most sensitive, namely the X-ray efficiency $f_X$, is varied within the range $[10^{-3},10^3]$. We allow for low values still not completely ruled out \citep{2025arXiv250321687D} since they produce the highest 21-cm fluctuations. This parameter is normalised so that a value of unity corresponds to the ratio of X-ray luminosity to star-formation rate $(L_X/\text{SFR})$ of local starburst galaxies with low metallicity high mass X-ray binaries \citep{fragos13a,fragos13b}. For each model, we fix the radio efficiency $f_r$ \citep[i.e. the ratio of low-frequency radio synchrotron luminosity to star-formation rate normalized to low-redshift galaxies;][]{2016MNRAS.462.1910H, 2018MNRAS.475.3010G} to be the maximum allowed according to the observed intensity of the extragalactic background, based on the measurements by ARCADE-2 \citep{fixsen11} and LWA-1 \citep{dowell18}. At $z=20.3$, these constraints yield $f_r \sim 140\,000$ following the methodology of \citet{sikder24}.

The right panel of Fig.~\ref{fig:exo_models} shows a set of models derived from the \texttt{Loreli II} simulations \citep{meriot2025} obtained using the LICORICE radiative transfer code \citep{semelin2007lyman,baek2009simulated,baek2010reionization,semelin2016detailed,semelin201721ssd}. While the original  \texttt{Loreli II} simulations are standard models of the early universe, a $\times 2000$ boost to the star formation rate (SFR) was applied compared to the original simulations. This leads to a star formation rate density (SFRD) at $z\sim20$ in the $2 \times 10^{-6} - 8 \times 10^{-3}$ M$_\odot$ yr$^{-1}$ cMpc$^{-3}$ range, that matches those presented, for instance, by \citet{cang2024}. Then, similarly to \texttt{21cmSPACE} models, high redshift galaxies were considered the source of intense radio emissions, creating a fluctuating radio background that dominates over the CMB. The radio background was computed using a modified version of the \texttt{SPINTER} code \citep{semelin2023}, following \citet{sikder2024}, and using a radio efficiency parameter of  $f_{r} = 10^3$ or $ f_{r} = 10^4$. We assumed negligible X-ray heating (i.e. $f_X=0$) and ionisation. For a given value of $f_r$, the different power spectra shown in Fig.~\ref{fig:exo_models} correspond to different pairs of minimum halo mass for star formation and gas conversion time scale, spanning the full range described in \citep{meriot2025}, and resulting in the range of SFRD mentioned above.

While our goal here is not to compare the two sets of models but rather to show a diversity of possible signals, we briefly comment on the observed differences between the \texttt{21cmSPACE} and \texttt{LICORICE} models. While their power spectra exhibit similar slopes and span a similar range of amplitudes, the \texttt{LICORICE} models reach somewhat higher amplitudes, even surpassing the NT04 upper limits at large scales. A plausible explanation is that, while both sets of models reach a similar SFRD, the emissivity is not distributed in the same way. \texttt{LICORICE} models can only form stars in halos more massive than $10^8$ M$_\odot$ (thus considering molecular cooling to be ineffective), while \texttt{21cmSPACE} models allow for star formation in halos down to $10^6$ M$_\odot$, although Lyman-Werner feedback is allowed to limit the efficiency in the least massive halos. This results in a more diffuse emissivity in \texttt{21cmSPACE}, for both the Wouthuysen-Field coupling and the radio background, and thus in a decreased power spectrum amplitude. 

As we can see in Fig.~\ref{fig:exo_models}, some models exceed the NT04 upper-limits at large scales only when stringent conditions are met: a very strong radio emission by high redshift galaxies, a high SFRD to generate sufficient Wouthuysen-Field coupling and an efficient suppression of star formation in halos below $10^8$ M$_\odot$ to generate a strongly fluctuating emissivity field. These models have a corresponding global signal in the $3000$ mK range, much stronger than the EDGES signal, and at a higher redshift. An improvement of a factor of $\gtrsim 10$ in the upper limits will allow us to constrain models with a power spectrum amplitude similar to the EDGES signal.

\subsection{Limitations and future work}
As hundreds of hours of observation with NenuFAR are integrated in the future, there are a few key areas where the current processing pipeline could be improved.
\paragraph*{Local RFI sources:}The current approach to mitigating contamination from local RFI sources in NenuFAR utilises the non-stationarity of the target field with respect to terrestrial RFI sources to flag their contribution in the $u\varv$ plane. However, the RFI power continues to be present in the rest of the $u\varv$ plane, but at a lower level due to suppression by the gridding kernel. Detailed modelling of the spectral and temporal behaviour of the local RFI sources in NenuFAR has already been investigated \citep{munshi2025near}, and approaches to subtract these near-field models can be further developed. The NenuFAR team has also improved the shielding of one of the identified RFI sources, which has been seen to reduce the contamination in more recent NenuFAR observations.
\paragraph*{Off-axis source contamination:}While the current approach of a carefully chosen target field significantly improves the A-team contamination compared to previous analyses, the signature of the residuals from the strongest A-team source (Cas A) can still be seen in the gridded $u\varv$ plane (Fig.~\ref{fig:ps_rfi}). Additionally, the 3C subtraction in this analysis followed a conservative approach to avoid overfitting, which results in 3C sources in the primary beam grating lobes dominating the power spectra after sky model subtraction. The inclusion of a primary beam model in calibration could improve the off-axis source subtraction, where the calibration solutions do not need to describe the complete beam variations but only need to capture deviations of the simulated beam from the true beam. The solutions can thus be constrained better due to the lower number of free parameters. The NenuFAR MA primary beam is analogue steered, and inactive antennas within an MA could drastically change the primary beam pattern and impact gain calibration \citep{brackenhoff2025robust}. Thus, it could be important to have prior knowledge of the broken antennas for each MA for effective inclusion of the primary beam in calibration.
\paragraph*{Faster processing mode:}Currently the computational time required by the entire calibration and sky-model subtraction pipeline implemented in \texttt{Nenuflow} (Fig.~\ref{fig:flowchart}) is about $12\,$h per night of observation (with duration $\approx 10\,$h) when run in a parallelised manner over 15 computational nodes of the \texttt{DAWN} cluster, running up to 5 processes parallelly in each node (one per CPU). A large fraction of this runtime ($\approx 61\%$) is spent on the imaging and sky model iteration step. In the future, we are planning to use a deep intrinsic sky model from these calibrated four nights of data, effectively reducing the total runtime to $\approx 4.7\,$h per night. In practice, the processing of a single night of NenuFAR observations takes longer than the runtime of $12\,$h, since the first A-team subtraction step needs to be repeated multiple times to identify faulty stations based on calibration solutions (right panel of Fig.~\ref{fig:rfi_stats}). This step could be automated in the future by using outlier detection algorithms on the calibration solutions.
\paragraph*{Excess variance bias correction:}The excess variance in the data above the thermal noise level has not been subtracted in this analysis, to set conservative upper limits on the 21-cm power spectrum. In forthcoming work, we will investigate the nature of the (in)coherence of the excess noise in greater detail, since making a bias correction for the incoherent component (as for the thermal noise) could lead to a considerably stronger upper limit.

\section{Conclusions}\label{sec:conclusion}
In this paper, we have analysed four nights of observations of the NT04 field (RA = $7\,$h 20 min, Dec = $35^{\circ}$) with NenuFAR to set upper limits on the 21-cm power spectrum from the Cosmic Dawn at two redshift bins centred at $z=20.3$ and $z=17.0$. Our main conclusions and insights are summarised below.

\paragraph*{Choice of a new field:}We have switched from the NCP field analysed by \citetalias{munshi2024first} to the NT04 target field, selected carefully through a survey of multiple fields. 
Firstly, the new field selection is optimised to reduce PSF sidelobe leakage from A-team sources, one of the main contributors to the excess variance observed by \citetalias{munshi2024first}. Secondly, switching to a field other than the NCP improves our ability to mitigate power from terrestrial RFI and strong off-axis sources. RFI manifests as a vertical line in the $u\varv$ plane \citep{munshi2025near}, which we mitigate to an extent by flagging the specific region in the $u\varv$ plane. Off-axis sources are most problematic when the phase centre is at a low elevation due to the foreground wedge being steeper \citep{munshi2025beyond}, and this foreground contamination at high $k_{\parallel}$ modes can be reduced by implementing an elevation threshold in the power spectrum estimation. Each of these steps decreases the contamination at the relevant $k_{\perp},k_{\parallel}$ modes by more than an order of magnitude. Compared to the NCP field used in \citetalias{munshi2024first}, the foreground wedge in the NT04 field is significantly better confined, reducing leakage into the EoR window by up to two orders of magnitude (Section~\ref{sec:comparison_ncp}).

\paragraph*{Improved A-team subtraction:}Since the PSF sidelobe contamination from A-team sources is one of the primary causes of the excess variance in NenuFAR data, we have implemented two key improvements in the A-team subtraction step of the calibration pipeline. Firstly, the data is divided into small time segments and calibrated against apparent sky models of sources above a flux cut for each segment. Secondly, the implementation of adaptive solution intervals in calibration improves the residuals in the direction of strong A-team sources. We organise all calibration and sky-model construction steps into a robust pipeline \texttt{Nenuflow}, which performs these tasks in a heavily parallelised fashion. This is particularly important as we look into the future, when we plan to integrate hundreds of hours of this new field, where computational time will become an important constraint.

\paragraph*{Power spectrum upper limits:}After sky model subtraction and residual foreground removal, we set upper limits on the 21-cm signal power spectrum at two redshift bins. At $z=20.3$, the best 2$\sigma$ upper limit, derived from 26.1 h of observations, is $\Delta^{2}_{21} < 4.6 \times 10^5 \, \textrm{mK}^{2}$ at $k=0.038$ $h\, \mathrm{cMpc}^{-1}$, which is the strongest Cosmic Dawn power spectrum limit to date and improve upon previous results by over an order of magnitude. At $z=17.0$, the best 2$\sigma$ limit, from 23.6 h of observations, is $\Delta^{2}_{21} < 5.0 \times 10^6 \, \textrm{mK}^{2}$ at $k=0.041$ $h\, \mathrm{cMpc}^{-1}$. While this limit is less stringent than the one at $z=20.3$, it is still deeper than Cosmic Dawn power spectrum limits obtained from all other instruments. A comparison of the upper limits at $z=20.3$ against simulated exotic 21-cm signals indicates that only the most extreme models, which predict a global signal significantly stronger than the absorption trough observed by EDGES, are currently excluded by these upper limits. An order-of-magnitude improvement in these limits would allow us to probe models with signal strengths comparable to the EDGES detection.
\paragraph*{Coherence of excess variance:}The new redshift bin we analysed at $z=17.0$ is severely affected by RFI and exhibits significantly stronger excess variance than $z=20.3$. Investigation of the nature of the excess variance at both redshift bins suggests that while the excess in the $z=20.3$ bin is largely incoherent, the excess in the $z=17.0$ bin has a coherent component that begins to dominate the power in the EoR window after several hours of integration. The redshift bin centred at $z=20.3$ thus offers a better window into 21-cm cosmology studies for the NenuFAR Cosmic Dawn KSP, and will be the primary focus of future efforts of the KSP.

\vspace{0.5cm}
This paper demonstrates how improved observation and analysis strategies can help address the challenges of 21-cm power spectrum detection from the Cosmic Dawn. The derived power spectrum limits, for the first time, allow us to reach astrophysically relevant 21-cm signal levels during the Cosmic Dawn, demonstrating the capability of NenuFAR to provide strong constraints on the 21-cm signal from the Cosmic Dawn as more observations are integrated in the future.

\section*{Acknowledgements}
SM, JKC, LVEK, SAB and SG acknowledge the financial support from the European Research Council (ERC) under the European Union’s Horizon 2020 research and innovation programme (Grant agreement No. 884760, "CoDEX”). FGM acknowledges support from the I-DAWN project, funded by the DIM-ORIGINS programme. RB and SS acknowledge the support of the Israel Science Foundation (grant no.\ 1078/24). AB acknowledges financial support from the INAF initiative "IAF Astronomy Fellowships in Italy" (grant name MEGASKAT). RG acknowledges support from SERB, DST Ramanujan Fellowship no. RJF/2022/000141. AKS acknowledges support from National Science Foundation (grant no. 2206602). This paper is based on data obtained using the NenuFAR radiotelescope. NenuFAR has benefitted from the following funding sources : CNRS-INSU, Observatoire de Paris, Station de Radioastronomie de Nançay, Observatoire des Sciences de l'Univers de la Région Centre, Région Centre-Val de Loire, Université d'Orléans, DIM-ACAV and DIM-ACAV+ de la Région Ile de France, Agence Nationale de la Recherche. We acknowledge the Nançay Data Center resources used for data reduction and storage.

\section*{Data availability}
The data underlying this article will be shared on reasonable request to the corresponding author.

\bibliographystyle{mnras}
\bibliography{example}

\begin{thebibliography}{}
\makeatletter
\relax
\def\mn@urlcharsother{\let\do\@makeother \do\$\do\&\do\#\do\^\do\_\do\%\do\~}
\def\mn@doi{\begingroup\mn@urlcharsother \@ifnextchar [ {\mn@doi@} {\mn@doi@[]}}
\def\mn@doi@[#1]#2{\def\@tempa{#1}\ifx\@tempa\@empty \href {http://dx.doi.org/#2} {doi:#2}\else \href {http://dx.doi.org/#2} {#1}\fi \endgroup}
\def\mn@eprint#1#2{\mn@eprint@#1:#2::\@nil}
\def\mn@eprint@arXiv#1{\href {http://arxiv.org/abs/#1} {{\tt arXiv:#1}}}
\def\mn@eprint@dblp#1{\href {http://dblp.uni-trier.de/rec/bibtex/#1.xml} {dblp:#1}}
\def\mn@eprint@#1:#2:#3:#4\@nil{\def\@tempa {#1}\def\@tempb {#2}\def\@tempc {#3}\ifx \@tempc \@empty \let \@tempc \@tempb \let \@tempb \@tempa \fi \ifx \@tempb \@empty \def\@tempb {arXiv}\fi \@ifundefined {mn@eprint@\@tempb}{\@tempb:\@tempc}{\expandafter \expandafter \csname mn@eprint@\@tempb\endcsname \expandafter{\@tempc}}}

\bibitem[\protect\citeauthoryear{Abdurashidova et~al.,}{Abdurashidova et~al.}{2022a}]{abdurashidova2022hera}
Abdurashidova Z.,  et~al., 2022a, ApJ, 924, 51

\bibitem[\protect\citeauthoryear{Abdurashidova et~al.,}{Abdurashidova et~al.}{2022b}]{abdurashidova2022first}
Abdurashidova Z.,  et~al., 2022b, ApJ, 925, 221

\bibitem[\protect\citeauthoryear{Acharya et~al.,}{Acharya et~al.}{2024}]{acharya202421}
Acharya A.,  et~al., 2024, MNRAS, 527, 7835

\bibitem[\protect\citeauthoryear{Adams et~al.,}{Adams et~al.}{2023}]{adams2023improved}
Adams T.,  et~al., 2023, ApJ, 945, 124

\bibitem[\protect\citeauthoryear{Baek, Di~Matteo, Semelin, Combes  \& Revaz}{Baek et~al.}{2009}]{baek2009simulated}
Baek S.,  Di~Matteo P.,  Semelin B.,  Combes F.,   Revaz Y.,  2009, Astronomy \& Astrophysics, 495, 389

\bibitem[\protect\citeauthoryear{Baek, Semelin, Di~Matteo, Revaz  \& Combes}{Baek et~al.}{2010}]{baek2010reionization}
Baek S.,  Semelin B.,  Di~Matteo P.,  Revaz Y.,   Combes F.,  2010, Astronomy \& Astrophysics, 523, A4

\bibitem[\protect\citeauthoryear{Barkana}{Barkana}{2018}]{barkana2018possible}
Barkana R.,  2018, Nature, 555, 71

\bibitem[\protect\citeauthoryear{Barry et~al.,}{Barry et~al.}{2019}]{barry2019improving}
Barry N.,  et~al., 2019, ApJ, 884, 1

\bibitem[\protect\citeauthoryear{Bennett \& Simth}{Bennett \& Simth}{1962}]{bennett1962preparation}
Bennett A.,  Simth F.,  1962, MNRAS, 125, 75

\bibitem[\protect\citeauthoryear{Berlin, Hooper, Krnjaic  \& McDermott}{Berlin et~al.}{2018}]{berlin2018severely}
Berlin A.,  Hooper D.,  Krnjaic G.,   McDermott S.~D.,  2018, Physical review letters, 121, 011102

\bibitem[\protect\citeauthoryear{Bonaldi et~al.,}{Bonaldi et~al.}{2025}]{bonaldi2025square}
Bonaldi A.,  et~al., 2025, arXiv preprint arXiv:2503.11740

\bibitem[\protect\citeauthoryear{Bowman, Rogers, Monsalve, Mozdzen  \& Mahesh}{Bowman et~al.}{2018}]{bowman2018absorption}
Bowman J.~D.,  Rogers A.~E.,  Monsalve R.~A.,  Mozdzen T.~J.,   Mahesh N.,  2018, Nature, 555, 67

\bibitem[\protect\citeauthoryear{Brackenhoff et~al.,}{Brackenhoff et~al.}{2024}]{brackenhoff2024ionospheric}
Brackenhoff S.,  et~al., 2024, MNRAS, 533, 632

\bibitem[\protect\citeauthoryear{Brackenhoff et~al.,}{Brackenhoff et~al.}{2025}]{brackenhoff2025robust}
Brackenhoff S.,  et~al., 2025, arXiv preprint arXiv:2504.02483

\bibitem[\protect\citeauthoryear{Bull et~al.,}{Bull et~al.}{2024}]{bull2024rhino}
Bull P.,  et~al., 2024, arXiv preprint arXiv:2410.00076

\bibitem[\protect\citeauthoryear{{Cang}, {Mesinger}, {Murray}, {Breitman}, {Qin}  \& {Trotta}}{{Cang} et~al.}{2024}]{cang2024}
{Cang} J.,  {Mesinger} A.,  {Murray} S.~G.,  {Breitman} D.,  {Qin} Y.,   {Trotta} R.,  2024, \mn@doi [arXiv e-prints] {10.48550/arXiv.2411.08134}, \href {https://ui.adsabs.harvard.edu/abs/2024arXiv241108134C} {p. arXiv:2411.08134}

\bibitem[\protect\citeauthoryear{Ceccotti et~al.,}{Ceccotti et~al.}{2025a}]{ceccotti2025first}
Ceccotti E.,  et~al., 2025a, arXiv preprint arXiv:2504.18534

\bibitem[\protect\citeauthoryear{Ceccotti et~al.,}{Ceccotti et~al.}{2025b}]{ceccotti2025spectral}
Ceccotti E.,  et~al., 2025b, arXiv preprint arXiv:2502.18459

\bibitem[\protect\citeauthoryear{Charles, Kern, Bernardi, Bester, Smirnov, Fagnoni  \& Acedo}{Charles et~al.}{2023}]{charles2023use}
Charles N.,  Kern N.,  Bernardi G.,  Bester L.,  Smirnov O.,  Fagnoni N.,   Acedo E. d.~L.,  2023, MNRAS, 522, 1009

\bibitem[\protect\citeauthoryear{Charles et~al.,}{Charles et~al.}{2024}]{charles2024mitigating}
Charles N.,  et~al., 2024, MNRAS, 534, 3349

\bibitem[\protect\citeauthoryear{Chege et~al.,}{Chege et~al.}{2024}]{chege2024impact}
Chege J.,  et~al., 2024, A\&A, 692, A211

\bibitem[\protect\citeauthoryear{Chokshi, Barry, Line, Jordan, Pindor  \& Webster}{Chokshi et~al.}{2024}]{chokshi2024necessity}
Chokshi A.,  Barry N.,  Line J.,  Jordan C.,  Pindor B.,   Webster R.,  2024, MNRAS, 534, 2475

\bibitem[\protect\citeauthoryear{Cohen, Lane, Cotton, Kassim, Lazio, Perley, Condon  \& Erickson}{Cohen et~al.}{2007}]{cohen2007vla}
Cohen A.,  Lane W.,  Cotton W.,  Kassim N.,  Lazio T.,  Perley R.,  Condon J.,   Erickson W.,  2007, The Astronomical Journal, 134, 1245

\bibitem[\protect\citeauthoryear{Condon, Cotton, Greisen, Yin, Perley, Taylor  \& Broderick}{Condon et~al.}{1998}]{condon1998nrao}
Condon J.~J.,  Cotton W.,  Greisen E.,  Yin Q.,  Perley R.~A.,  Taylor G.,   Broderick J.,  1998, The Astronomical Journal, 115, 1693

\bibitem[\protect\citeauthoryear{Datta, Mellema, Mao, Iliev, Shapiro  \& Ahn}{Datta et~al.}{2012}]{datta2012light}
Datta K.~K.,  Mellema G.,  Mao Y.,  Iliev I.~T.,  Shapiro P.~R.,   Ahn K.,  2012, MNRAS, 424, 1877

\bibitem[\protect\citeauthoryear{Datta, Jensen, Majumdar, Mellema, Iliev, Mao, Shapiro  \& Ahn}{Datta et~al.}{2014}]{datta2014light}
Datta K.~K.,  Jensen H.,  Majumdar S.,  Mellema G.,  Iliev I.~T.,  Mao Y.,  Shapiro P.~R.,   Ahn K.,  2014, MNRAS, 442, 1491

\bibitem[\protect\citeauthoryear{{Dhandha} et~al.,}{{Dhandha} et~al.}{2025}]{2025arXiv250321687D}
{Dhandha} J.,  et~al., 2025, \mn@doi [arXiv e-prints] {10.48550/arXiv.2503.21687}, \href {https://ui.adsabs.harvard.edu/abs/2025arXiv250321687D} {p. arXiv:2503.21687}

\bibitem[\protect\citeauthoryear{Di~Tommaso, Chatzou, Floden, Barja, Palumbo  \& Notredame}{Di~Tommaso et~al.}{2017}]{di2017nextflow}
Di~Tommaso P.,  Chatzou M.,  Floden E.~W.,  Barja P.~P.,  Palumbo E.,   Notredame C.,  2017, Nature biotechnology, 35, 316

\bibitem[\protect\citeauthoryear{Dowell \& Taylor}{Dowell \& Taylor}{2018a}]{dowell2018radio}
Dowell J.,  Taylor G.~B.,  2018a, The Astrophysical Journal Letters, 858, L9

\bibitem[\protect\citeauthoryear{{Dowell} \& {Taylor}}{{Dowell} \& {Taylor}}{2018b}]{dowell18}
{Dowell} J.,  {Taylor} G.~B.,  2018b, \mn@doi [\apjl] {10.3847/2041-8213/aabf86}, \href {https://ui.adsabs.harvard.edu/abs/2018ApJ...858L...9D} {858, L9}

\bibitem[\protect\citeauthoryear{Eastwood et~al.,}{Eastwood et~al.}{2019}]{eastwood201921}
Eastwood M.~W.,  et~al., 2019, AJ, 158, 84

\bibitem[\protect\citeauthoryear{Ewall-Wice et~al.,}{Ewall-Wice et~al.}{2016}]{ewall2016first}
Ewall-Wice A.,  et~al., 2016, MNRAS, 460, 4320

\bibitem[\protect\citeauthoryear{Ewall-Wice, Chang, Lazio, Dor{\'e}, Seiffert  \& Monsalve}{Ewall-Wice et~al.}{2018}]{ewall2018modeling}
Ewall-Wice A.,  Chang T.-C.,  Lazio J.,  Dor{\'e} O.,  Seiffert M.,   Monsalve R.,  2018, ApJ, 868, 63

\bibitem[\protect\citeauthoryear{Feng \& Holder}{Feng \& Holder}{2018a}]{feng2018enhanced}
Feng C.,  Holder G.,  2018a, The Astrophysical Journal Letters, 858, L17

\bibitem[\protect\citeauthoryear{{Feng} \& {Holder}}{{Feng} \& {Holder}}{2018b}]{feng18}
{Feng} C.,  {Holder} G.,  2018b, \mn@doi [\apjl] {10.3847/2041-8213/aac0fe}, \href {https://ui.adsabs.harvard.edu/abs/2018ApJ...858L..17F} {858, L17}

\bibitem[\protect\citeauthoryear{Fialkov \& Barkana}{Fialkov \& Barkana}{2019}]{fialkov2019signature}
Fialkov A.,  Barkana R.,  2019, MNRAS, 486, 1763

\bibitem[\protect\citeauthoryear{{Fialkov}, {Barkana}, {Tseliakhovich}  \& {Hirata}}{{Fialkov} et~al.}{2012}]{fialkov12}
{Fialkov} A.,  {Barkana} R.,  {Tseliakhovich} D.,   {Hirata} C.~M.,  2012, \mn@doi [\mnras] {10.1111/j.1365-2966.2012.21318.x}, \href {https://ui.adsabs.harvard.edu/abs/2012MNRAS.424.1335F} {424, 1335}

\bibitem[\protect\citeauthoryear{{Fialkov}, {Barkana}, {Visbal}, {Tseliakhovich}  \& {Hirata}}{{Fialkov} et~al.}{2013}]{fialkov13}
{Fialkov} A.,  {Barkana} R.,  {Visbal} E.,  {Tseliakhovich} D.,   {Hirata} C.~M.,  2013, \mn@doi [\mnras] {10.1093/mnras/stt650}, \href {https://ui.adsabs.harvard.edu/abs/2013MNRAS.432.2909F} {432, 2909}

\bibitem[\protect\citeauthoryear{{Fialkov}, {Barkana}  \& {Visbal}}{{Fialkov} et~al.}{2014}]{fialkov14}
{Fialkov} A.,  {Barkana} R.,   {Visbal} E.,  2014, \mn@doi [\nat] {10.1038/nature12999}, \href {https://ui.adsabs.harvard.edu/abs/2014Natur.506..197F} {506, 197}

\bibitem[\protect\citeauthoryear{Fialkov, Barkana  \& Cohen}{Fialkov et~al.}{2018}]{fialkov2018constraining}
Fialkov A.,  Barkana R.,   Cohen A.,  2018, Physical review letters, 121, 011101

\bibitem[\protect\citeauthoryear{{Fixsen} et~al.,}{{Fixsen} et~al.}{2011}]{fixsen11}
{Fixsen} D.~J.,  et~al., 2011, \mn@doi [\apj] {10.1088/0004-637X/734/1/5}, \href {https://ui.adsabs.harvard.edu/abs/2011ApJ...734....5F} {734, 5}

\bibitem[\protect\citeauthoryear{{Fragos} et~al.,}{{Fragos} et~al.}{2013a}]{fragos13a}
{Fragos} T.,  et~al., 2013a, \mn@doi [\apj] {10.1088/0004-637X/764/1/41}, \href {https://ui.adsabs.harvard.edu/abs/2013ApJ...764...41F} {764, 41}

\bibitem[\protect\citeauthoryear{{Fragos}, {Lehmer}, {Naoz}, {Zezas}  \& {Basu-Zych}}{{Fragos} et~al.}{2013b}]{fragos13b}
{Fragos} T.,  {Lehmer} B.~D.,  {Naoz} S.,  {Zezas} A.,   {Basu-Zych} A.,  2013b, \mn@doi [\apjl] {10.1088/2041-8205/776/2/L31}, \href {https://ui.adsabs.harvard.edu/abs/2013ApJ...776L..31F} {776, L31}

\bibitem[\protect\citeauthoryear{Furlanetto, Oh  \& Briggs}{Furlanetto et~al.}{2006}]{furlanetto2006cosmology}
Furlanetto S.~R.,  Oh S.~P.,   Briggs F.~H.,  2006, Physics reports, 433, 181

\bibitem[\protect\citeauthoryear{Gan et~al.,}{Gan et~al.}{2022a}]{gan2022assessing}
Gan H.,  et~al., 2022a, arXiv preprint arXiv:2209.07854

\bibitem[\protect\citeauthoryear{Gan et~al.,}{Gan et~al.}{2022b}]{gan2022statistical}
Gan H.,  et~al., 2022b, A\&A, 663, A9

\bibitem[\protect\citeauthoryear{Gan, Mertens, Koopmans, Offringa, Pandey, Gehlot  et~al.}{Gan et~al.}{2023}]{gan2023assessing}
Gan H.,  Mertens F.,  Koopmans L.,  Offringa A.,  Pandey V.,  Gehlot B.,   et~al., 2023, A\&A, 669, A20

\bibitem[\protect\citeauthoryear{Garsden et~al.,}{Garsden et~al.}{2021}]{garsden202121}
Garsden H.,  et~al., 2021, MNRAS, 506, 5802

\bibitem[\protect\citeauthoryear{Garsden et~al.,}{Garsden et~al.}{2024}]{garsden2024demonstration}
Garsden H.,  et~al., 2024, MNRAS, 535, 3218

\bibitem[\protect\citeauthoryear{Gehlot et~al.,}{Gehlot et~al.}{2019}]{gehlot2019first}
Gehlot B.,  et~al., 2019, MNRAS, 488, 4271

\bibitem[\protect\citeauthoryear{Gehlot et~al.,}{Gehlot et~al.}{2020}]{gehlot2020aartfaac}
Gehlot B.,  et~al., 2020, MNRAS, 499, 4158

\bibitem[\protect\citeauthoryear{Gehlot et~al.,}{Gehlot et~al.}{2024}]{gehlot2024transient}
Gehlot B.,  et~al., 2024, A\&A, 681, A71

\bibitem[\protect\citeauthoryear{Ghara et~al.,}{Ghara et~al.}{2020}]{ghara2020constraining}
Ghara R.,  et~al., 2020, MNRAS, 493, 4728

\bibitem[\protect\citeauthoryear{Ghara et~al.,}{Ghara et~al.}{2025}]{ghara2025constraints}
Ghara R.,  et~al., 2025, arXiv preprint arXiv:2505.00373

\bibitem[\protect\citeauthoryear{Greig, Trott, Barry, Mutch, Pindor, Webster  \& Wyithe}{Greig et~al.}{2021}]{greig2021exploring}
Greig B.,  Trott C.~M.,  Barry N.,  Mutch S.~J.,  Pindor B.,  Webster R.~L.,   Wyithe J. S.~B.,  2021, MNRAS, 500, 5322

\bibitem[\protect\citeauthoryear{{G{\"u}rkan} et~al.,}{{G{\"u}rkan} et~al.}{2018}]{2018MNRAS.475.3010G}
{G{\"u}rkan} G.,  et~al., 2018, \mn@doi [\mnras] {10.1093/mnras/sty016}, \href {https://ui.adsabs.harvard.edu/abs/2018MNRAS.475.3010G} {475, 3010}

\bibitem[\protect\citeauthoryear{{Hardcastle} et~al.,}{{Hardcastle} et~al.}{2016}]{2016MNRAS.462.1910H}
{Hardcastle} M.~J.,  et~al., 2016, \mn@doi [\mnras] {10.1093/mnras/stw1763}, \href {https://ui.adsabs.harvard.edu/abs/2016MNRAS.462.1910H} {462, 1910}

\bibitem[\protect\citeauthoryear{H{\"o}fer et~al.,}{H{\"o}fer et~al.}{2025}]{hofer2025impact}
H{\"o}fer C.,  et~al., 2025, arXiv preprint arXiv:2504.03554

\bibitem[\protect\citeauthoryear{{Hutter}}{{Hutter}}{2018}]{2018MNRAS.477.1549H}
{Hutter} A.,  2018, \mn@doi [\mnras] {10.1093/mnras/sty683}, \href {https://ui.adsabs.harvard.edu/abs/2018MNRAS.477.1549H} {477, 1549}

\bibitem[\protect\citeauthoryear{{Intema}, {Jagannathan}, {Mooley}  \& {Frail}}{{Intema} et~al.}{2017}]{2017A&A...598A..78I}
{Intema} H.~T.,  {Jagannathan} P.,  {Mooley} K.~P.,   {Frail} D.~A.,  2017, \mn@doi [\aap] {10.1051/0004-6361/201628536}, \href {https://ui.adsabs.harvard.edu/abs/2017A&A...598A..78I} {598, A78}

\bibitem[\protect\citeauthoryear{Kern, Parsons, Dillon, Lanman, Fagnoni  \& de Lera~Acedo}{Kern et~al.}{2019}]{kern2019mitigating}
Kern N.~S.,  Parsons A.~R.,  Dillon J.~S.,  Lanman A.~E.,  Fagnoni N.,   de Lera~Acedo E.,  2019, ApJ, 884, 105

\bibitem[\protect\citeauthoryear{Kern et~al.,}{Kern et~al.}{2020}]{kern2020mitigating}
Kern N.~S.,  et~al., 2020, ApJ, 888, 70

\bibitem[\protect\citeauthoryear{Kolopanis et~al.,}{Kolopanis et~al.}{2019}]{kolopanis2019simplified}
Kolopanis M.,  et~al., 2019, ApJ, 883, 133

\bibitem[\protect\citeauthoryear{Koopmans et~al.,}{Koopmans et~al.}{2015}]{koopmans2015cosmic}
Koopmans L.,  et~al., 2015, arXiv preprint arXiv:1505.07568

\bibitem[\protect\citeauthoryear{Li et~al.,}{Li et~al.}{2019}]{li2019first}
Li W.,  et~al., 2019, ApJ, 887, 141

\bibitem[\protect\citeauthoryear{Liu, Outmezguine, Redigolo  \& Volansky}{Liu et~al.}{2019}]{liu2019reviving}
Liu H.,  Outmezguine N.~J.,  Redigolo D.,   Volansky T.,  2019, Physical Review D, 100, 123011

\bibitem[\protect\citeauthoryear{{Liu} et~al.,}{{Liu} et~al.}{2025}]{liu25}
{Liu} B.,  et~al., 2025, \mn@doi [arXiv e-prints] {10.48550/arXiv.2504.00535}, \href {https://ui.adsabs.harvard.edu/abs/2025arXiv250400535L} {p. arXiv:2504.00535}

\bibitem[\protect\citeauthoryear{Lloyd}{Lloyd}{1982}]{lloyd1982least}
Lloyd S.,  1982, IEEE transactions on information theory, 28, 129

\bibitem[\protect\citeauthoryear{Loh, coutouly, Girard, EmilieMauduit  \& Viou}{Loh et~al.}{2023}]{alan_loh_2023_7994526}
Loh A.,  coutouly Girard J.~N.,  EmilieMauduit  Viou C.,  2023, AlanLoh/nenupy: 2023 release, \mn@doi{10.5281/zenodo.7994526}, \url {https://doi.org/10.5281/zenodo.7994526}

\bibitem[\protect\citeauthoryear{MacQueen}{MacQueen}{1967}]{macqueen1967some}
MacQueen J.,  1967, in Proceedings of the Fifth Berkeley Symposium on Mathematical Statistics and Probability, Volume 1: Statistics. pp 281--298

\bibitem[\protect\citeauthoryear{{Magg} et~al.,}{{Magg} et~al.}{2022}]{magg22}
{Magg} M.,  et~al., 2022, \mn@doi [\mnras] {10.1093/mnras/stac1664}, \href {https://ui.adsabs.harvard.edu/abs/2022MNRAS.514.4433M} {514, 4433}

\bibitem[\protect\citeauthoryear{{Meriot}, {Semelin}  \& {Cornu}}{{Meriot} et~al.}{2024}]{meriot2025}
{Meriot} R.,  {Semelin} B.,   {Cornu} D.,  2024, \mn@doi [arXiv e-prints] {10.48550/arXiv.2411.03093}, \href {https://ui.adsabs.harvard.edu/abs/2024arXiv241103093M} {p. arXiv:2411.03093}

\bibitem[\protect\citeauthoryear{Mertens, Ghosh  \& Koopmans}{Mertens et~al.}{2018}]{mertens2018statistical}
Mertens F.,  Ghosh A.,   Koopmans L.,  2018, MNRAS, 478, 3640

\bibitem[\protect\citeauthoryear{Mertens et~al.,}{Mertens et~al.}{2020}]{mertens2020improved}
Mertens F.~G.,  et~al., 2020, MNRAS, 493, 1662

\bibitem[\protect\citeauthoryear{{Mertens}, {Semelin}  \& {Koopmans}}{{Mertens} et~al.}{2021}]{2021sf2a.conf..211M}
{Mertens} F.~G.,  {Semelin} B.,   {Koopmans} L.~V.~E.,  2021, in {Siebert} A.,  et~al., eds, SF2A-2021: Proceedings of the Annual meeting of the French Society of Astronomy and Astrophysics. pp 211--214 (\mn@eprint {arXiv} {2109.10055})

\bibitem[\protect\citeauthoryear{Mertens, Bobin  \& Carucci}{Mertens et~al.}{2024}]{mertens2024retrieving}
Mertens F.~G.,  Bobin J.,   Carucci I.~P.,  2024, Monthly Notices of the Royal Astronomical Society, 527, 3517

\bibitem[\protect\citeauthoryear{Mertens et~al.,}{Mertens et~al.}{2025}]{mertens2025deeper}
Mertens F.,  et~al., 2025, arXiv preprint arXiv:2503.05576

\bibitem[\protect\citeauthoryear{Mesinger, Furlanetto  \& Cen}{Mesinger et~al.}{2011}]{mesinger201121cmfast}
Mesinger A.,  Furlanetto S.,   Cen R.,  2011, MNRAS, 411, 955

\bibitem[\protect\citeauthoryear{Mesinger, Greig  \& Sobacchi}{Mesinger et~al.}{2016}]{mesinger2016evolution}
Mesinger A.,  Greig B.,   Sobacchi E.,  2016, MNRAS, 459, 2342

\bibitem[\protect\citeauthoryear{Mevius et~al.,}{Mevius et~al.}{2022}]{mevius2022numerical}
Mevius M.,  et~al., 2022, MNRAS, 509, 3693

\bibitem[\protect\citeauthoryear{{Mirocha} \& {Furlanetto}}{{Mirocha} \& {Furlanetto}}{2019}]{mirocha19}
{Mirocha} J.,  {Furlanetto} S.~R.,  2019, \mn@doi [\mnras] {10.1093/mnras/sty3260}, \href {https://ui.adsabs.harvard.edu/abs/2019MNRAS.483.1980M} {483, 1980}

\bibitem[\protect\citeauthoryear{Monsalve et~al.,}{Monsalve et~al.}{2024}]{monsalve2024mapper}
Monsalve R.,  et~al., 2024, MNRAS, 530, 4125

\bibitem[\protect\citeauthoryear{Morales, Hazelton, Sullivan  \& Beardsley}{Morales et~al.}{2012}]{morales2012four}
Morales M.~F.,  Hazelton B.,  Sullivan I.,   Beardsley A.,  2012, ApJ, 752, 137

\bibitem[\protect\citeauthoryear{Mouri~Sardarabadi \& Koopmans}{Mouri~Sardarabadi \& Koopmans}{2019}]{mouri2019quantifying}
Mouri~Sardarabadi A.,  Koopmans L.,  2019, MNRAS, 483, 5480

\bibitem[\protect\citeauthoryear{Mu{\~n}oz \& Loeb}{Mu{\~n}oz \& Loeb}{2018}]{munoz2018small}
Mu{\~n}oz J.~B.,  Loeb A.,  2018, Nature, 557, 684

\bibitem[\protect\citeauthoryear{Munshi et~al.,}{Munshi et~al.}{2024}]{munshi2024first}
Munshi S.,  et~al., 2024, A\&A, 681, A62

\bibitem[\protect\citeauthoryear{Munshi et~al.,}{Munshi et~al.}{2025a}]{munshi2025beyond}
Munshi S.,  et~al., 2025a, A\&A, 693, A276

\bibitem[\protect\citeauthoryear{Munshi et~al.,}{Munshi et~al.}{2025b}]{munshi2025near}
Munshi S.,  et~al., 2025b, A\&A, 697, A203

\bibitem[\protect\citeauthoryear{Murray~Steven, Bradley, Andrei, Mu{\~n}oz~Julian, Yuxiang, Jaehong  \& Watkinson~Catherine}{Murray~Steven et~al.}{2020}]{murray202021cmfast}
Murray~Steven G.,  Bradley G.,  Andrei M.,  Mu{\~n}oz~Julian B.,  Yuxiang Q.,  Jaehong P.,   Watkinson~Catherine A.,  2020, Open Source Softw, 5, 2582

\bibitem[\protect\citeauthoryear{Nunhokee et~al.,}{Nunhokee et~al.}{2025}]{nunhokee2025limits}
Nunhokee C.,  et~al., 2025, arXiv preprint arXiv:2505.09097

\bibitem[\protect\citeauthoryear{Offringa}{Offringa}{2016}]{offringa2016compression}
Offringa A.,  2016, A\&A, 595, A99

\bibitem[\protect\citeauthoryear{Offringa, Van De~Gronde  \& Roerdink}{Offringa et~al.}{2012}]{offringa2012morphological}
Offringa A.,  Van De~Gronde J.,   Roerdink J.,  2012, A\&A, 539, A95

\bibitem[\protect\citeauthoryear{Offringa et~al.,}{Offringa et~al.}{2014}]{offringa2014wsclean}
Offringa A.,  et~al., 2014, MNRAS, 444, 606

\bibitem[\protect\citeauthoryear{Offringa, Mertens, Van~der Tol, Veenboer, Gehlot, Koopmans  \& Mevius}{Offringa et~al.}{2019}]{offringa2019precision}
Offringa A.,  Mertens F.,  Van~der Tol S.,  Veenboer B.,  Gehlot B.,  Koopmans L.,   Mevius M.,  2019, A\&A, 631, A12

\bibitem[\protect\citeauthoryear{Paciga et~al.,}{Paciga et~al.}{2013}]{paciga2013simulation}
Paciga G.,  et~al., 2013, MNRAS, 433, 639

\bibitem[\protect\citeauthoryear{Pandey, Koopmans, Tiesinga, Albers  \& Koers}{Pandey et~al.}{2020}]{pandey2020integrated}
Pandey V.,  Koopmans L.,  Tiesinga E.,  Albers W.,   Koers H.,  2020, Astronomical Data Analysis Software and Systems XXIX, 527, 473

\bibitem[\protect\citeauthoryear{Patil et~al.,}{Patil et~al.}{2016}]{patil2016systematic}
Patil A.~H.,  et~al., 2016, MNRAS, 463, 4317

\bibitem[\protect\citeauthoryear{Patil et~al.,}{Patil et~al.}{2017}]{patil2017upper}
Patil A.,  et~al., 2017, ApJ, 838, 65

\bibitem[\protect\citeauthoryear{Philip et~al.,}{Philip et~al.}{2019}]{philip2019probing}
Philip L.,  et~al., 2019, Journal of Astronomical Instrumentation, 8, 1950004

\bibitem[\protect\citeauthoryear{{Planck Collaboration} et~al.,}{{Planck Collaboration} et~al.}{2014}]{2014A&A...571A..16P}
{Planck Collaboration} et~al., 2014, \mn@doi [\aap] {10.1051/0004-6361/201321591}, \href {https://ui.adsabs.harvard.edu/abs/2014A&A...571A..16P} {571, A16}

\bibitem[\protect\citeauthoryear{{Planck Collaboration} et~al.}{{Planck Collaboration} et~al.}{2016}]{planck2016planck}
{Planck Collaboration} et~al., 2016, A\&A, 594, A13

\bibitem[\protect\citeauthoryear{Pritchard \& Loeb}{Pritchard \& Loeb}{2012}]{pritchard201221}
Pritchard J.~R.,  Loeb A.,  2012, Reports on Progress in Physics, 75, 086901

\bibitem[\protect\citeauthoryear{{Reis}, {Fialkov}  \& {Barkana}}{{Reis} et~al.}{2020}]{Reis2020}
{Reis} I.,  {Fialkov} A.,   {Barkana} R.,  2020, \mn@doi [\mnras] {10.1093/mnras/staa3091}, \href {https://ui.adsabs.harvard.edu/abs/2020MNRAS.499.5993R} {499, 5993}

\bibitem[\protect\citeauthoryear{{Rengelink}, {Tang}, {de Bruyn}, {Miley}, {Bremer}, {Roettgering}  \& {Bremer}}{{Rengelink} et~al.}{1997}]{1997A&AS..124..259R}
{Rengelink} R.~B.,  {Tang} Y.,  {de Bruyn} A.~G.,  {Miley} G.~K.,  {Bremer} M.~N.,  {Roettgering} H.~J.~A.,   {Bremer} M.~A.~R.,  1997, \mn@doi [\aaps] {10.1051/aas:1997358}, \href {https://ui.adsabs.harvard.edu/abs/1997A&AS..124..259R} {124, 259}

\bibitem[\protect\citeauthoryear{Semelin}{Semelin}{2016}]{semelin2016detailed}
Semelin B.,  2016, Monthly Notices of the Royal Astronomical Society, 455, 962

\bibitem[\protect\citeauthoryear{Semelin, Combes  \& Baek}{Semelin et~al.}{2007}]{semelin2007lyman}
Semelin B.,  Combes F.,   Baek S.,  2007, Astronomy \& Astrophysics, 474, 365

\bibitem[\protect\citeauthoryear{Semelin, Eames, Bolgar  \& Caillat}{Semelin et~al.}{2017}]{semelin201721ssd}
Semelin B.,  Eames E.,  Bolgar F.,   Caillat M.,  2017, Monthly Notices of the Royal Astronomical Society, 472, 4508

\bibitem[\protect\citeauthoryear{{Semelin} et~al.,}{{Semelin} et~al.}{2023}]{semelin2023}
{Semelin} B.,  et~al., 2023, \mn@doi [\aap] {10.1051/0004-6361/202244722}, \href {https://ui.adsabs.harvard.edu/abs/2023A&A...672A.162S} {672, A162}

\bibitem[\protect\citeauthoryear{{Sikder}, {Barkana}, {Fialkov}  \& {Reis}}{{Sikder} et~al.}{2024a}]{sikder2024}
{Sikder} S.,  {Barkana} R.,  {Fialkov} A.,   {Reis} I.,  2024a, \mn@doi [\mnras] {10.1093/mnras/stad3847}, \href {https://ui.adsabs.harvard.edu/abs/2024MNRAS.52710975S} {527, 10975}

\bibitem[\protect\citeauthoryear{{Sikder}, {Barkana}  \& {Fialkov}}{{Sikder} et~al.}{2024b}]{sikder24}
{Sikder} S.,  {Barkana} R.,   {Fialkov} A.,  2024b, \mn@doi [\apjl] {10.3847/2041-8213/ad5c5f}, \href {https://ui.adsabs.harvard.edu/abs/2024ApJ...970L..25S} {970, L25}

\bibitem[\protect\citeauthoryear{Singh et~al.,}{Singh et~al.}{2022}]{singh2022detection}
Singh S.,  et~al., 2022, Nature Astronomy, pp 1--11

\bibitem[\protect\citeauthoryear{Sokolowski et~al.,}{Sokolowski et~al.}{2015}]{sokolowski2015bighorns}
Sokolowski M.,  et~al., 2015, Publications of the Astronomical Society of Australia, 32, e004

\bibitem[\protect\citeauthoryear{Speagle}{Speagle}{2020}]{speagle2020dynesty}
Speagle J.~S.,  2020, MNRAS, 493, 3132

\bibitem[\protect\citeauthoryear{Trott et~al.,}{Trott et~al.}{2020}]{trott2020deep}
Trott C.~M.,  et~al., 2020, MNRAS, 493, 4711

\bibitem[\protect\citeauthoryear{Van{\'\i}{\v{c}}ek}{Van{\'\i}{\v{c}}ek}{1969}]{vanivcek1969approximate}
Van{\'\i}{\v{c}}ek P.,  1969, Astrophysics and Space Science, 4, 387

\bibitem[\protect\citeauthoryear{{Visbal}, {Barkana}, {Fialkov}, {Tseliakhovich}  \& {Hirata}}{{Visbal} et~al.}{2012}]{visbal12}
{Visbal} E.,  {Barkana} R.,  {Fialkov} A.,  {Tseliakhovich} D.,   {Hirata} C.~M.,  2012, \mn@doi [\nat] {10.1038/nature11177}, \href {https://ui.adsabs.harvard.edu/abs/2012Natur.487...70V} {487, 70}

\bibitem[\protect\citeauthoryear{Yoshiura et~al.,}{Yoshiura et~al.}{2021}]{yoshiura2021new}
Yoshiura S.,  et~al., 2021, MNRAS, 505, 4775

\bibitem[\protect\citeauthoryear{{Yung}, {Somerville}  \& {Iyer}}{{Yung} et~al.}{2025}]{2025arXiv250418618Y}
{Yung} L.~Y.~A.,  {Somerville} R.~S.,   {Iyer} K.~G.,  2025, \mn@doi [arXiv e-prints] {10.48550/arXiv.2504.18618}, \href {https://ui.adsabs.harvard.edu/abs/2025arXiv250418618Y} {p. arXiv:2504.18618}

\bibitem[\protect\citeauthoryear{{Zahn}, {Mesinger}, {McQuinn}, {Trac}, {Cen}  \& {Hernquist}}{{Zahn} et~al.}{2011}]{2011MNRAS.414..727Z}
{Zahn} O.,  {Mesinger} A.,  {McQuinn} M.,  {Trac} H.,  {Cen} R.,   {Hernquist} L.~E.,  2011, \mn@doi [\mnras] {10.1111/j.1365-2966.2011.18439.x}, \href {https://ui.adsabs.harvard.edu/abs/2011MNRAS.414..727Z} {414, 727}

\bibitem[\protect\citeauthoryear{Zarka, Girard, Tagger  \& Denis}{Zarka et~al.}{2012}]{zarka2012lss}
Zarka P.,  Girard J.,  Tagger M.,   Denis L.,  2012, in SF2A-2012: Proceedings of the Annual meeting of the French Society of Astronomy and Astrophysics. pp 687--694

\bibitem[\protect\citeauthoryear{Zarka et~al.,}{Zarka et~al.}{2015}]{zarka2015nenufar}
Zarka P.,  et~al., 2015, in 2015 International Conference on Antenna Theory and Techniques (ICATT). pp~1--6

\bibitem[\protect\citeauthoryear{Zarka et~al.,}{Zarka et~al.}{2020}]{zarka2020low}
Zarka P.,  et~al., 2020, in URSI GASS 2020.

\bibitem[\protect\citeauthoryear{de Lera~Acedo et~al.,}{de~Lera~Acedo et~al.}{2022}]{de2022reach}
de Lera~Acedo E.,  et~al., 2022, Nature Astronomy, 6, 984

\bibitem[\protect\citeauthoryear{van Haarlem et~al.,}{van Haarlem et~al.}{2013}]{van2013lofar}
van Haarlem M.~P.,  et~al., 2013, A\&A, 556, A2

\makeatother
\end{thebibliography}
%%%%%%%%%%%%%%%%%%%%%%%%%%%%%%%%%%%%%%%%%%%%%%%%%%
\cleardoublepage
\newpage
\appendix
\section{Statistics based segment flagging}\label{sec:std_residuals}
\begin{figure}
    \centering
    \includegraphics[width=\columnwidth]{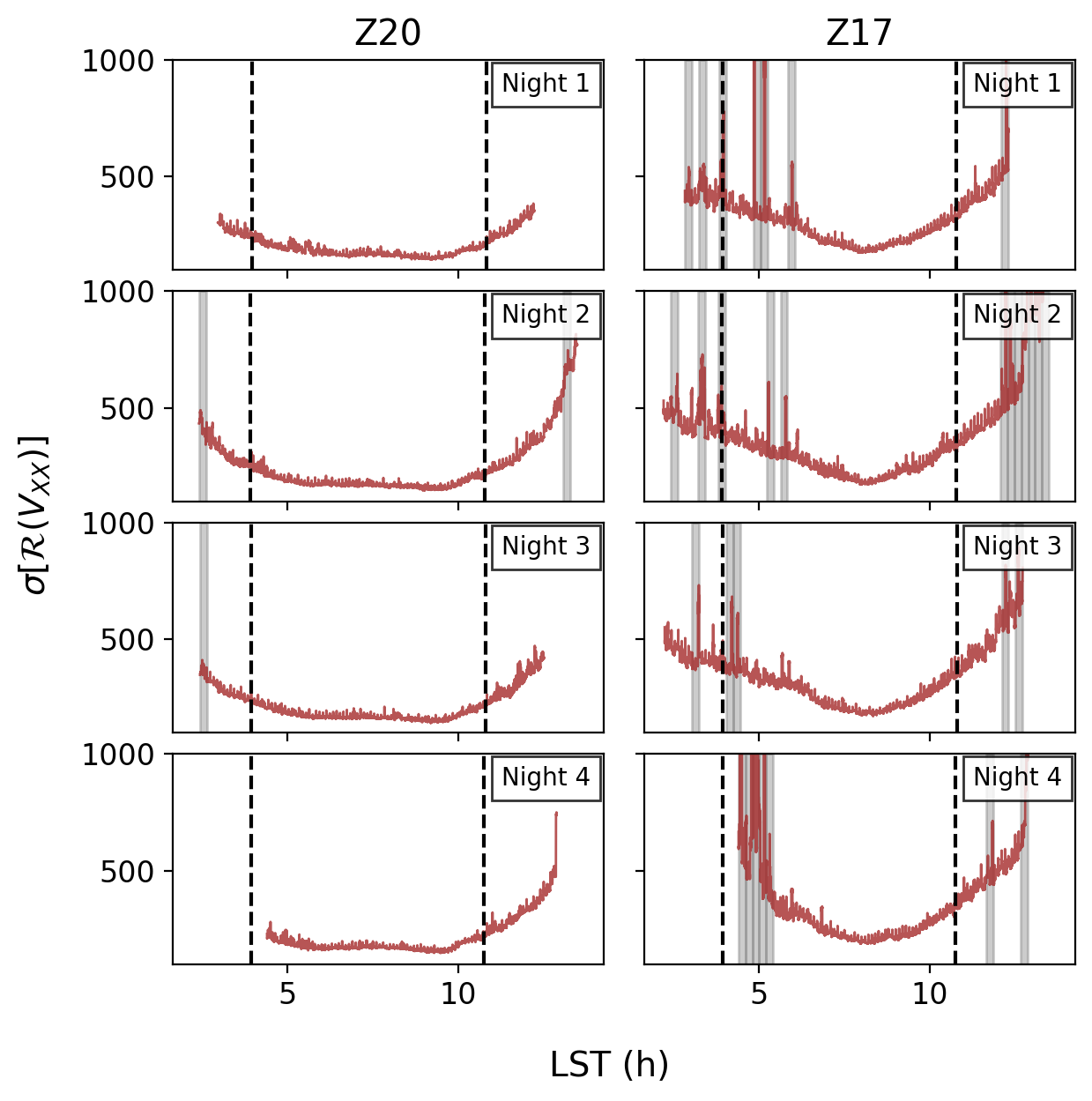}
    \caption{Standard deviation of residual visibilities after A-team subtraction and DI correction. The vertical dashed lines indicate the LSTs where the phase centre crosses $50^{\circ}$ elevation. The grey bands indicate the time segments that are flagged before power spectrum estimation.}
    \label{fig:std_residuals}
\end{figure}

Before power spectrum estimation, we perform a round of flagging of a selection of 12 min segments based on visibility statistics. We compute the standard deviation of the residual visibilities after A-team subtraction and DI correction as a function of time. Fig.~\ref{fig:std_residuals} shows this for all four nights of the two spectral windows. The decrease in variance during the middle of the observation is caused by the increased sensitivity as the target field moves closer to the zenith. The vertical dashed lines indicate the LST range where the target field is above $50^{\circ}$ elevation, which is ultimately used in upper limits estimation. We find that Z20 is significantly cleaner than Z17, which exhibits strong peaks in the residual standard deviation at different time segments for each night. We first detrend the standard deviation curve using a running mean and identify and flag the time segments containing 3$\sigma$ outliers. The time segments that were flagged in this step are indicated with grey vertical bands. Flagging these time segments completely before power spectrum estimation significantly reduces the excess variance for Z17.

\section{Posterior distribution of GPR hyperparameters}\label{sec:gpr_post}
\begin{figure}
    \centering
    \includegraphics[width=\columnwidth]{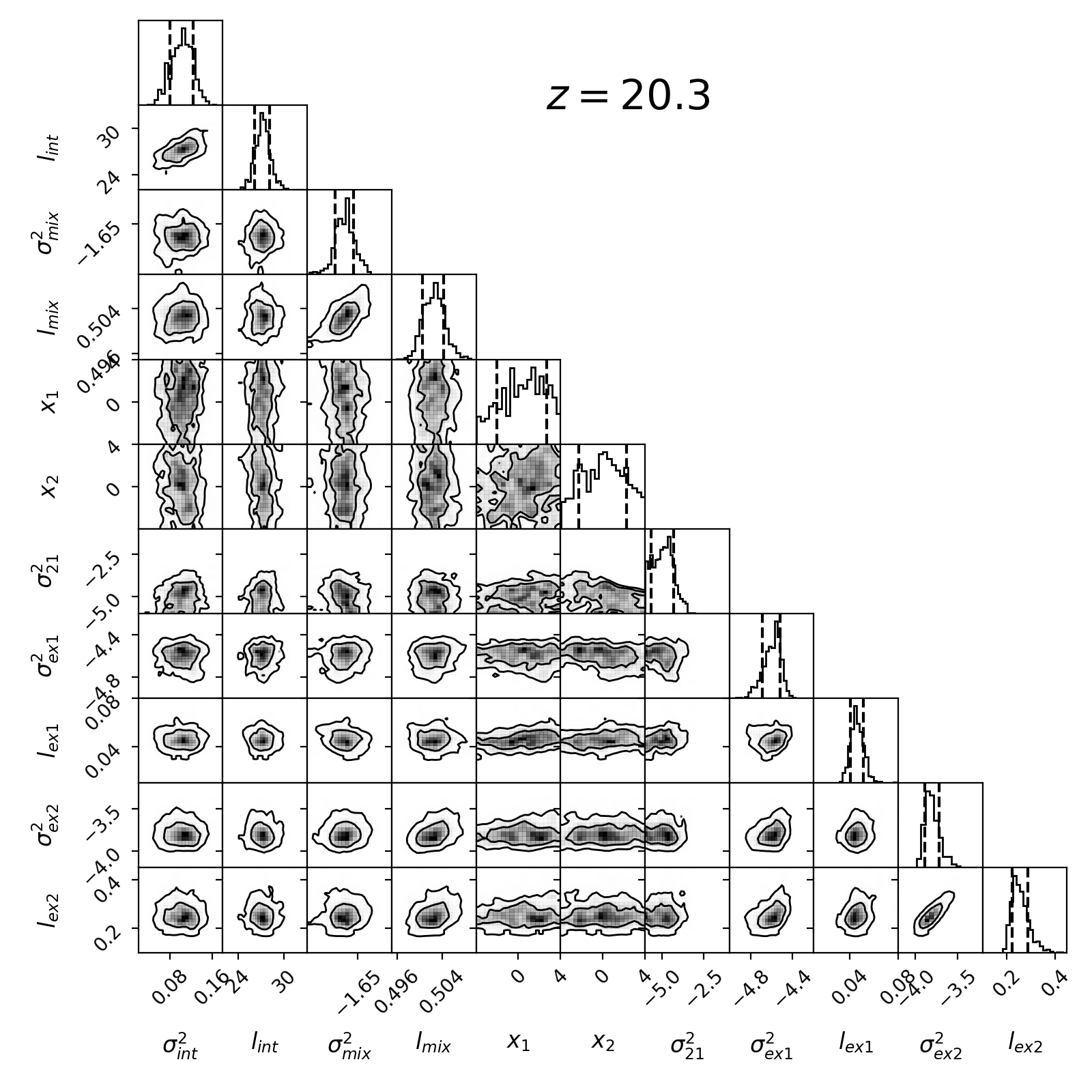}
    \includegraphics[width=\columnwidth]{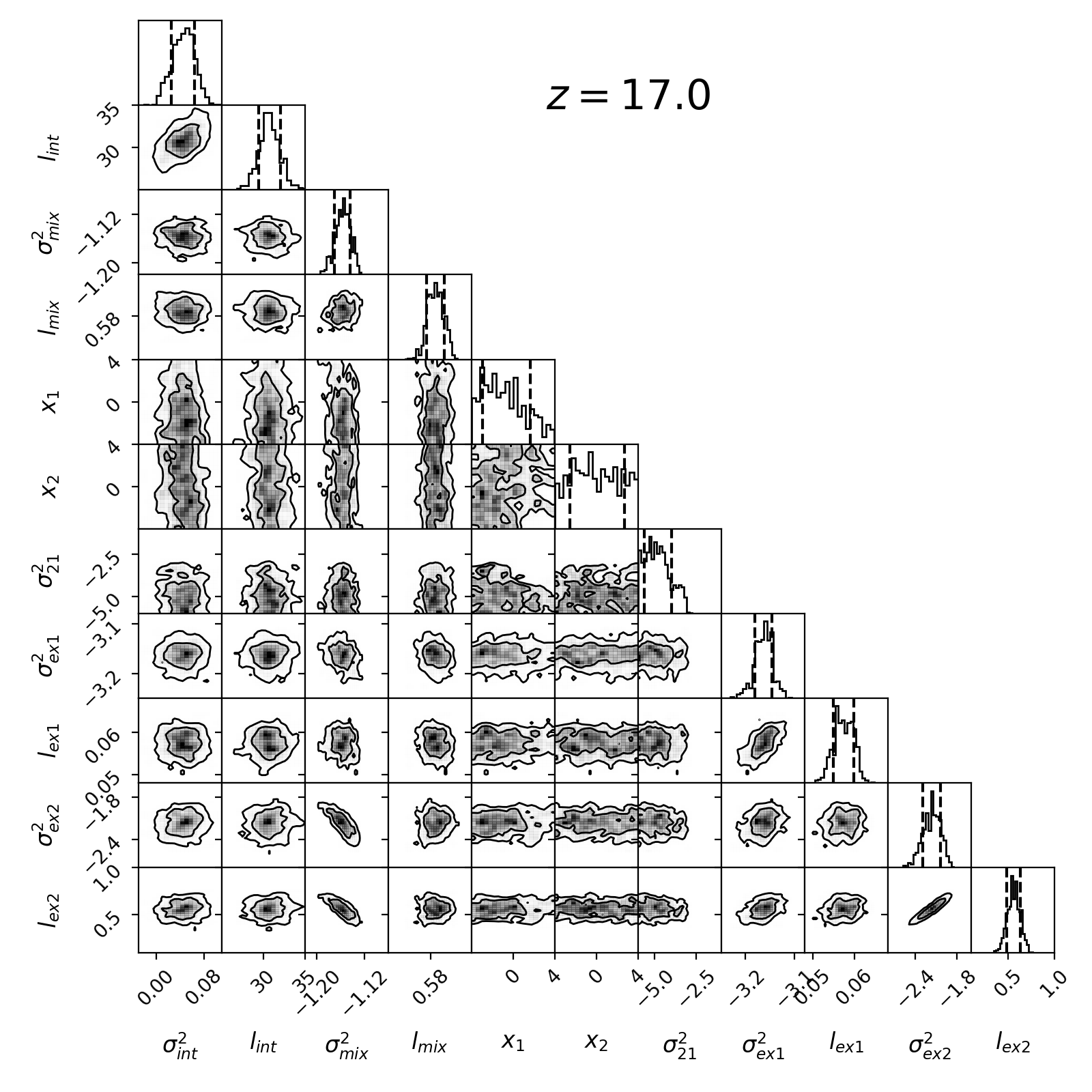}
    \caption{Corner plots showing the posterior distribution of the Gaussian process model hyperparameters for the two redshift bins. The contours in the 2D histograms show the 68\% and 95\% confidence intervals, and the vertical dashed lines in the 1D histograms indicate the 68\% confidence intervals.}
    \label{fig:corner}
\end{figure}

The posterior distribution of the hyperparameters in the Gaussian process covariance model used in both redshift bins are presented in Fig.~\ref{fig:corner}.

% Don't change these lines
\bsp	% typesetting comment
\label{lastpage}
\end{document}